\documentclass[aps,prc,superscriptaddress,showpacs,floatfix]{revtex4}

\usepackage{amssymb}
\usepackage{amsmath}
\usepackage{graphicx}
\usepackage[normalem]{ulem}
\usepackage{color}
\usepackage{multirow}
\usepackage{verbatim}
\begin{document}

\title{
Two-particle angular correlations in $pp$ and p-Pb collisions at LHC energies from a multi-phase transport model
}

\author{Liu-Yao Zhang}
\affiliation{Shanghai Institute of Applied Physics, Chinese Academy of Sciences, Shanghai 201800, China}
\affiliation{University of Chinese Academy of Sciences, Beijing 100049, China}
\author{Jin-Hui Chen}
\affiliation{Shanghai Institute of Applied Physics, Chinese Academy of Sciences, Shanghai 201800, China}
\author {Zi-Wei Lin}
   \affiliation{Key Laboratory of Quarks and Lepton Physics (MOE) and
     Institute of Particle Physics, Central China Normal University, Wuhan 430079, 
     China}
   \affiliation{Department of Physics, East Carolina University,
     Greenville, North Carolina 27858, USA}
\author{Yu-Gang Ma}
\affiliation{Shanghai Institute of Applied Physics, Chinese Academy of Sciences, Shanghai 201800, China}
\author{Song Zhang}
\affiliation{Shanghai Institute of Applied Physics, Chinese Academy of Sciences, Shanghai 201800, China}

\date{\today}

\begin{abstract}

We apply a multi-phase transport (AMPT) model to study two-particle angular correlations in $pp$ collisions at $\sqrt{s}= 7$ TeV. Besides being able to describe the angular correlation functions of meson-meson pairs, a large improvement for the angular correlations of baryon-baryon and antibaryon-antibaryon is achieved. We further find that the AMPT model with new quark coalescence provides an even better description on the anti-correlation feature of baryon-baryon correlations observed in the experiments. We also extend the study to p-Pb collisions at  $\sqrt{s}= 5.02$ TeV and obtained similar results. These results help us better understand the particle production mechanism in $pp$ and p-Pb collisions at LHC energies. 
\end{abstract}
\maketitle

\section{Introduction}
The Relativistic Heavy Ion Collider at BNL and the Large Hadron Collider at CERN create a hot and dense matter similar to the early universe microseconds after the Big Bang. Such facilities enable us to study strongly interacting matter at the extreme temperature in the laboratory~\cite{Shuryak:2017,Chen:2018}. Analysis of multi-particle correlations is a powerful tool in exploring the underlying mechanism of particle production in hot QCD matter~\cite{M.Aaboud:2017,C.Aidala:2017,L.Adamczyk:2015,C.Adare:2015,L.Adamczyk:2014,S.Chatrchyan:2011,V.Khachatryan:2010,B.Alver:2010_1,B.Alver:2010,B.I.Abelev:2009,R.E.Ansorge:1988}. For example, inclusive two-particle $\Delta\eta$-$\Delta\phi$ correlations have been found to include two components: direct two-particle correlations and an effective "long-range" correlation due to event-by-event fluctuations of the overall particle multiplicity~\cite{M.Aaboud:2017,C.Aidala:2017,L.Adamczyk:2015,C.Adare:2015,L.Adamczyk:2014,S.Chatrchyan:2011,V.Khachatryan:2010,B.Alver:2010_1,B.Alver:2010,B.I.Abelev:2009,R.E.Ansorge:1988}. For small systems such as $pp$ collisions, one physics mechanism underlying all correlations is the global conservation of energy and momentum as well as the net strangeness, baryon number, and electric charge \cite{J.Adam:2017}. Data in p-Pb \cite{S.Chatrchyan:2013_1,G.Aad:2013,B.Abelev:2013} and $pp$ collisions~\cite{B.Alver:2007,V.Khachatryan:2010,J.Adam:2017} show that two particles separated by many units of pseudorapidity prefer to have similar azimuthal angles, thus their correlation function is peaked at $\Delta\phi$ = 0. The mechanism of the peak is dominated by effects associated with the fragmentation of hard-scattered partons from the same minijet~\cite{J.Adam:2017}, resonance decays, and the femtoscopic correlation~\cite{E.P.Rogochaya:2017,Alice:2018-3}. Exactly the same phenomenon was observed in heavy-ion collisions \cite{S.Chatrchyan:2011}, where the anisotropic flows are believed to originate from the hydrodynamical evolution~\cite{CMS-2018} or the anisotropic parton escape in transport models~\cite{L.H:2016,Z.W.Lin:2016}.

Recently, the ALICE Collaboration measured two-particle correlations in $pp$ collisions for low $p_T$ particles (below 2.5 GeV/c) at $\sqrt{s}$ = 7 TeV; it found a pronounced near-side depression in baryon-baryon correlations, which did not show up in meson-meson or  baryon-antibaryon correlation functions~\cite{J.Adam:2017}. The ALICE Collaboration also  compared with several Monte Carlo (MC) model calculations, where the models are unable to reproduce even qualitatively the depletion in the data. This may suggest the need to modify the particle production mechanism or the fragmentation functions in the Monte Carlo models~\cite{J.Adam:2017}. Here we perform a study of two-particle correlations with a different model: a multi-phase transport (AMPT) model. We study the dynamical evolution of the  correlations in different stages including partonic interactions, hadronization, and hadronic interactions; we then investigate the underlying physics responsible for the depletion of the baryon-baryon correlations in the near side.

\section{Model and methodology}
Both the string melting version and default version of the AMPT model \cite{Z.W.Lin:2005} are applied in this work. The AMPT model was developed to simulate heavy-ion collisions in a wide colliding energy range from AGS to LHC. It consists of four main components: the initial conditions, partonic rescatterings, the conversion from partonic matter into hadronic matter, and hadronic interactions. The initial conditions, which include the spatial and momentum distributions of minijet partons and soft excited strings, are obtained from the HIJING model \cite{M.Gyulassy:1994}. In the string melting version \cite{Lin:2001zk}, excited strings are melt into quarks and antiquarks. Scatterings among the partons are modeled by Zhang's parton cascade model (ZPC) \cite{B.Zhang:1998}, which includes two-body scatterings with the cross section obtained from the perturbative QCD calculation with a screening mass. In the default AMPT model (denoted as AMPT-Default), most of the energy produced in the overlap volume of a heavy ion collision is in hadronic strings and thus not included in the parton cascade. In the string melting version of AMPT (denoted as AMPT-Melting), on the other hand, all excited hadronic strings in the overlap volume are converted into partons. In the AMPT-Default model, partons are recombined with their parent string when they stop interacting, and the resulting strings are converted to hadrons via the Lund string fragmentation \cite{B.Andersson:1983}. In the AMPT-Melting model, a spatial quark coalescence model is used to combine partons into hadrons. Dynamics of the subsequent hadronic matter is then described by the extended version of a relativistic transport (ART) model \cite{B.A.Li:1995}. So far the AMPT model has been often used in studies of heavy ion collisions. For example, it has been successful in describing multiple observables in relativistic heavy ion collisions at RHIC and LHC \cite{Z.W.Lin:2005, Z.W.Lin:2014,G.L.Ma:2016}, including pion HBT correlations \cite{Z.W.Lin:2002}, two-particle azimuthal angular correlations or longitudinal decorrelation~\cite{G.L.Ma:2006,J.Xu:2011,L.G.Pang:2015}. It has also been used to study the particle production mechanism~\cite{Y.J.Ye:2016,Liu:2017}. 

Many new phenomena have been observed at LHC energies, such as the large anisotropic flows developed at p-Pb collisions~\cite{S.Chatrchyan:2013,G.Aad:2013B,B.B.Abelev:2013} and even in high multiplicity $pp$ collisions~\cite{V.Khachatryan:2010,G.Aad:2016,V.Khachatryan:2016}, which indicate that a quark-gluon plasma have been developed in the small collision systems at LHC energies~\cite{L.H:2016,Song:2017,Adare:2017wlc,J.L.Nagle:2018}. In earlier studies, the AMPT version v1.26t5/v2.26t5 \cite{ampt} has been used to study $pp$ and p-Pb collisions at LHC energies. This version \cite{Z.W.Lin:2014,G.L.Ma:2016} describes well the long range azimuthal correlations present in small collision systems~\cite{G.L.Ma:2014,A.Bzdak:2014}. 
We use this version in our study, where the parton cross section is set to 1.5 mb~\cite{He:2017}.

In order to compare with experimental measurements of correlations between trigger and associated particles, we follow exactly the same analysis method as used by the ALICE Collaboration~\cite{J.Adam:2017}. The two-particle correlation function as a function of relative azimuthal angle $\Delta\phi$ and relative pseudorapidity $\Delta\eta$ between the particle pair of interest is defined as:
\begin{equation}
	C(\Delta\eta,\Delta\phi) = \frac{S(\Delta\eta,\Delta\phi)/N_{pairs}^{signal}}{B(\Delta\eta,\Delta\phi)/N_{pairs}^{mixed}},
\end{equation}
where $S$($\Delta\eta$,$\Delta\phi$) is the distribution of correlated pairs and $B$($\Delta\eta$,$\Delta\phi$) is the reference distribution reflecting the single-particle acceptance. $S$ is constructed from particle pairs coming from the same event: 
\begin{equation}
	S(\Delta\eta,\Delta\phi) = \frac{d^2N_{pairs}^{signal}}{d{\Delta\eta}d\Delta\phi},  
\end{equation}
where $N_{pairs}^{signal}$ is the number of particle pairs. $B$ is constructed using an event-mixing technique:
\begin{equation}
 B(\Delta\eta,\Delta\phi) = \frac{d^2N_{pairs}^{mixed}}{d{\Delta\eta}d\Delta\phi},
\end{equation} 
where $N_{pairs}^{mixed}$ is the number of particle pairs mixed from different events. In the AMPT model, particles from each event are combined with particles in the same event to build $S$($\Delta\eta$,$\Delta\phi$), while they are combined with particles from other events to build $B$($\Delta\eta$,$\Delta\phi$). Each event is mixed with 10 other events in this study to improve the statistical power of the reference estimation, and the impact parameter direction of the AMPT events is rotated randomly in the transverse plane for the $B$($\Delta\eta$,$\Delta\phi$) calculations. Another check by mixing event with similar event plane direction is carried out. The difference between different background reconstructions is negligible. In order to further investigate the correlation or anti-correlation of the particle pairs quantitatively, a one-dimensional (1-D) $\Delta\phi$ correlation function can be constructed from the 2-D correlation function by integrating over $\Delta\eta$ as
\begin{equation}
	C(\Delta\phi) = A\times\frac{\int S(\Delta\eta,\Delta\phi)d\Delta\eta}{\int B(\Delta\eta,\Delta\phi)d\Delta\eta},
\end{equation}
where the normalization constant $A$ is given by $N_{pairs}^{mixed}/N_{pairs}^{signal}$.

\section{Results \& discussions}
\subsection{Two particle correlations in AMPT model}
\begin{figure}[!htb]
	\centering
	\includegraphics[scale=0.20]{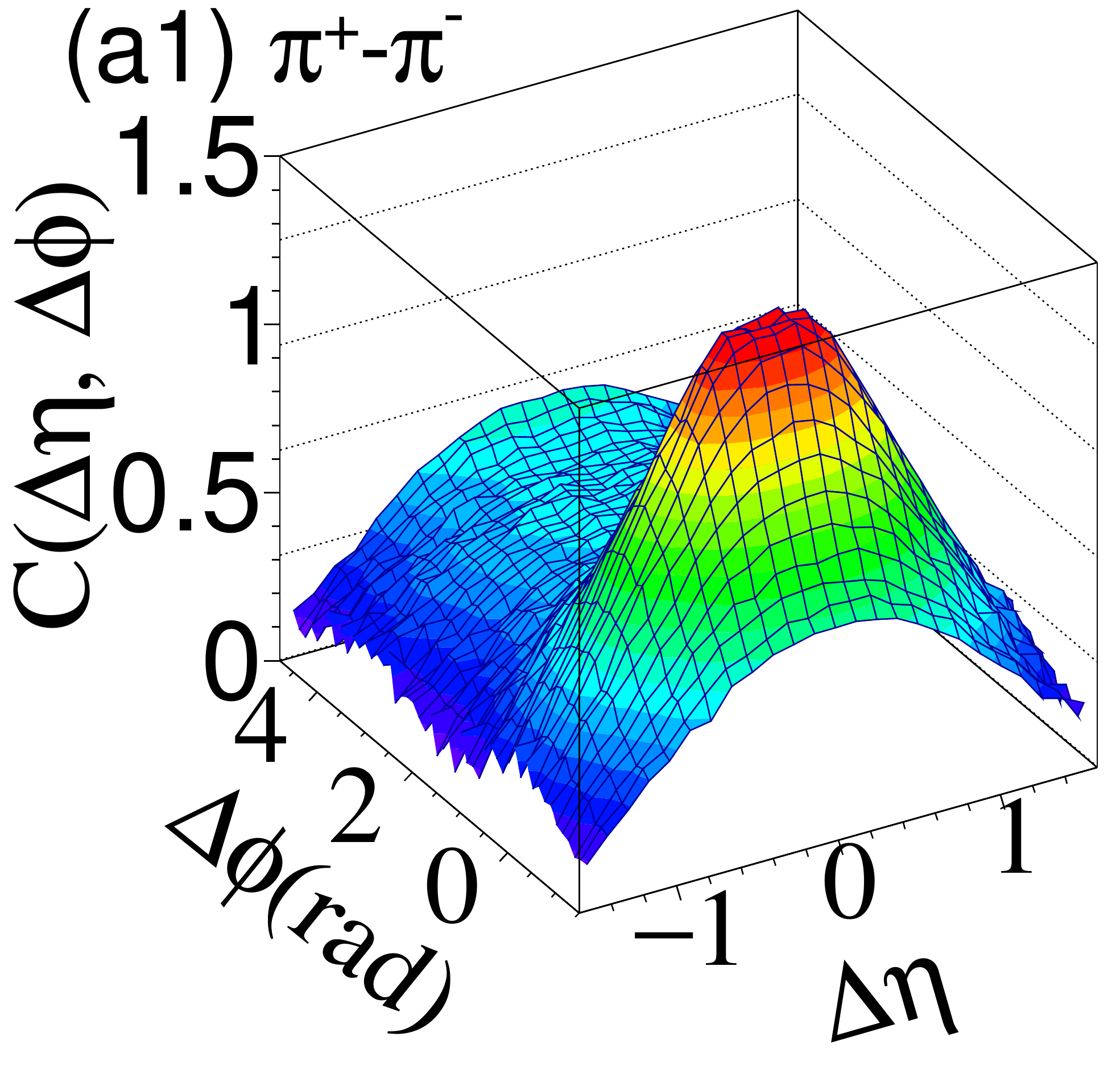}
	\includegraphics[scale=0.20]{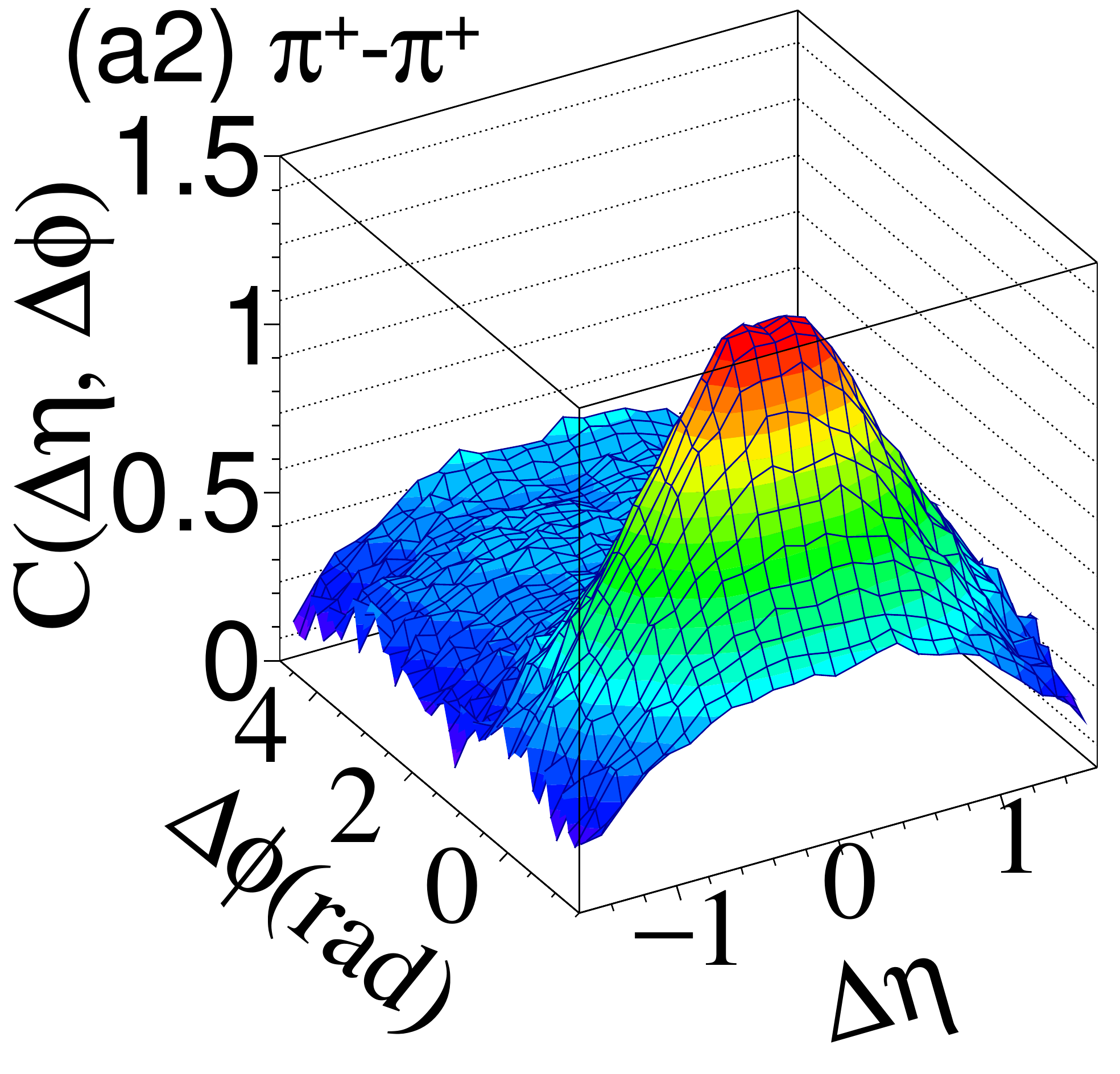}
	\includegraphics[scale=0.20]{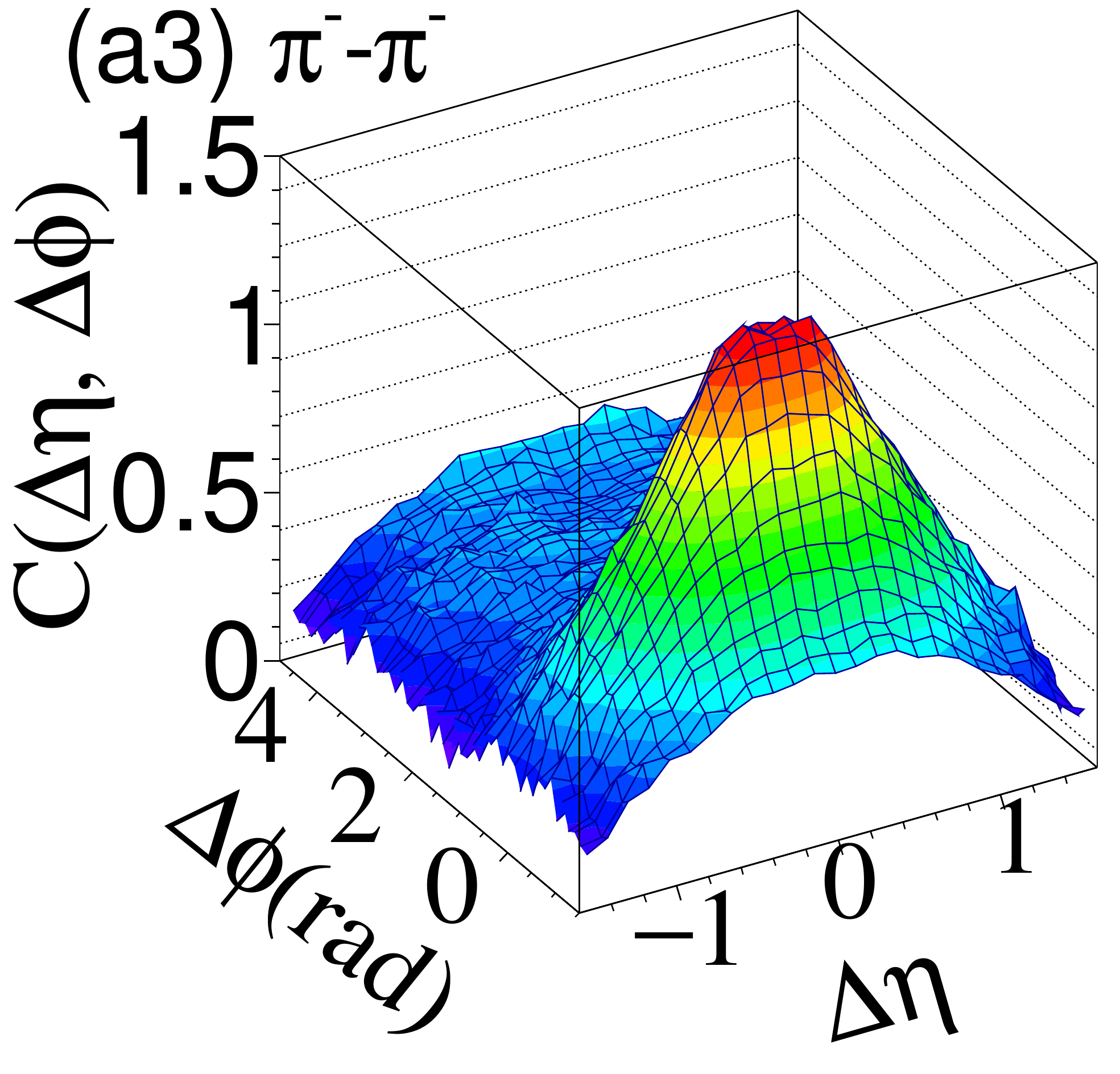}\\
	
	\includegraphics[scale=0.20]{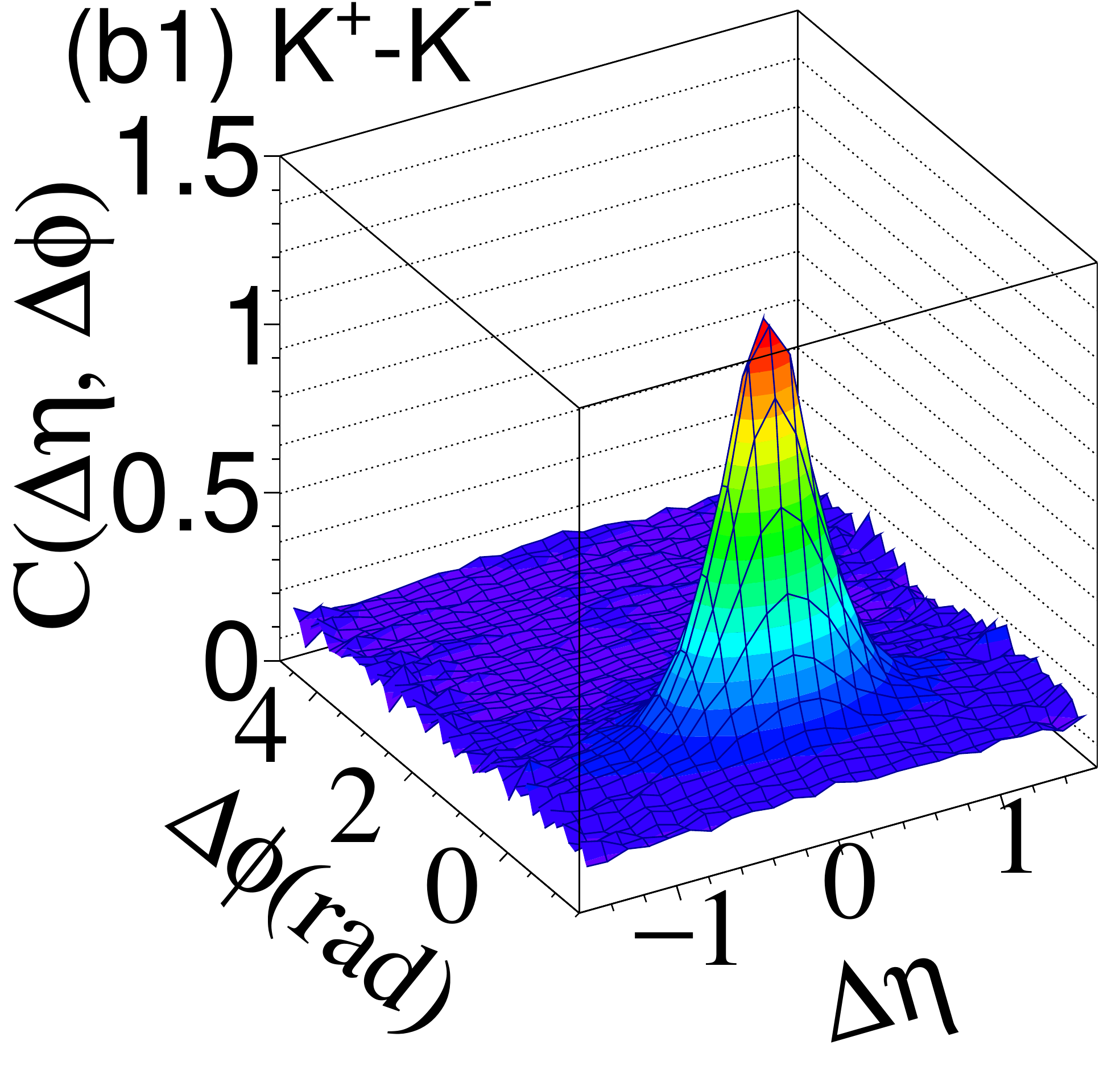}
	\includegraphics[scale=0.20]{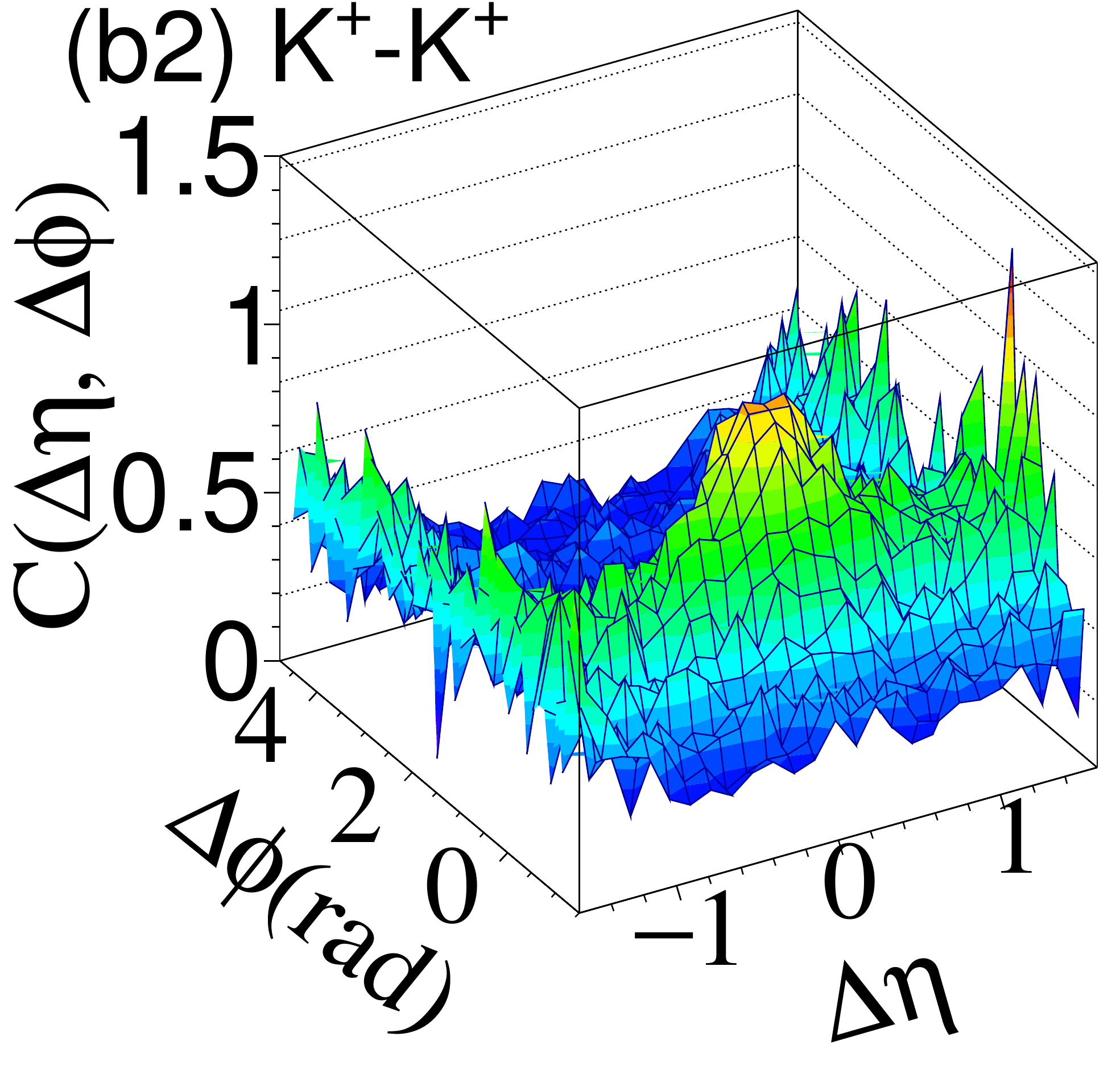}
	\includegraphics[scale=0.20]{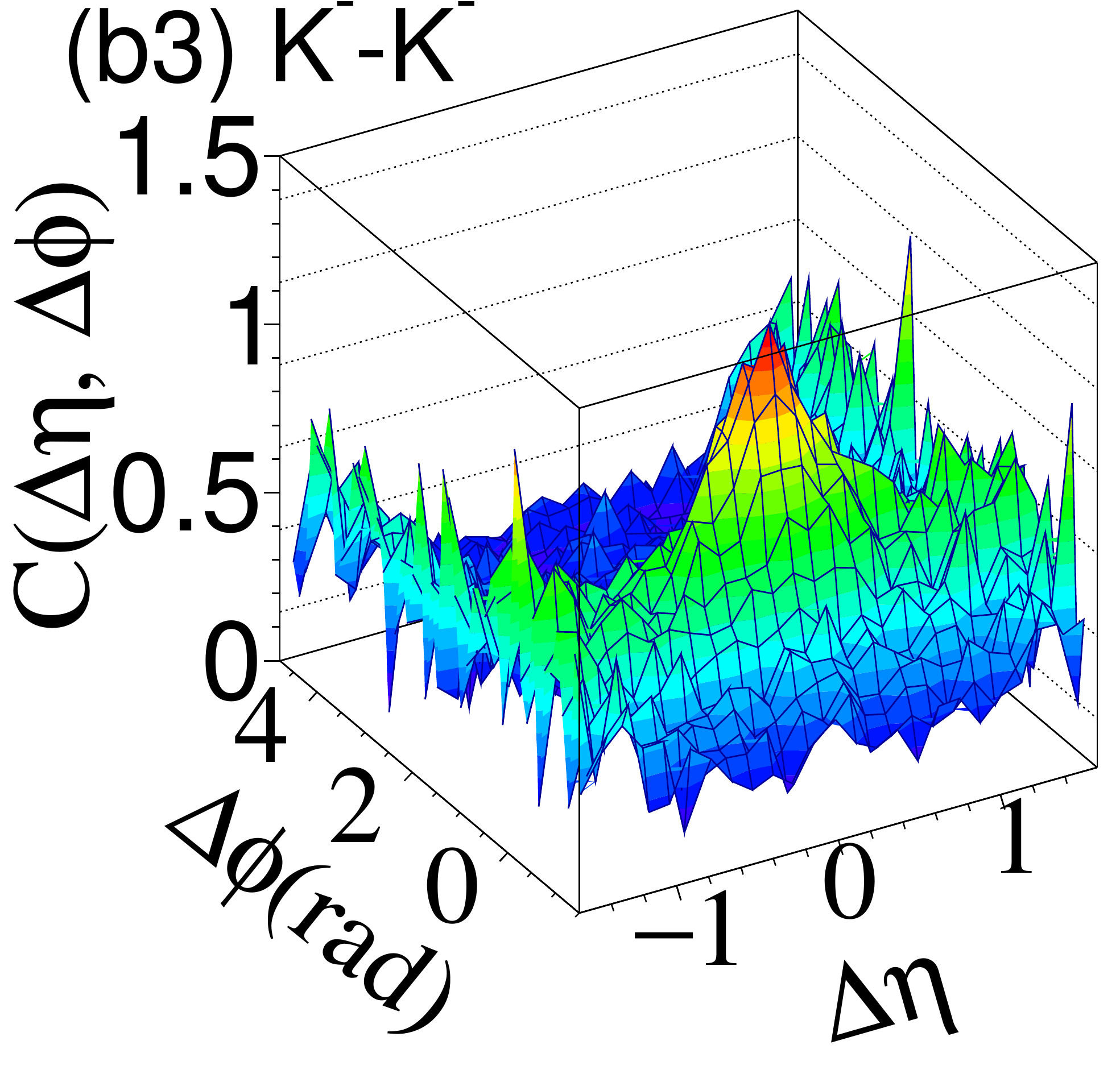}\\
	
	\includegraphics[scale=0.20]{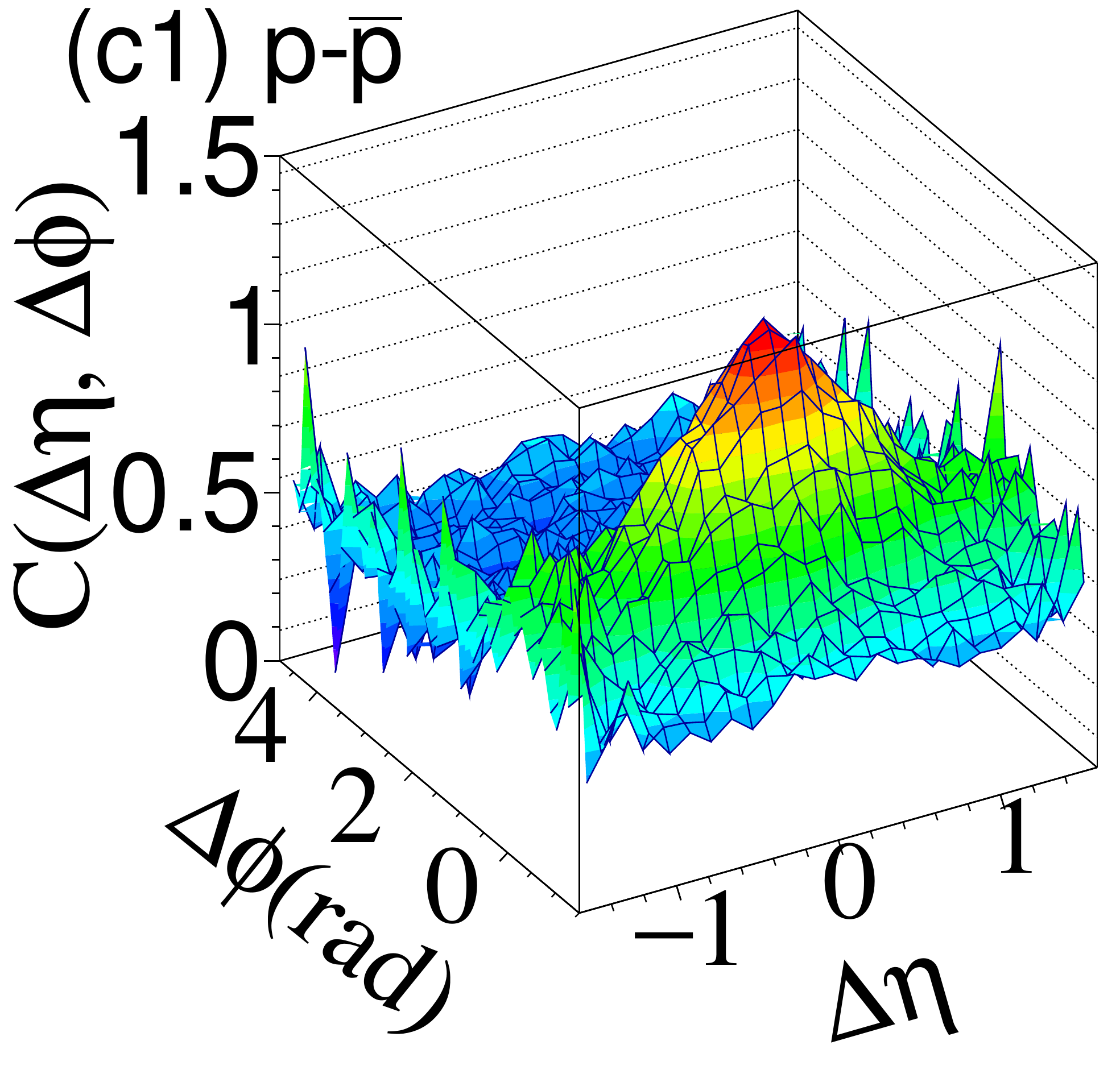}
	\includegraphics[scale=0.20]{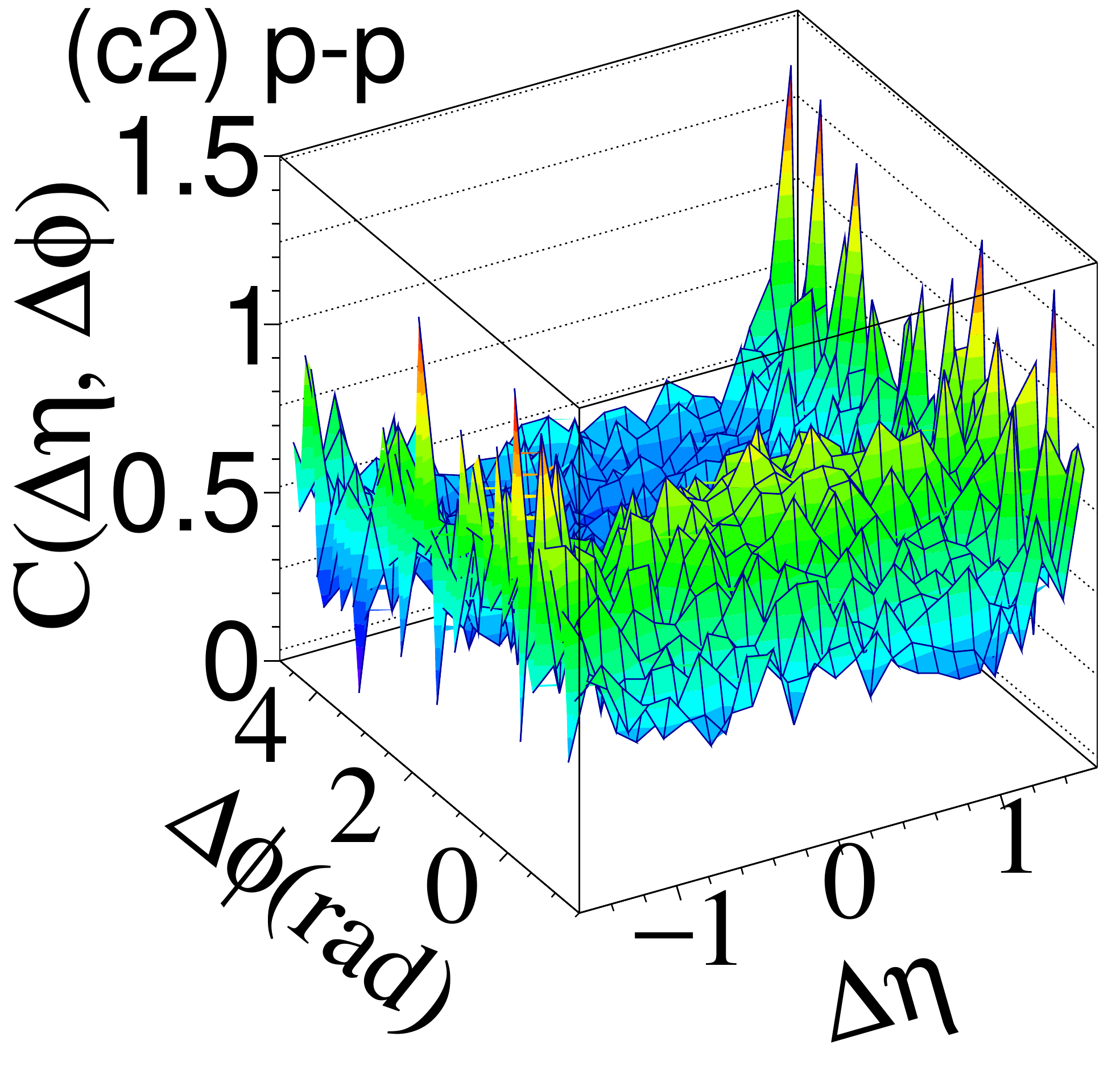}
	\includegraphics[scale=0.20]{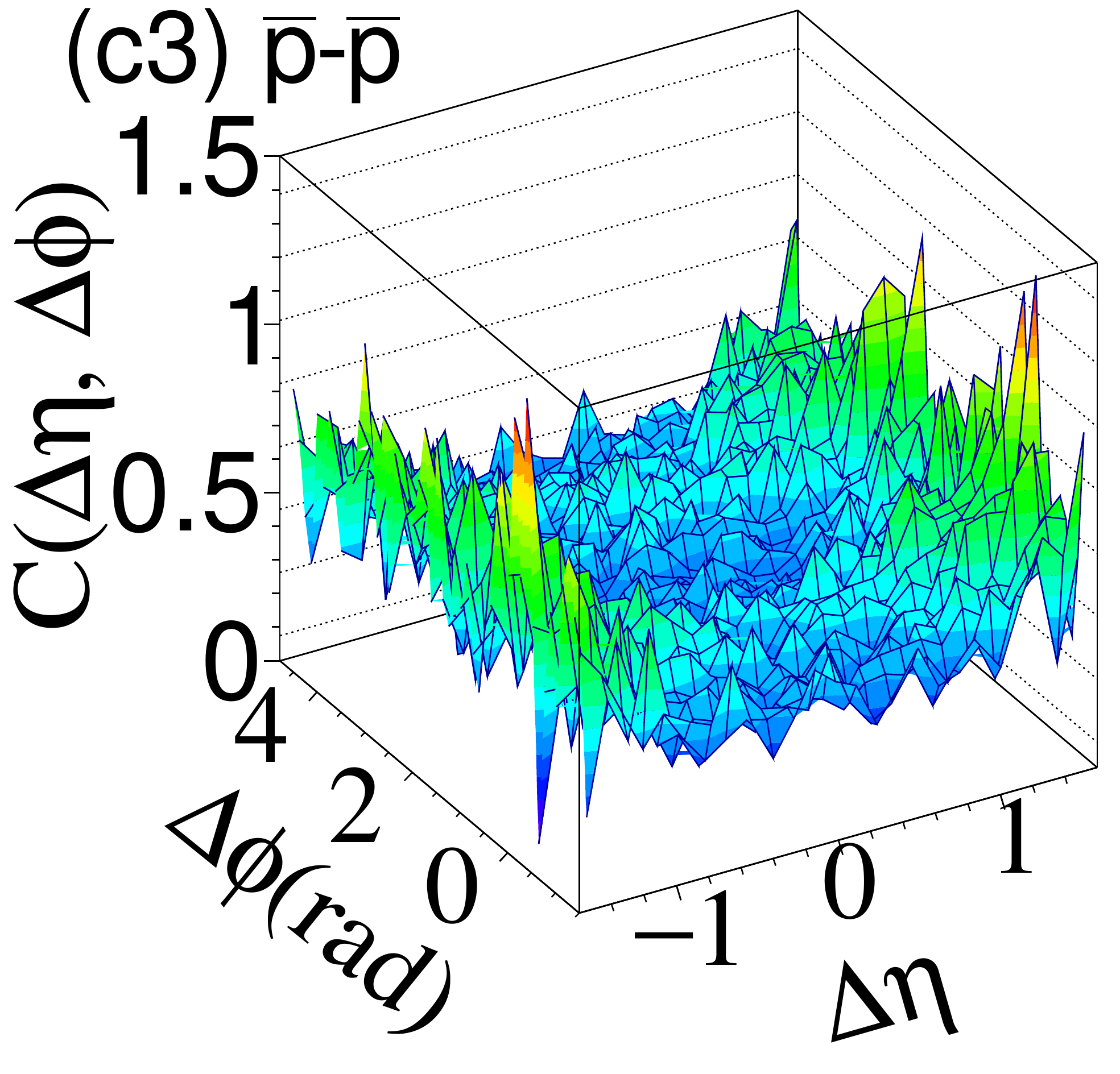}\\
   
        \includegraphics[scale=0.20]{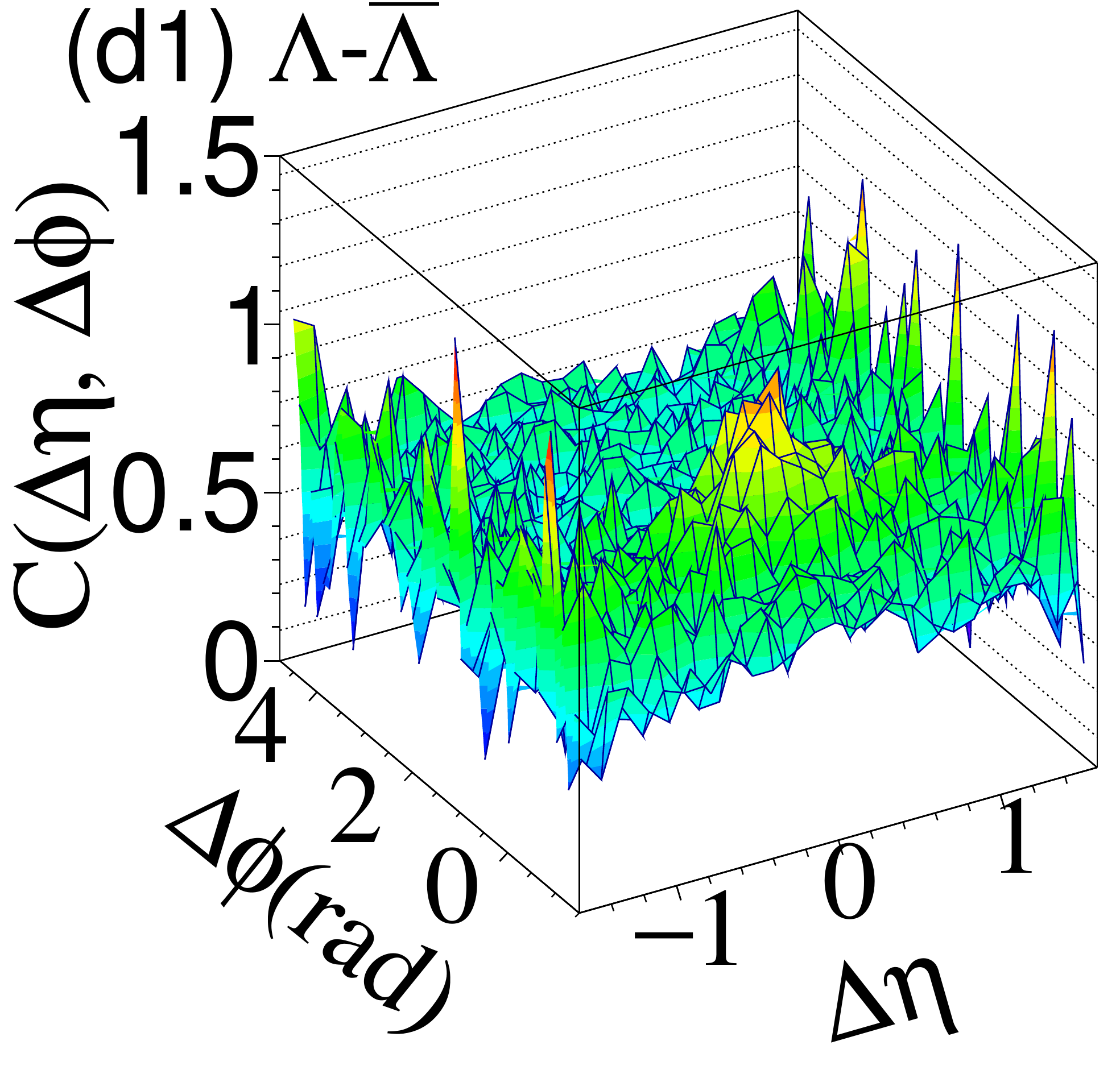}
        \includegraphics[scale=0.20]{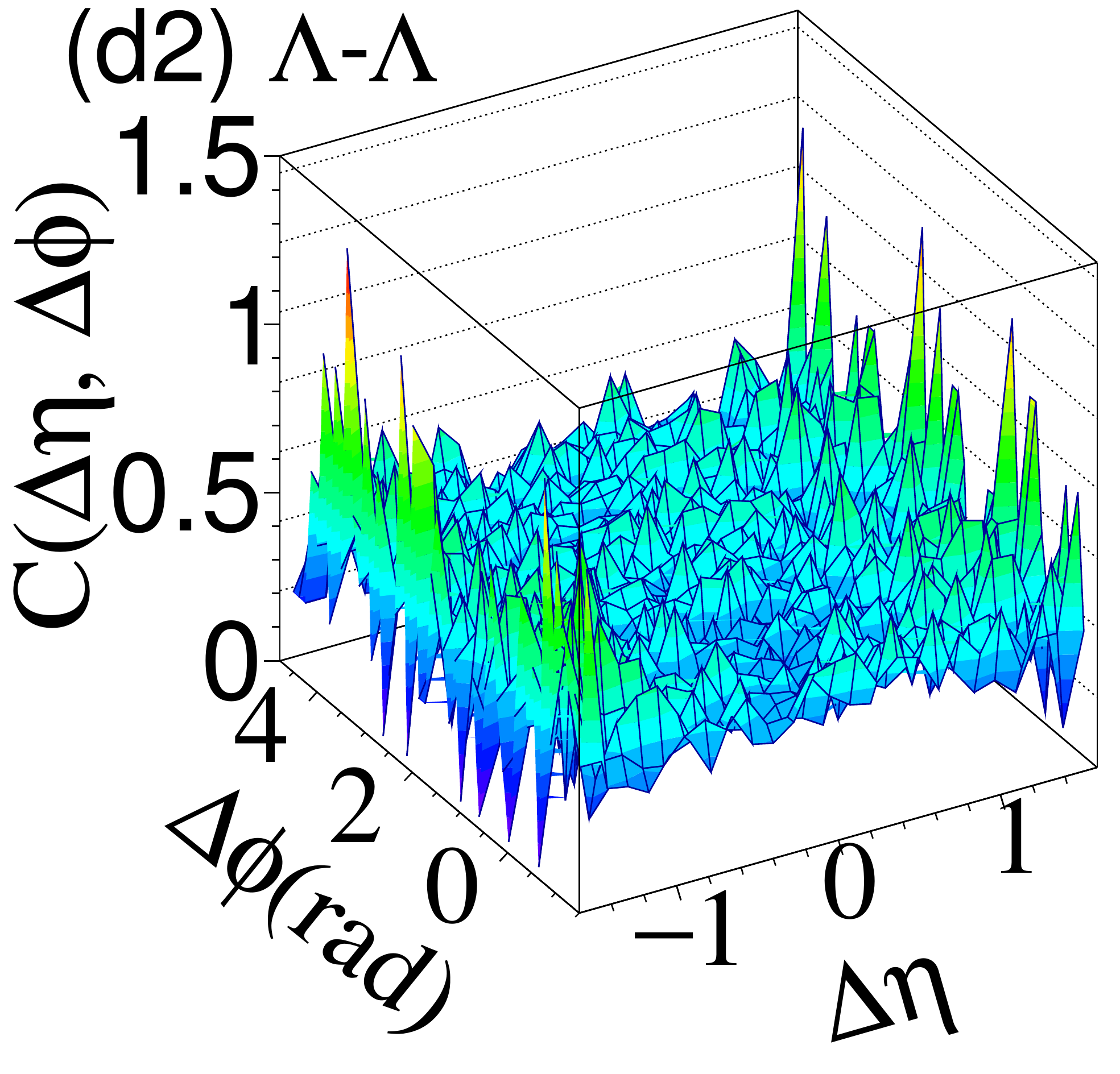}
        \includegraphics[scale=0.20]{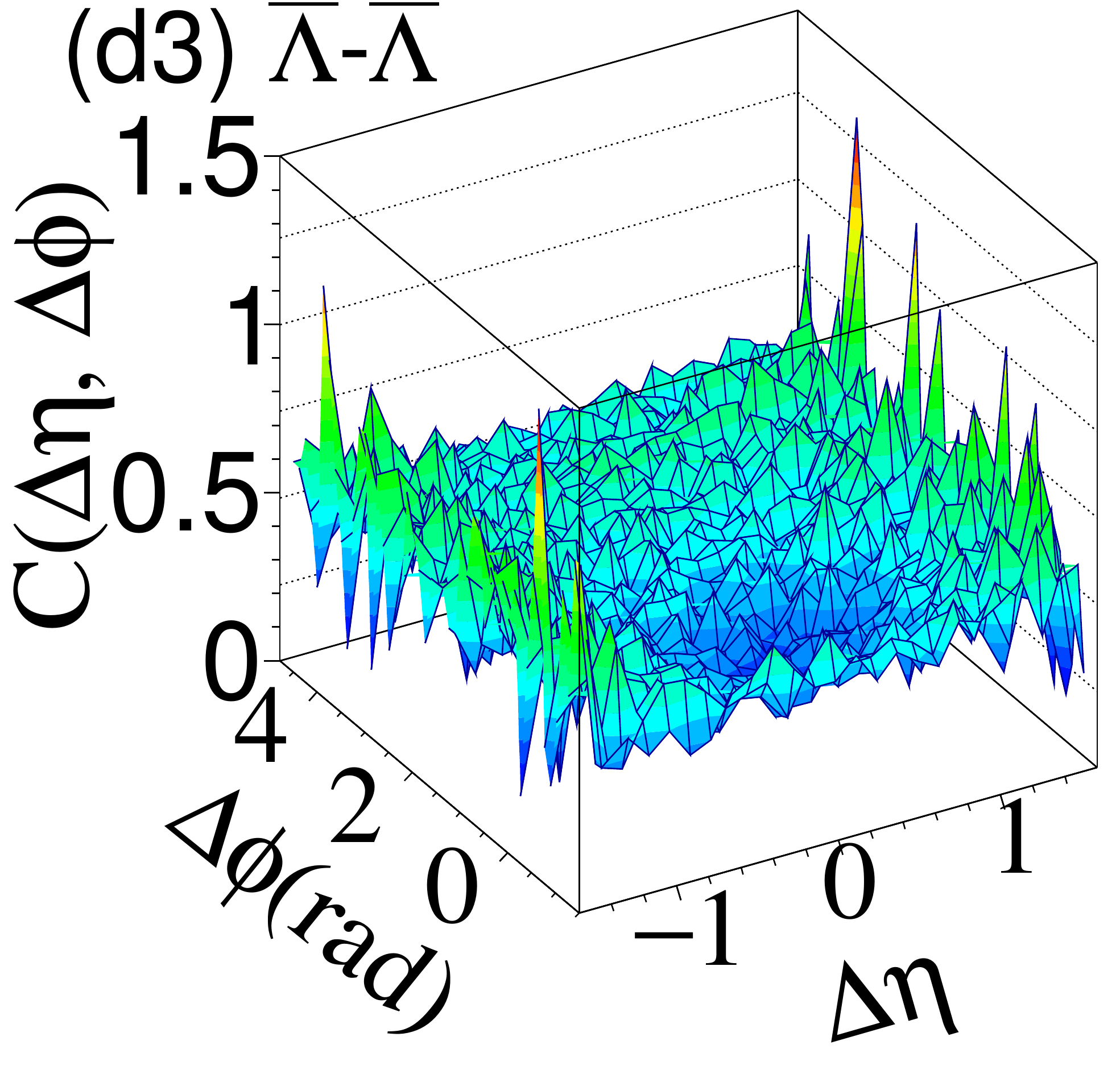}\\

	\caption{The $\Delta\phi$-$\Delta\eta$ correlation functions for different particle pairs from the string melting AMPT model for $pp$ collisions at $\sqrt{s}$ = 7 TeV. }
	\label{fig1:2D-CorF-melting}
\end{figure}

\begin{figure}[!htb]
	\centering
	\includegraphics[scale=0.20]{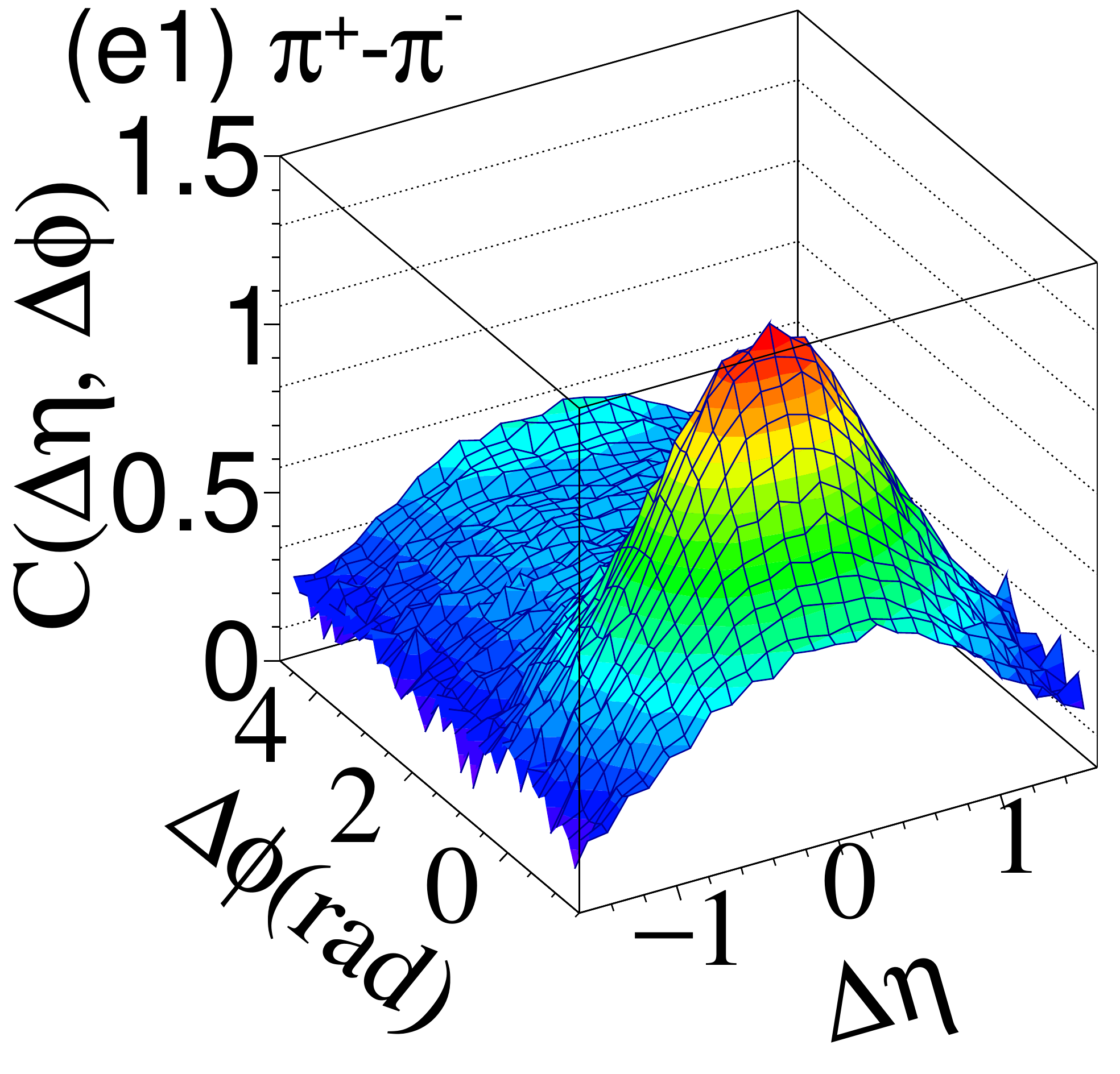}
	\includegraphics[scale=0.20]{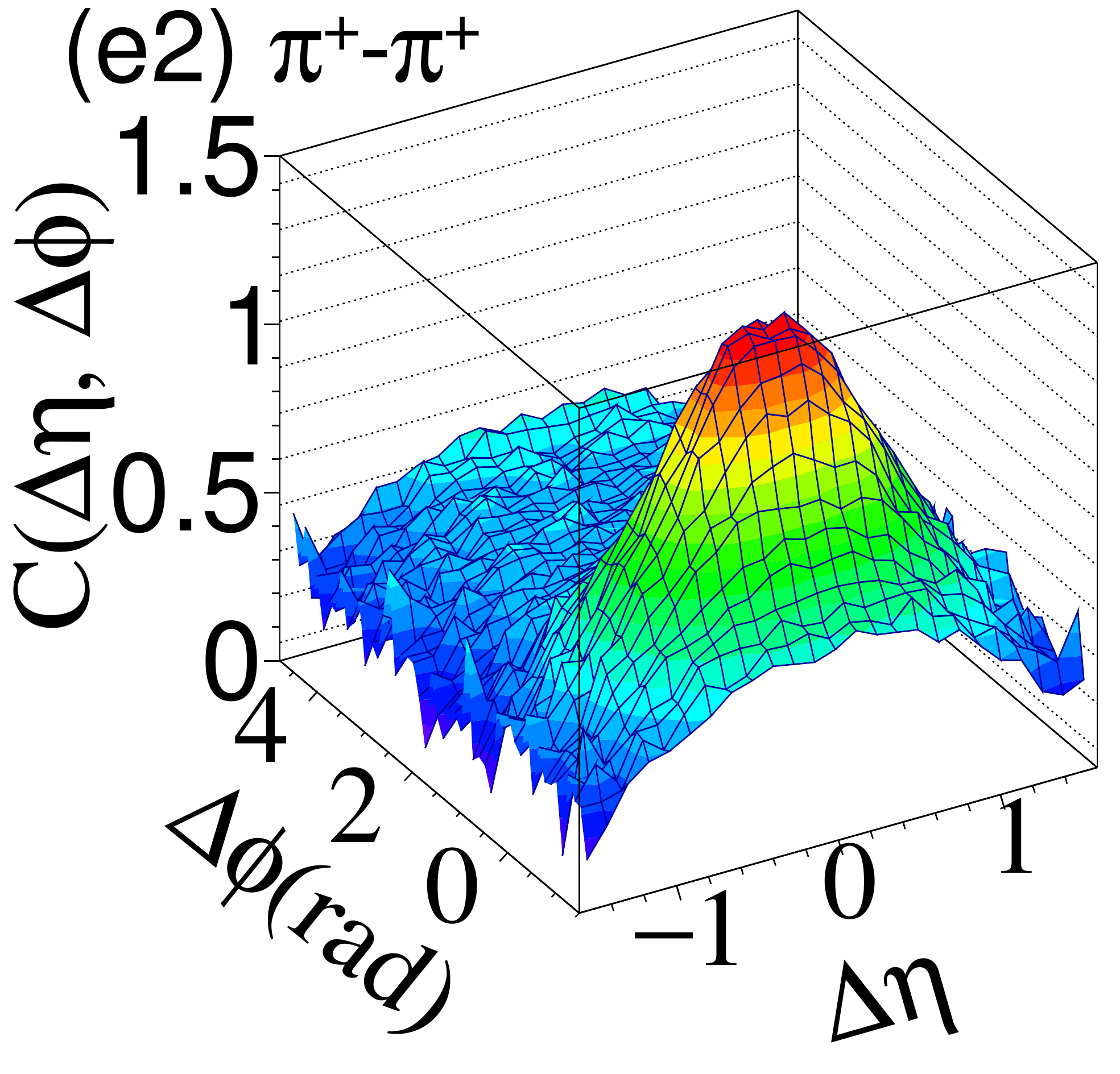}
	\includegraphics[scale=0.20]{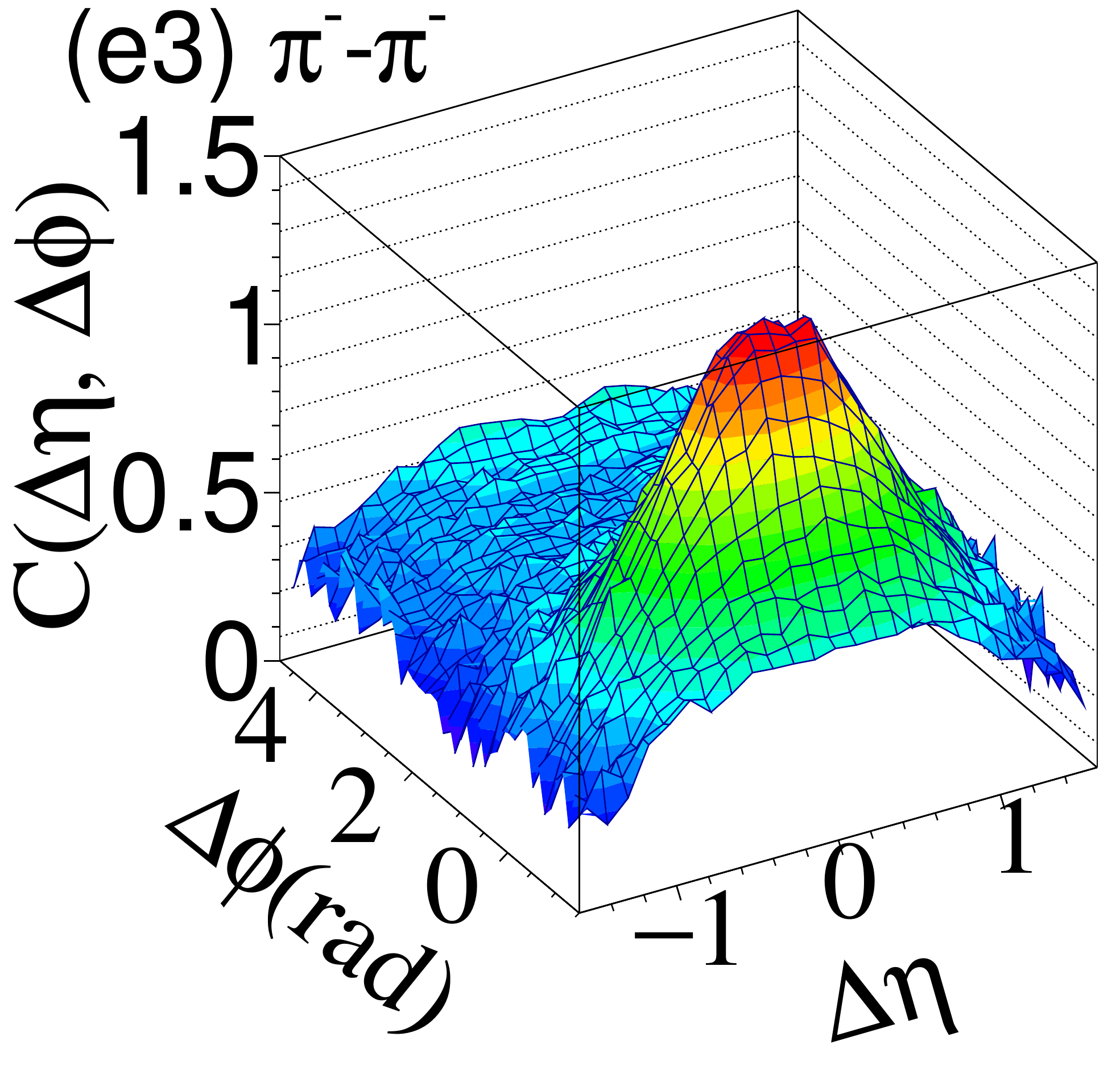}\\
	
         \includegraphics[scale=0.20]{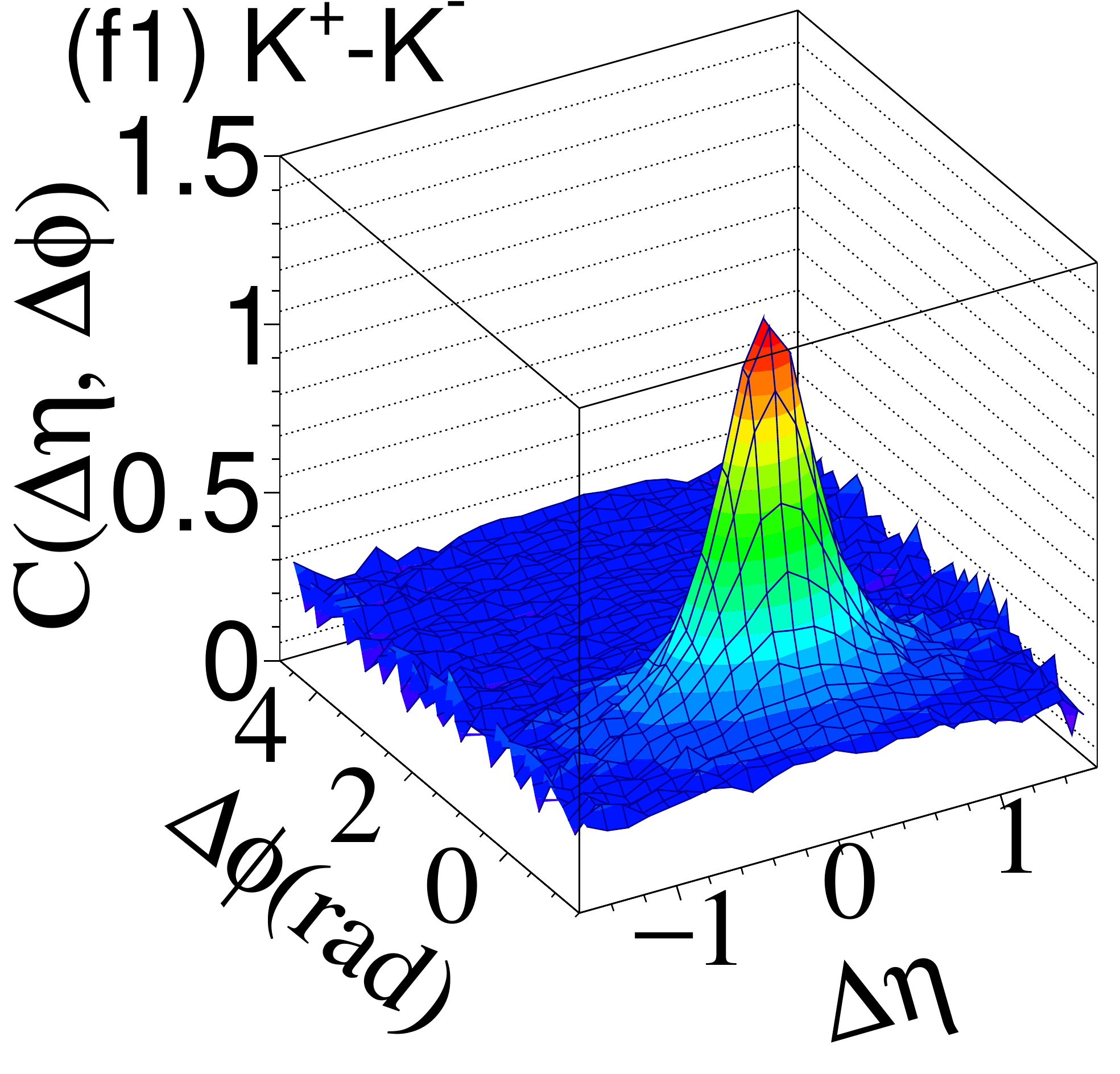}
	 \includegraphics[scale=0.20]{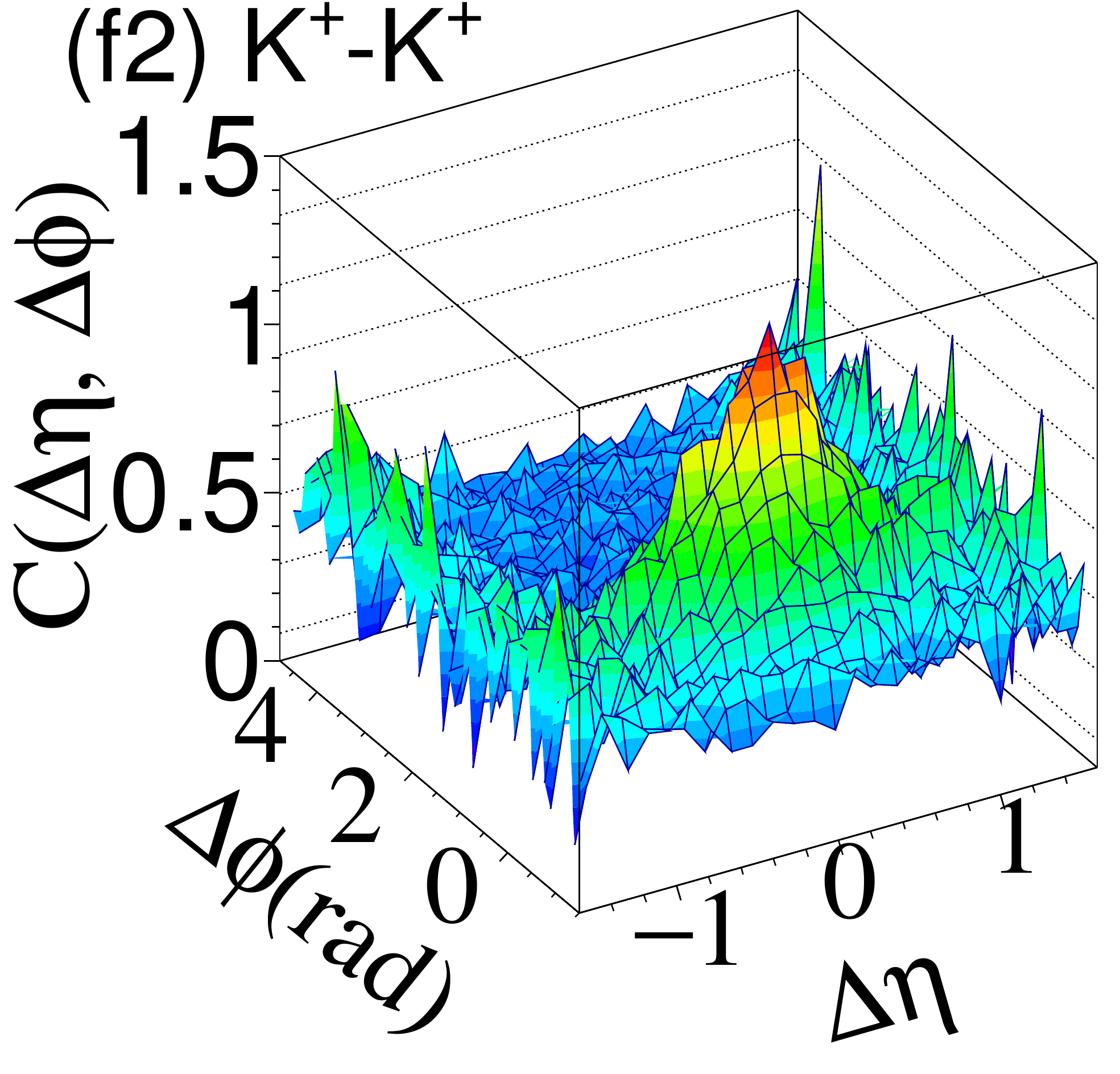}
	 \includegraphics[scale=0.20]{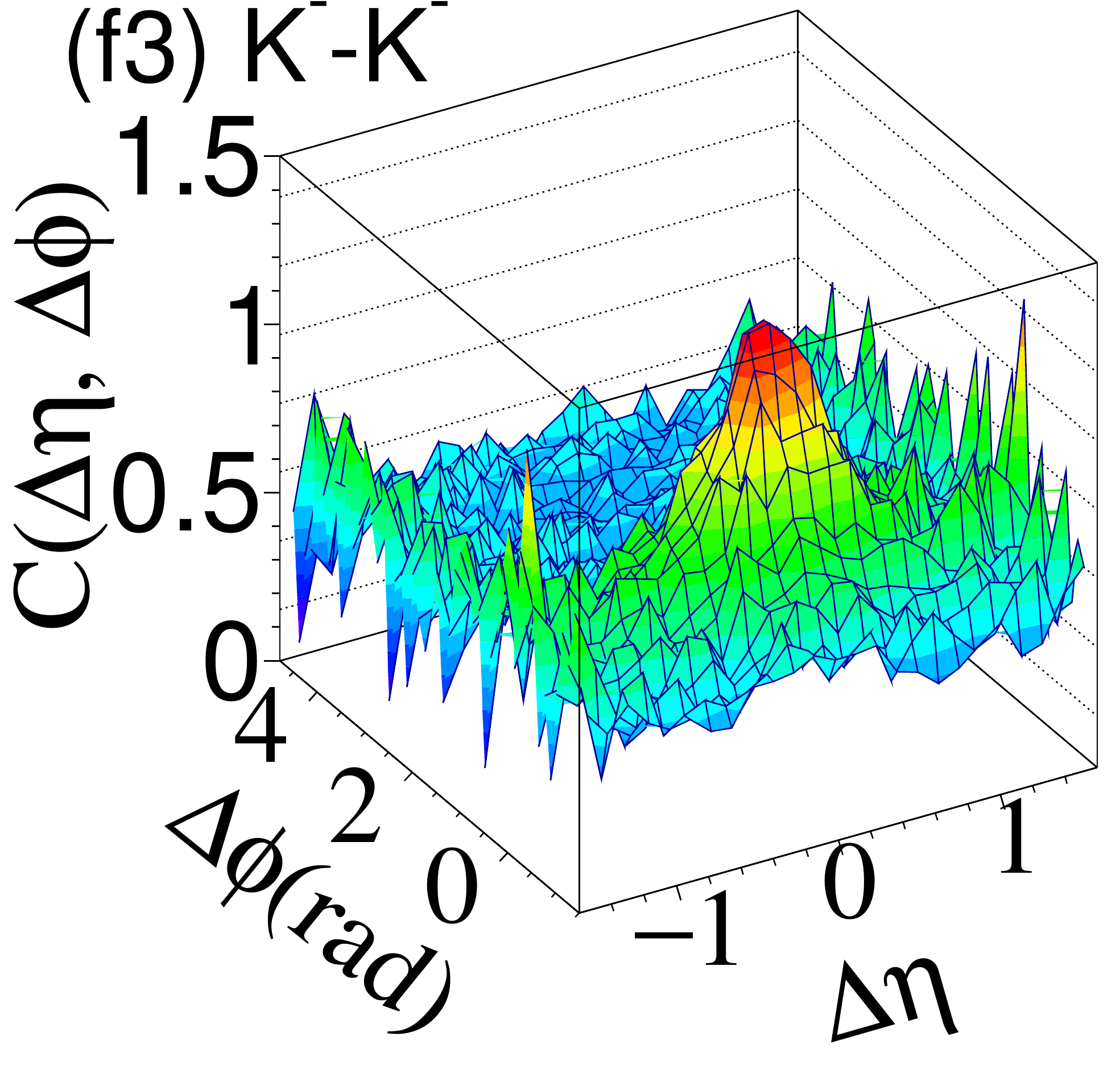}\\
          
          \includegraphics[scale=0.20]{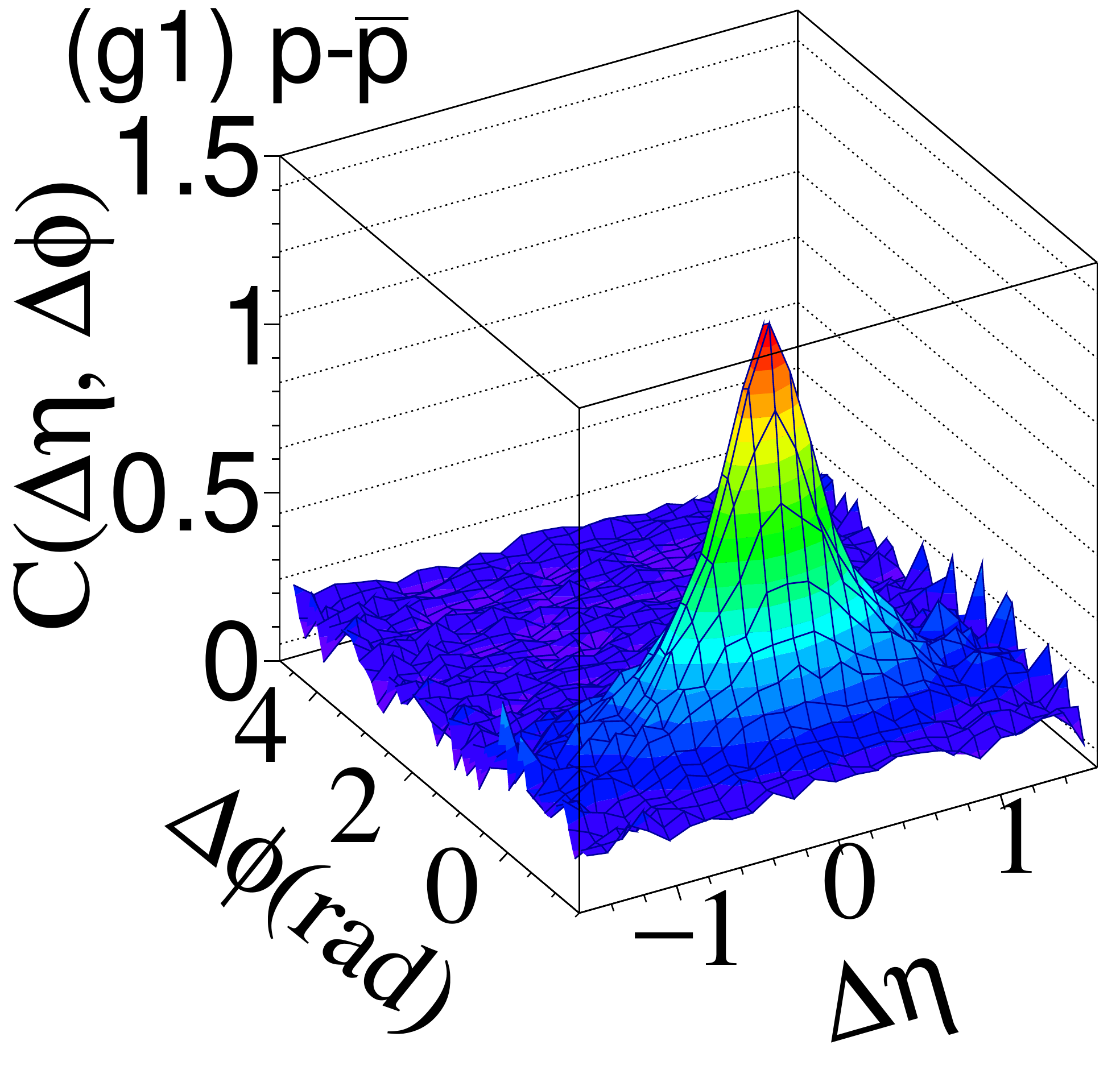}
	  \includegraphics[scale=0.20]{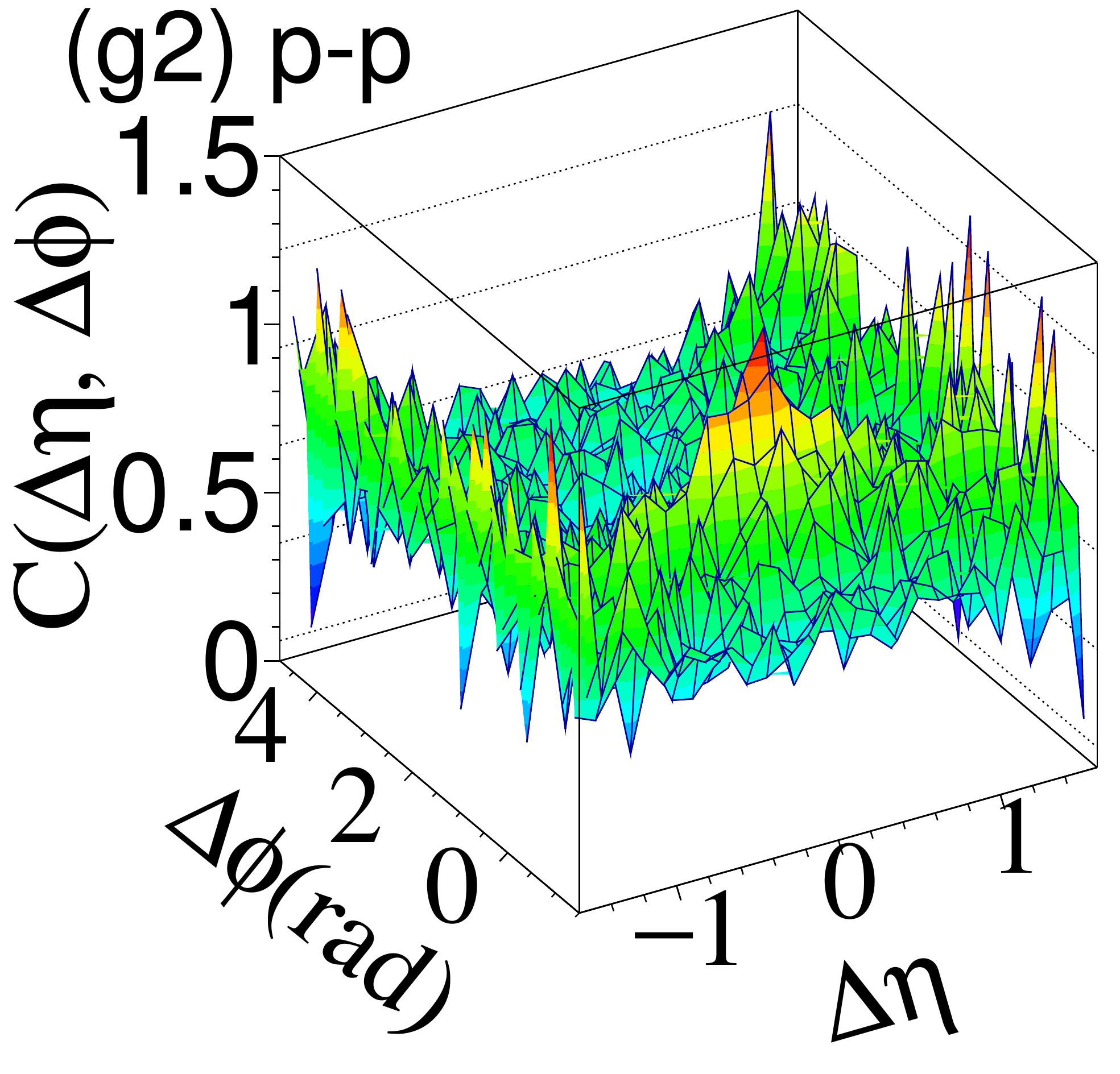}
	   \includegraphics[scale=0.20]{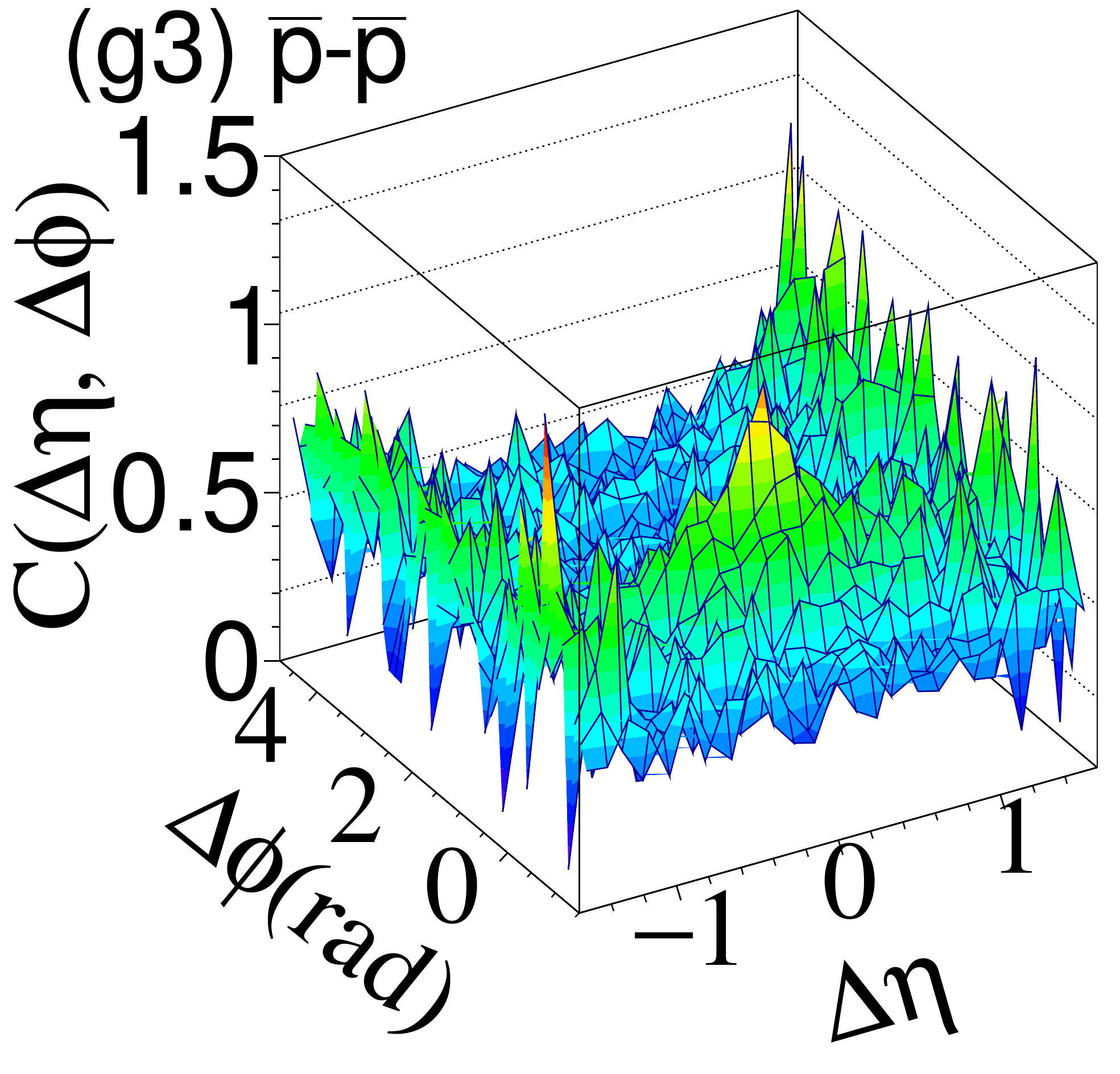}\\	 
	 
            \includegraphics[scale=0.20]{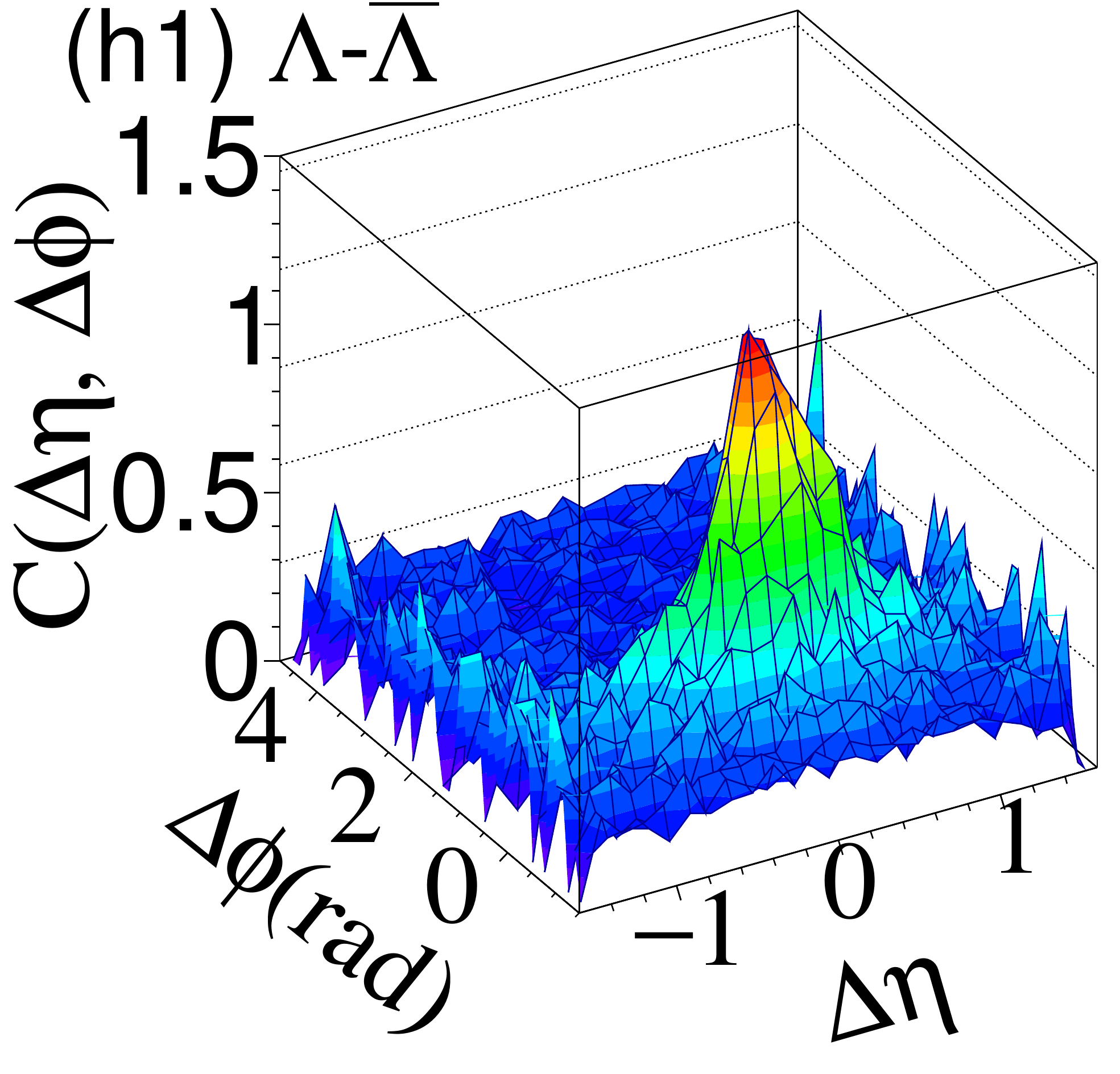}
	    \includegraphics[scale=0.20]{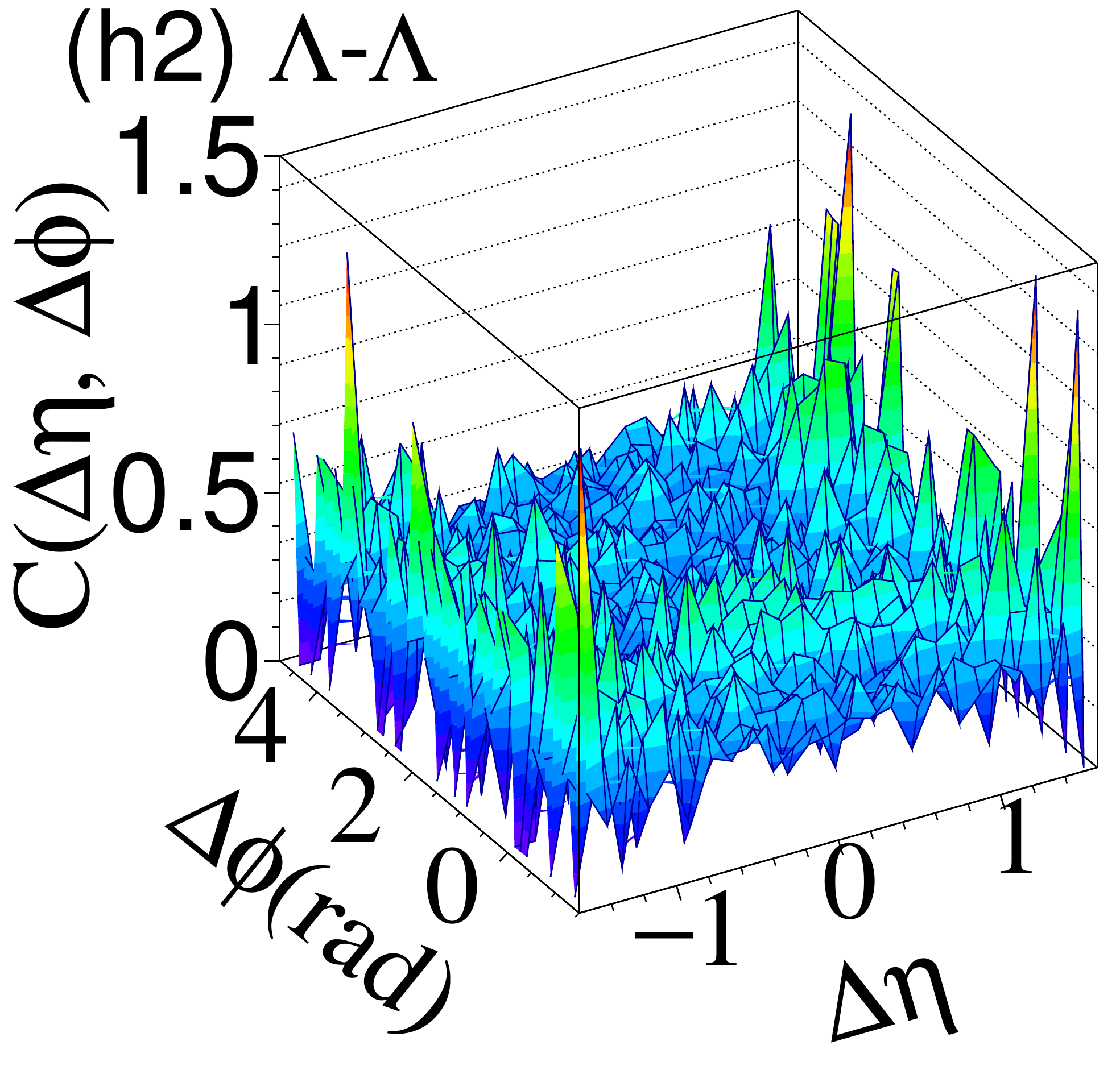}
	     \includegraphics[scale=0.20]{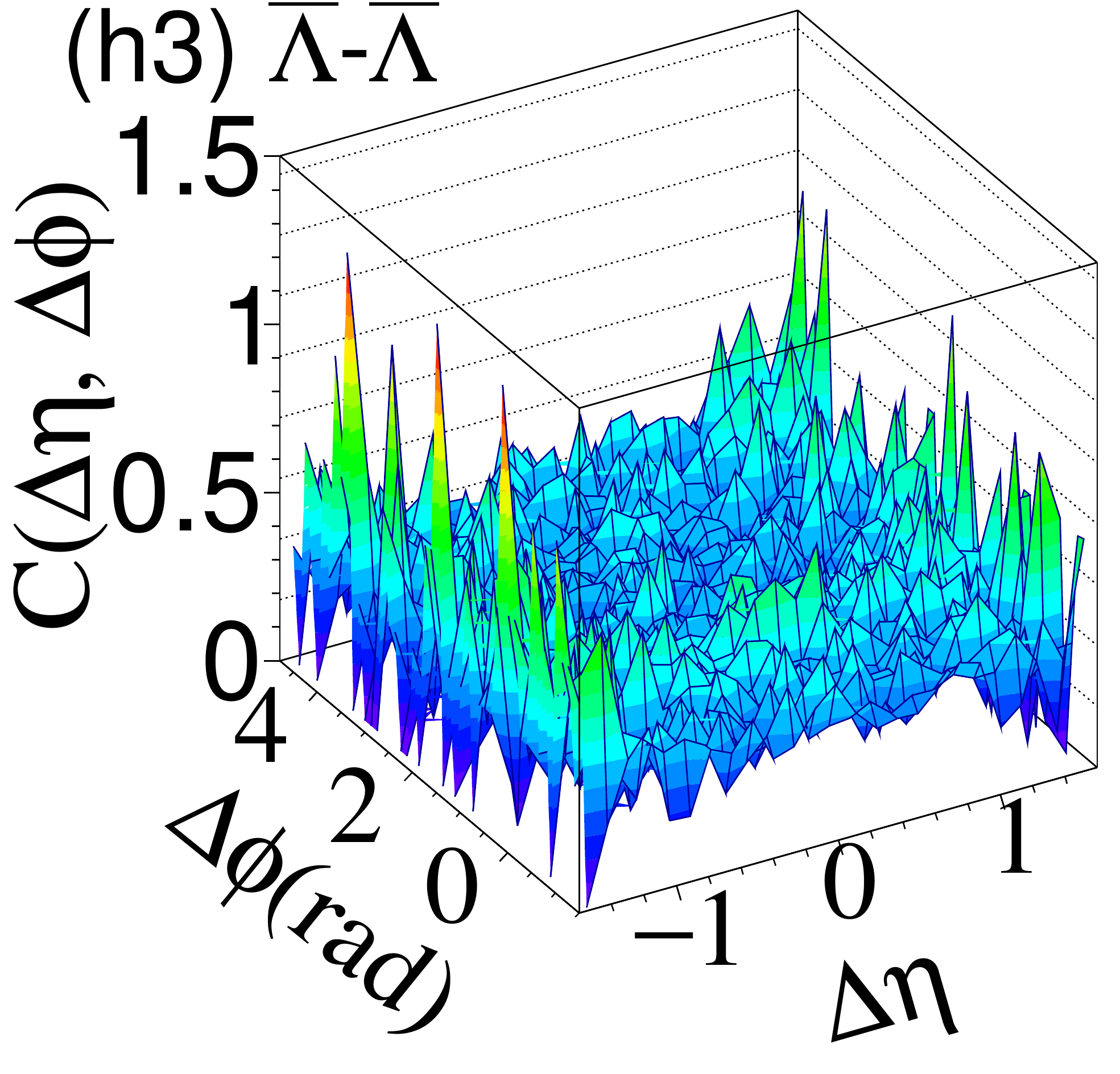}

\caption{Similar to Fig.~\ref{fig1:2D-CorF-melting} but from the default AMPT model.} 
\label{fig2:2D-CorF-default}
\end{figure}

We first use the AMPT model to calculate minimum bias $pp$ events at $\sqrt{s}$ = 7 TeV. The correlation functions for different particle pairs from AMPT-Melting are shown in Fig.~\ref{fig1:2D-CorF-melting}, and results from AMPT-Default are shown in Fig.~\ref{fig2:2D-CorF-default}. In these figures, we have applied the same kinematic selection criteria as used in the experiment measurement~\cite{J.Adam:2017} in order to directly compare with the data. They include a pseudorapidity range $|\eta|<0.8$ for all particles and a particle-type-dependent $p_T$ selection  due to the detector capability: $p_T >$ 0.5 GeV/c for p($\bar{p}$), $p_T >$ 0.3 GeV/c for $K^{\pm}$, $p_T >$ 0.2 GeV/c for $\pi^{\pm}$, and $p_T >$ 0.6 GeV/c for $\Lambda$($\bar{\Lambda}$). The particle-anti-particle correlation functions (panels a1, b1, c1, d1, e1, f1, g1, h1) in Fig.~\ref{fig1:2D-CorF-melting} and Fig.~\ref{fig2:2D-CorF-default} show a clear near-side peak structure, where baryon-antibaryon correlations are qualitatively similar to the meson's. The only difference is the magnitude and width of the near-side peak, where the magnitude is higher for mesons and lower for baryon-antibaryon. For the same particle pairs of mesons and baryons (c.f. Fig.~\ref{fig1:2D-CorF-melting} and Fig.~\ref{fig2:2D-CorF-default}), the near side peak may represent the contributions from mini-jet interactions with the medium.
 
%%%%%2-D Correlation functions

Similar to the measured correlation functions in $pp$ collisions at LHC energies, a distinct near-side peak at ($\Delta\eta$,$\Delta\phi$) $\sim$ (0,0) is observed for meson-meson pairs~\cite{V.Khachatryan:2010,V.Khachatryan:2016,G.Aad:2016}. As discussed in earlier studies~\cite{V.Khachatryan:2010,V.Khachatryan:2016,G.Aad:2016,J.Adam:2017}, the peak is a combination of several effects, such as the fragmentation of hard-scattered partons, higher mass resonance decays, femtoscopic correlations and Coulomb interaction among charge particles. The AMPT model includes most of the physics process except the femtoscopic effect, thus it reproduces the peak structure of the correlation functions. A more interesting feature is the pronounced depression of near-side distribution in the $\bar{p}$-$\bar{p}$ and $\bar{\Lambda}$-$\bar{\Lambda}$ correlation functions as shown in panel (c3) and (d3) of Fig.~\ref{fig1:2D-CorF-melting}. This depression structure only shows up in the AMPT-Melting model but not in the AMPT-Default version [c.f. Fig.~\ref{fig2:2D-CorF-default}]. It could be due to additional parton cascade and different hadronization process between the two versions of the AMPT model, as we shall discuss in more details next.
			
\begin{figure}[!htb]
 	\includegraphics[scale=0.32]{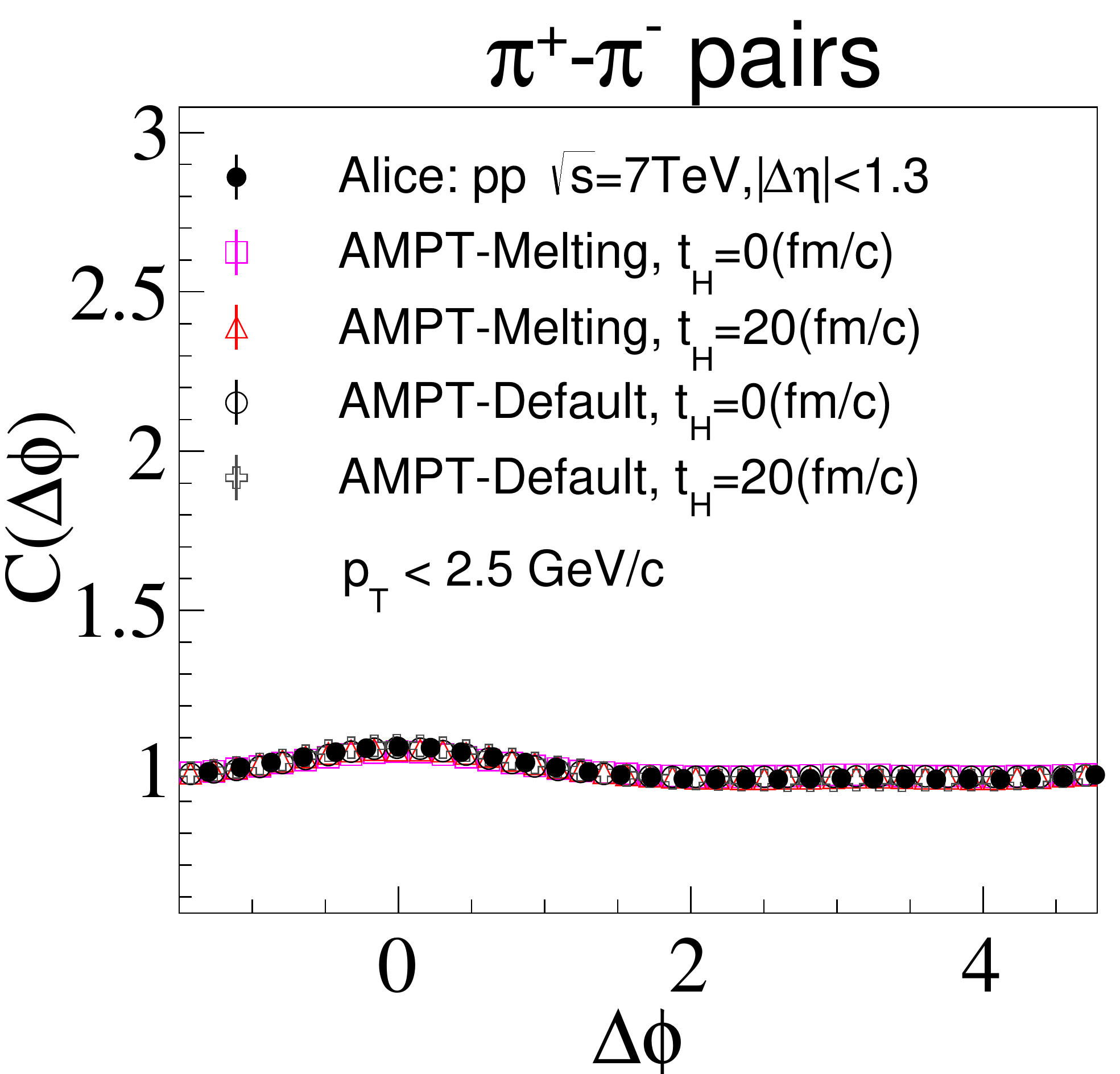}
 	\includegraphics[scale=0.32]{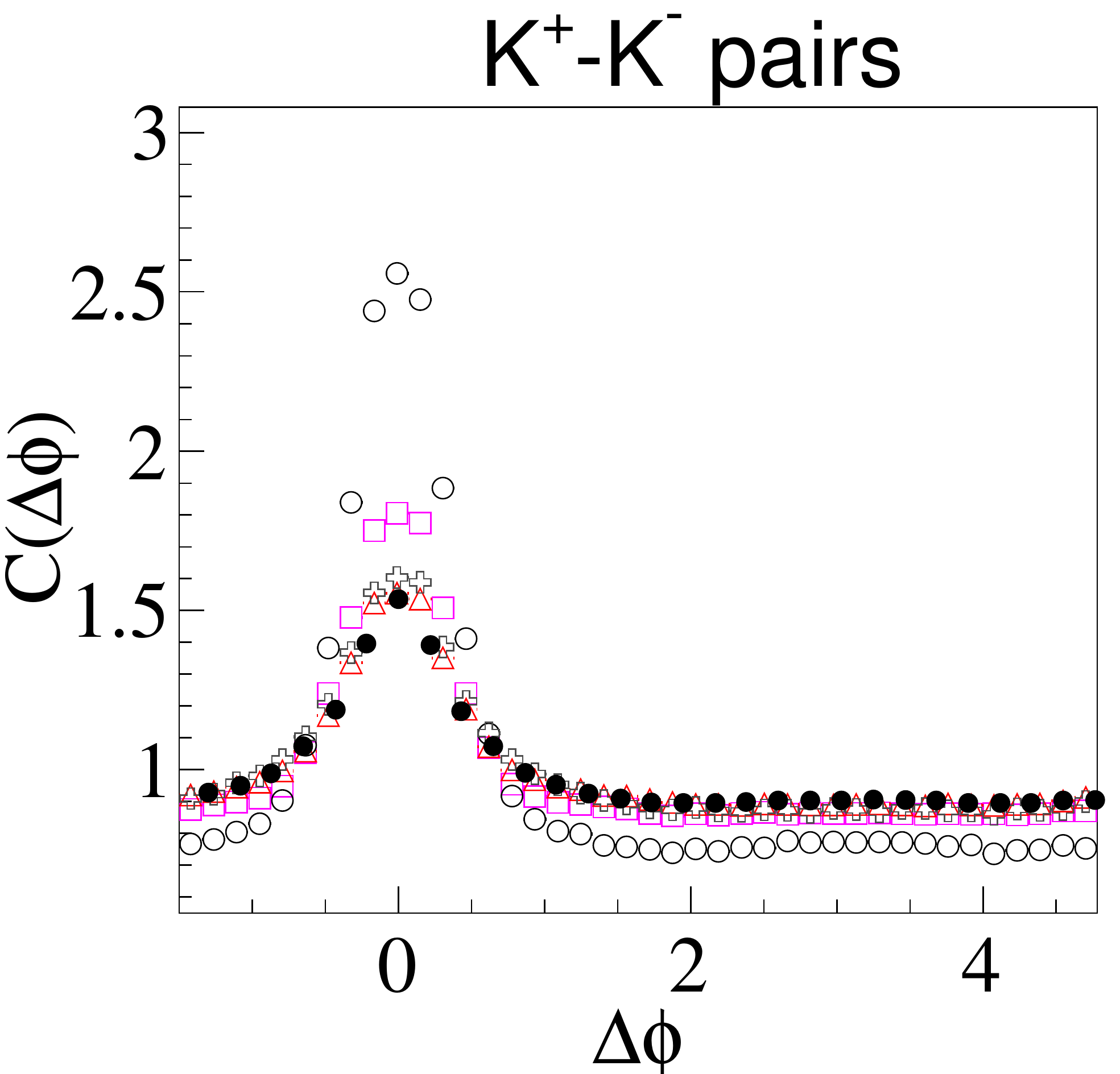}\\
	\includegraphics[scale=0.32]{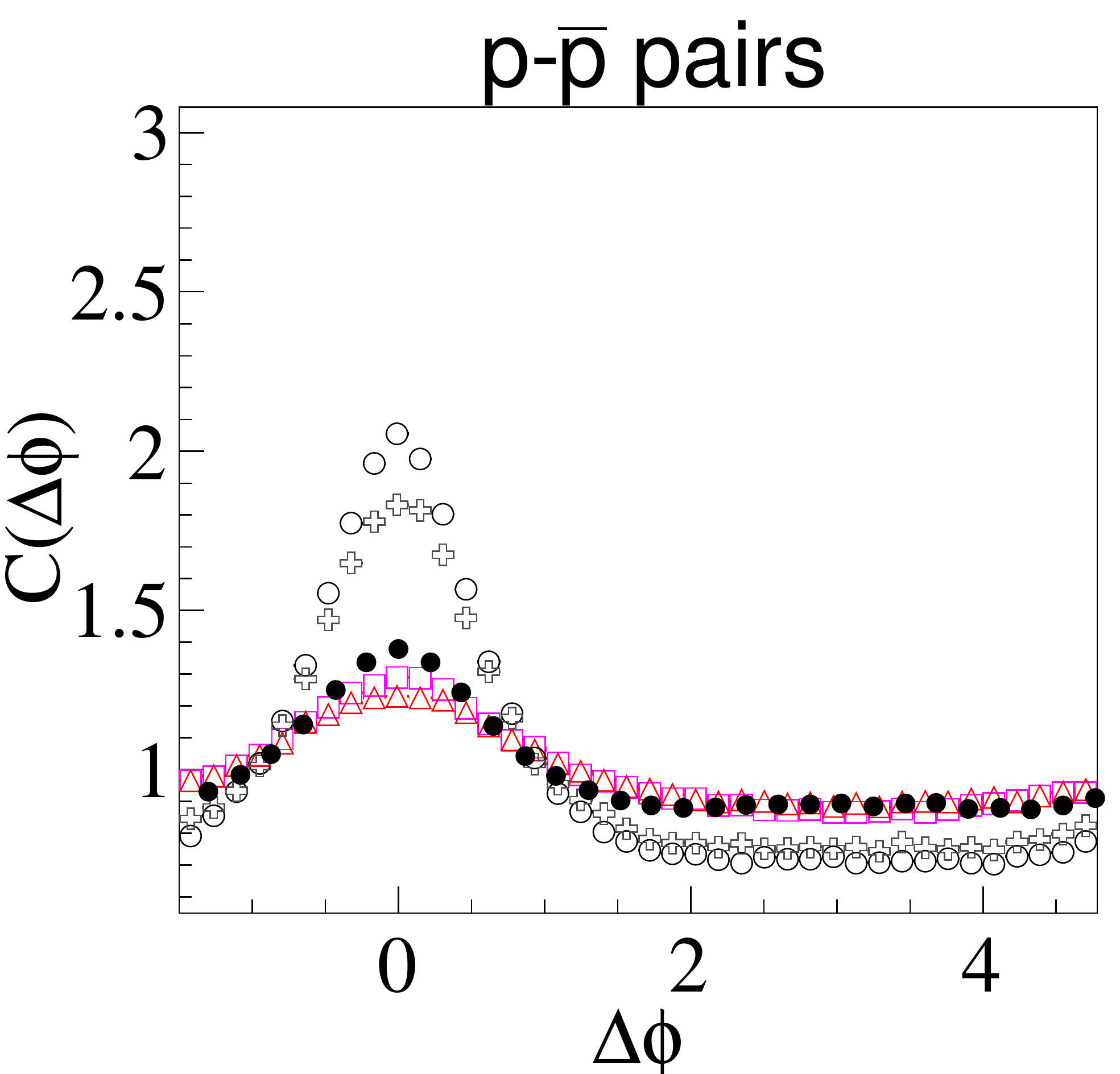}
	\includegraphics[scale=0.32]{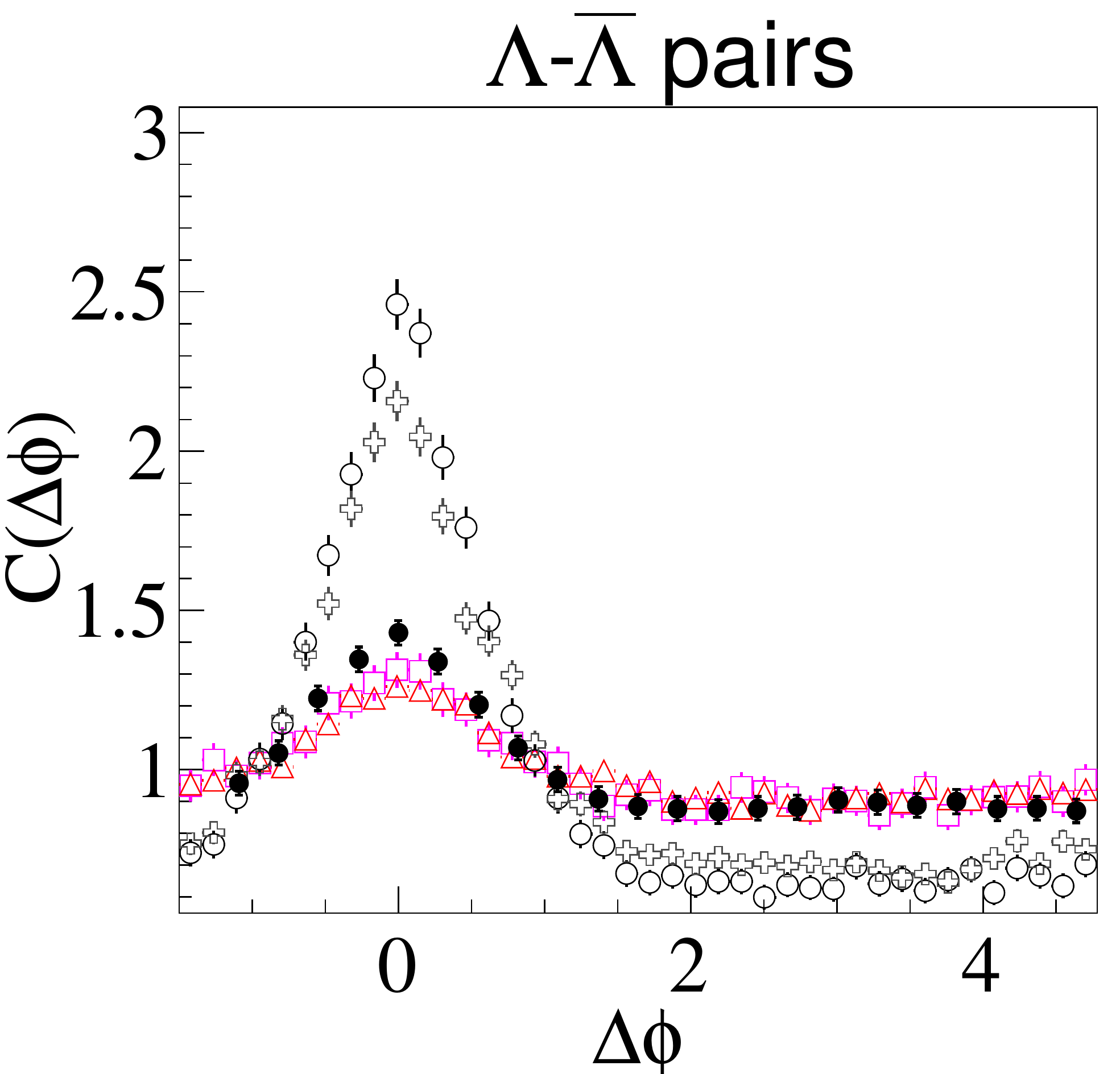}
  	\caption{One-dimensional $\Delta\phi$ correlation functions for $\pi^+$-$\pi^-$, $K^+$-$K^-$, $p$-$\bar{p}$, and  $\Lambda$-$\bar{\Lambda}$ in $pp$ collisions. Open symbols are AMPT results with t(hadron) = 0 fm/c (hadronic stage off) or t(hadron)=20 fm/c (hadronic stage on). Solid points are experimental data~\cite{J.Adam:2017}.}
    \label{fig3:1D-CorF-unlike-sign}
 \end{figure}

To further compare with the experimental data \cite{J.Adam:2017}, we project the 2-D correlation functions in Fig.~\ref{fig1:2D-CorF-melting} and Fig.~\ref{fig2:2D-CorF-default} over the $|\Delta\eta| < $1.3 window to the $\Delta\phi$ axis. Figure~\ref{fig3:1D-CorF-unlike-sign} shows the particle-anti-particle pair correlation functions along the $\Delta\phi$ axis in the AMPT model. 
We have chosen the hadronic rescattering time $t_H$ of 0 or 20 fm/c, with $t_H$ = 0 fm/c representing the results with the hadronic cascade turned off, to investigate the hadronic rescattering contributions to the correlation functions. In general, both AMPT-Default and AMPT-Melting can qualitatively describe the experimental data, including the near side peak structure and the away side flat distributions. It is expected that mini-jets gives a dominant contribution to the structure of the two-particle angular correlations in $pp$ collisions. Quantitatively, both versions of the AMPT model with sufficient hadronic interaction time ($t_H$ = 20 fm/c) better describe the $K^+$-$K^-$ correlation function than that with no hadronic interaction time, indicating that high mass resonance decays and hadronic scatterings contribute significantly to the correlation function. In addition, the string melting AMPT model provides a better description of the correlation function data of baryon-antibaryon pairs than the default AMPT model; this suggests that, in addition to hadronic interactions, partonic interactions and hadronization are also important for baryon correlation functions in $pp$ collisions at LHC energies. 

\begin{figure}[!htb]
	\includegraphics[scale=0.32]{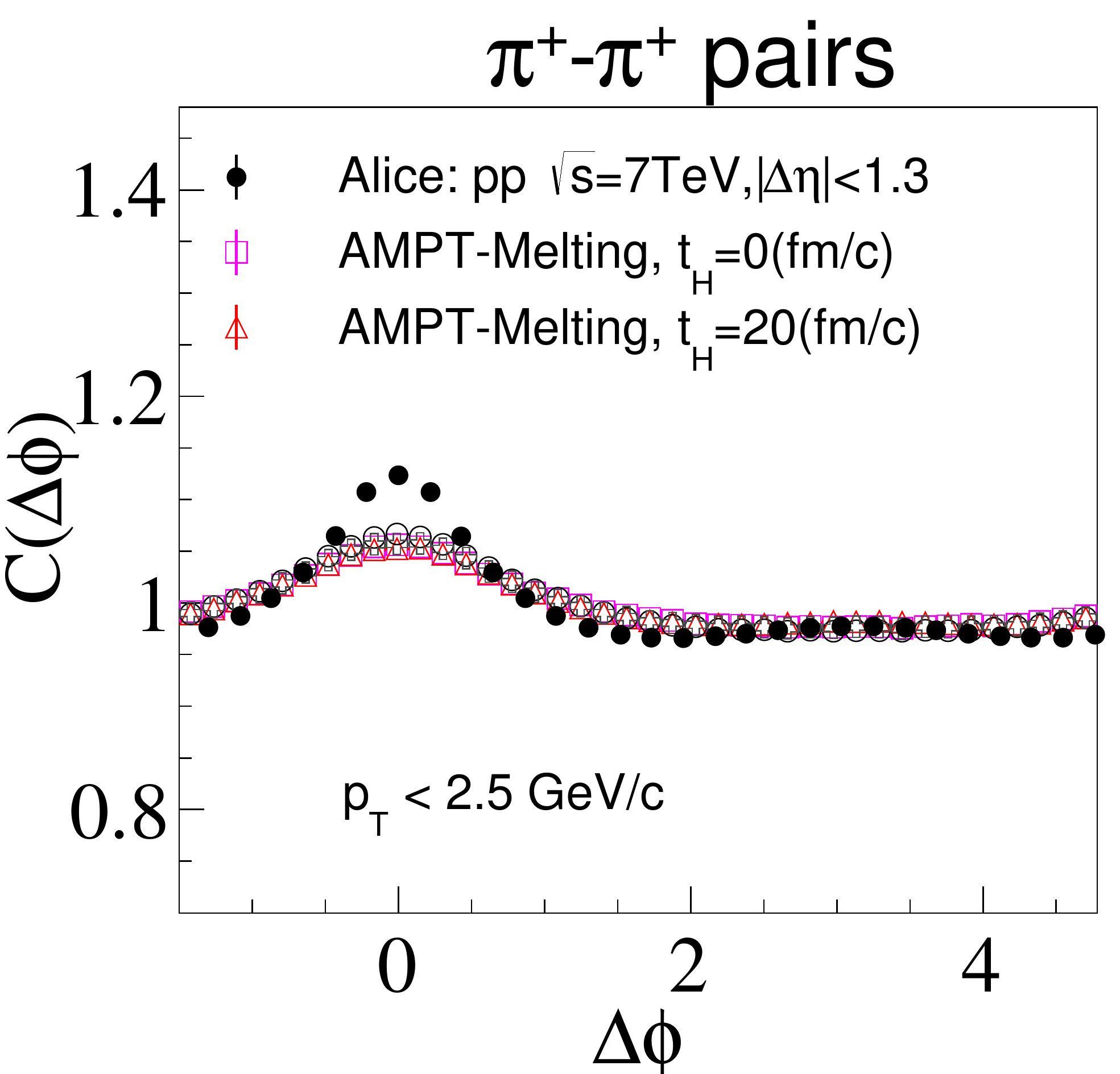}
	\includegraphics[scale=0.32]{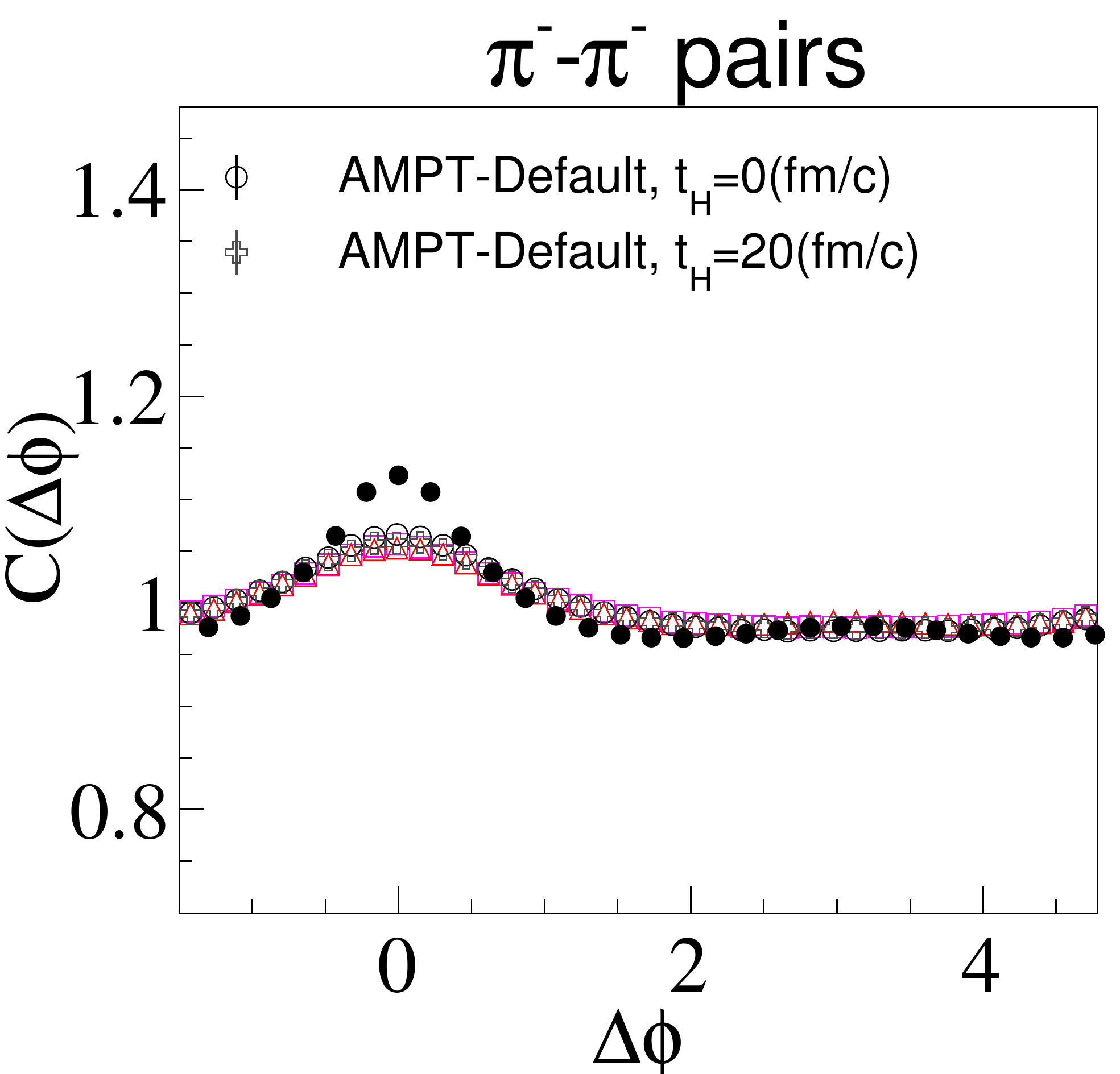}\\
	\includegraphics[scale=0.32]{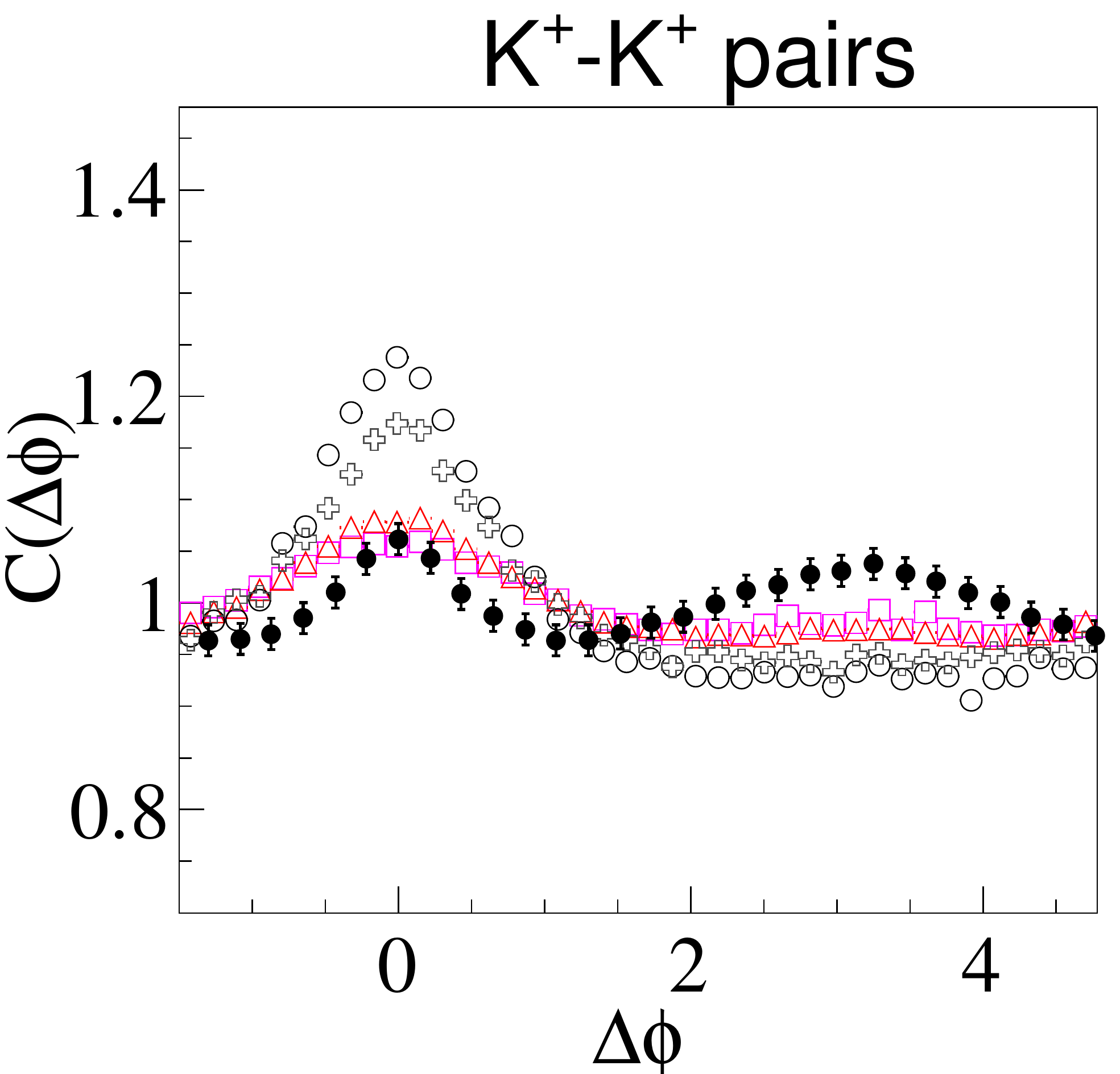}
	\includegraphics[scale=0.32]{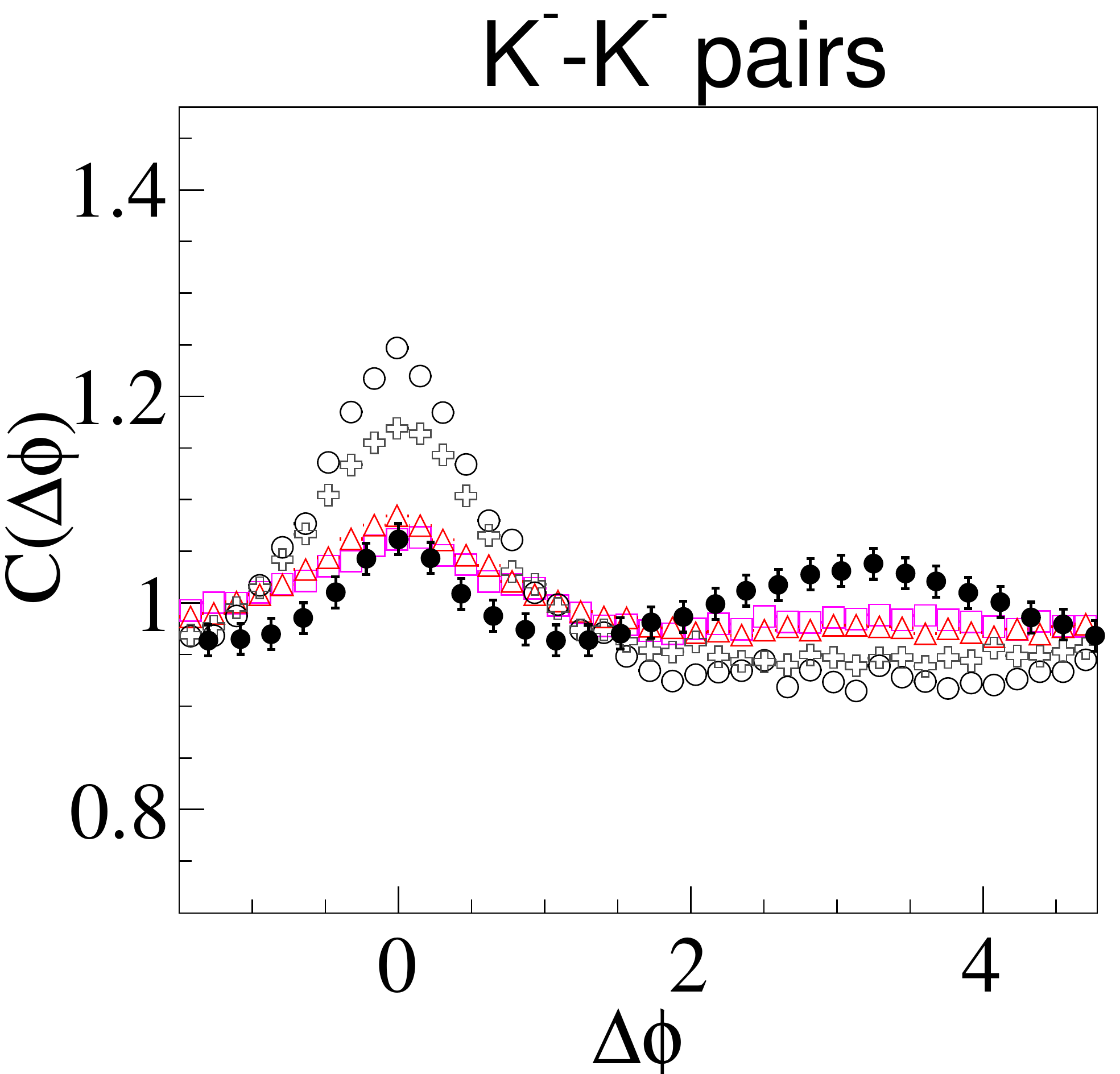}\\
	\includegraphics[scale=0.32]{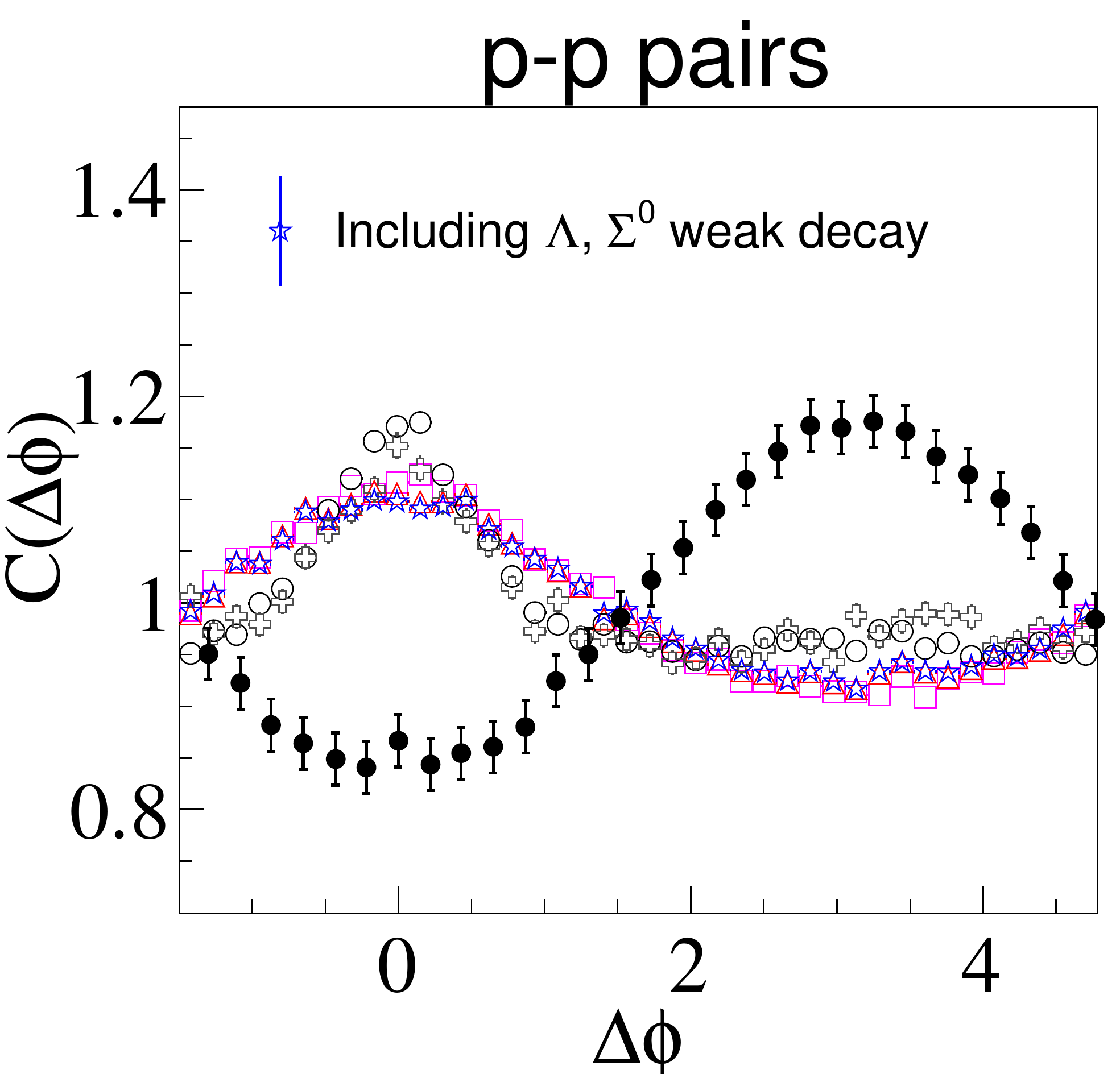}
	\includegraphics[scale=0.32]{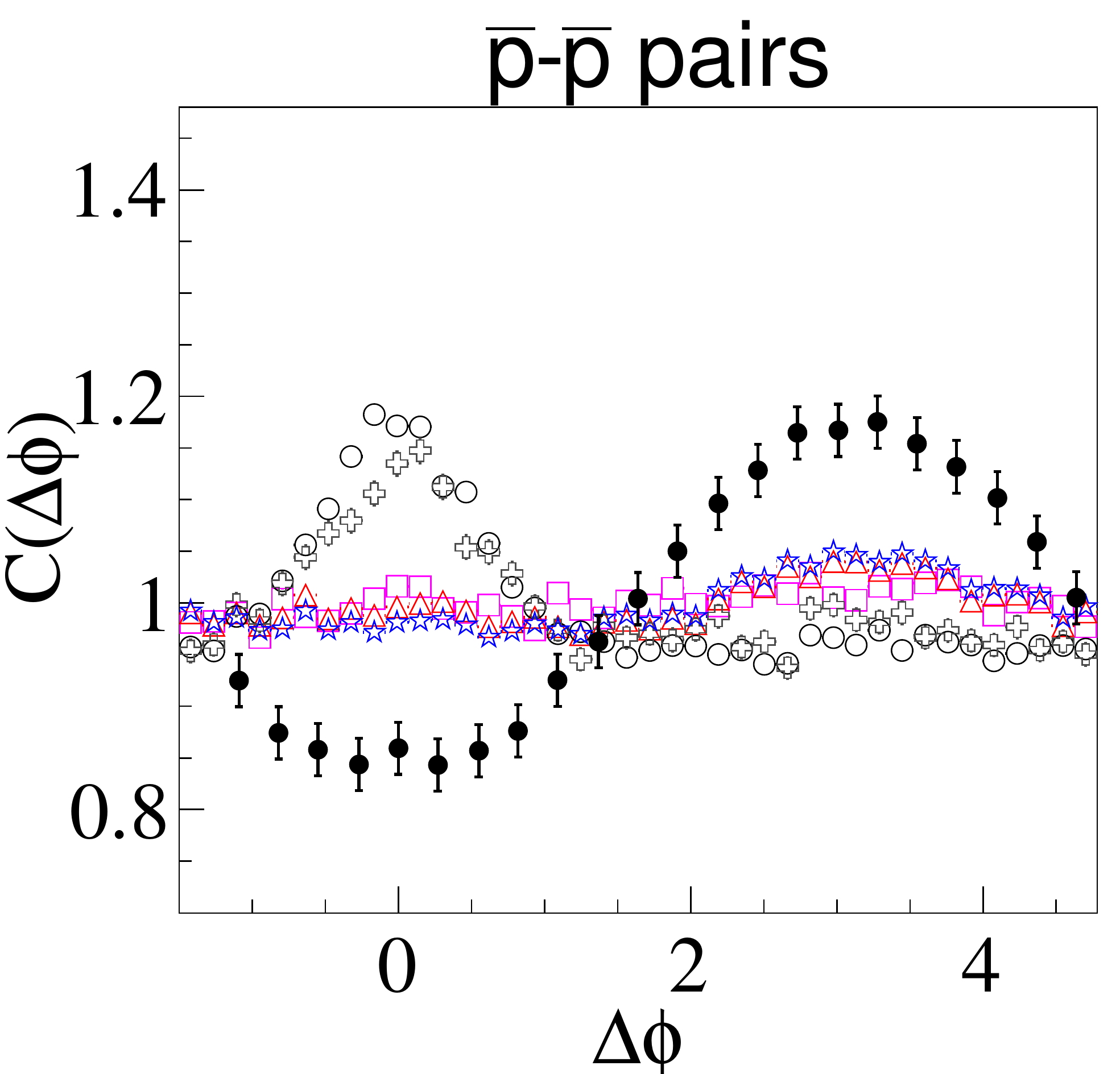}\\
	\includegraphics[scale=0.32]{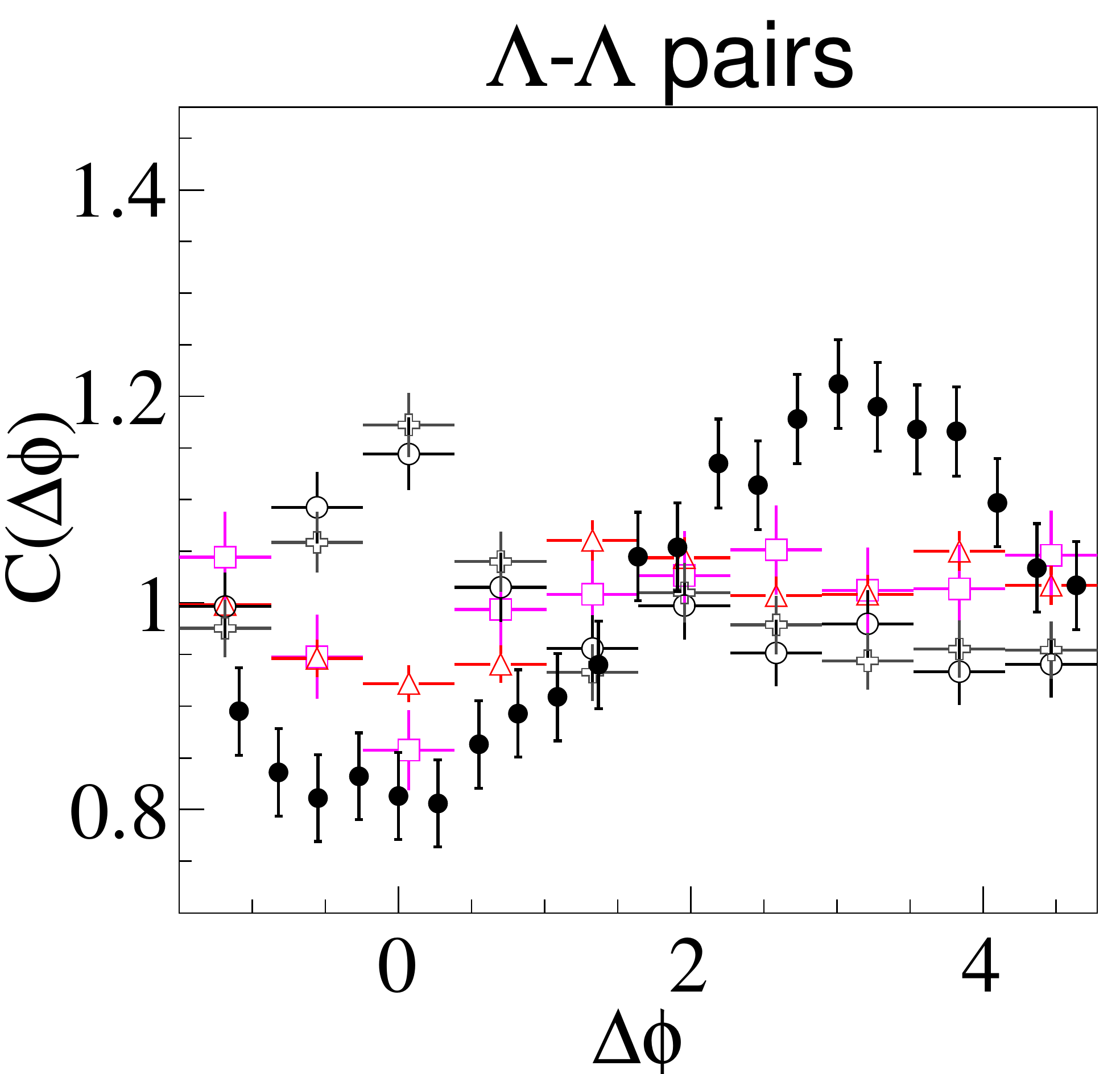}
	\includegraphics[scale=0.32]{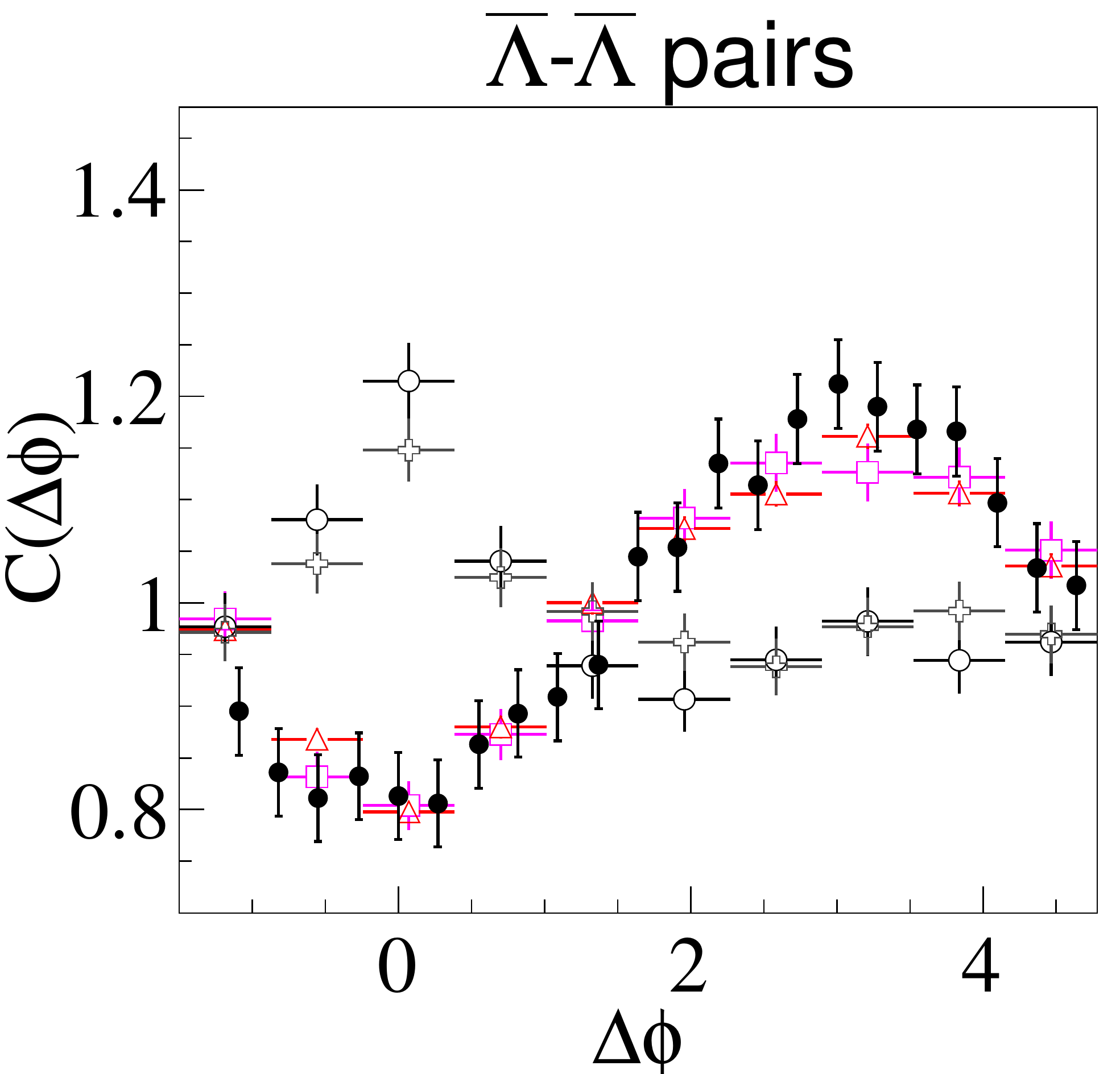}
	\caption{Similar to Fig.~\ref{fig3:1D-CorF-unlike-sign} but for correlation functions of pair with the same charge.}
	\label{fig4:1D-CorF-like-sign}
\end{figure}

We show the same-charge correlation functions from the AMPT model in Fig.~\ref{fig4:1D-CorF-like-sign}. For $\pi$-$\pi$ correlations, our model calculations describe the shape of the data reasonably well but have a lower magnitude for the near side peak. This is likely due to the abscence of quantum statistics effects in the AMPT model. Similar findings were reached based on the comparison of PYTHIA calculations to the experimental data by the ALICE Collaboration~\cite{J.Adam:2017}. 
On the proton-proton correlation function, results from AMPT-Melting and AMPT-Default both show no depression in the near side. 
However, the anti-proton-anti-proton correlation functions from the AMPT-Melting version 
are closer to the data than that from the AMPT-Default calculations. 
Since the proton-proton correlation function may suffer from resonance weak decay contributions, we have investigated this effect by forcing all the final $\Lambda$ and $\Sigma^0$ to decay in the model calculations. The results including these weak decays are shown as blue open stars in Fig.~\ref{fig4:1D-CorF-like-sign}, and we see that the effect is small. 

Regarding the $\Lambda$-$\Lambda$ and $\bar{\Lambda}$-$\bar{\Lambda}$ correlation functions, the AMPT-Melting version well describes the experimental data, including a clear depression structure on the near side and a strong enhancement on the away side. The results between $t_H$ = 0 fm/c and $t_H$ = 20 fm/c for $\Lambda$s are almost identical. It could be due to the fact that the current AMPT model does not contain resonance decay contribution to $\Lambda$ (other than from $\Sigma^0$ decays), or the role of hadronic scattering is tiny. We also find that results from the default AMPT model are similar to results from PYTHIA calculations~\cite{J.Adam:2017} and can not describe the experiment data.

Similar studies using the Monte Carlo event generators with different input parameters of PYTHIA~\cite{T.Sjostrand:2006} were done by the ALICE Collaboration~\cite{J.Adam:2017}. The results of MC models reproduce reasonably well the meson pair correlations but fail to reproduce the baryon correlations. For those MC model studies\cite{J.Adam:2017}, first of all, significant differences are also seen for baryon-antibaryon pairs, where the magnitude of the near-side peak is much higher in all MC models than the ALICE data. Here we have seen that the AMPT-Melting version reproduces the experimental results reasonably well [c.f. Fig.~\ref{fig3:1D-CorF-unlike-sign}]. Furthermore, no depression is observed for protons and $\Lambda$s for any of the MC models using by ALICE~\cite{J.Adam:2017}. Instead, a near-side peak is present for particle-particle pairs in the PYTHIA results~\cite{J.Adam:2017}. In our calculations, an interesting observation is that a depression in the near side is present in Fig.\ref{fig4:1D-CorF-like-sign} on $\Lambda$-$\Lambda$ and $\bar{\Lambda}$-$\bar{\Lambda}$ correlations. However, no depression is observed for proton-proton or $\bar{p}$-$\bar{p}$ correlations in from our AMPT calculations; instead a near-side peak is present in proton-proton correlations.

\subsection{Two particle correlations in AMPT model with new quark coalescence}
\begin{figure}[!htb]
	\includegraphics[scale=0.24]{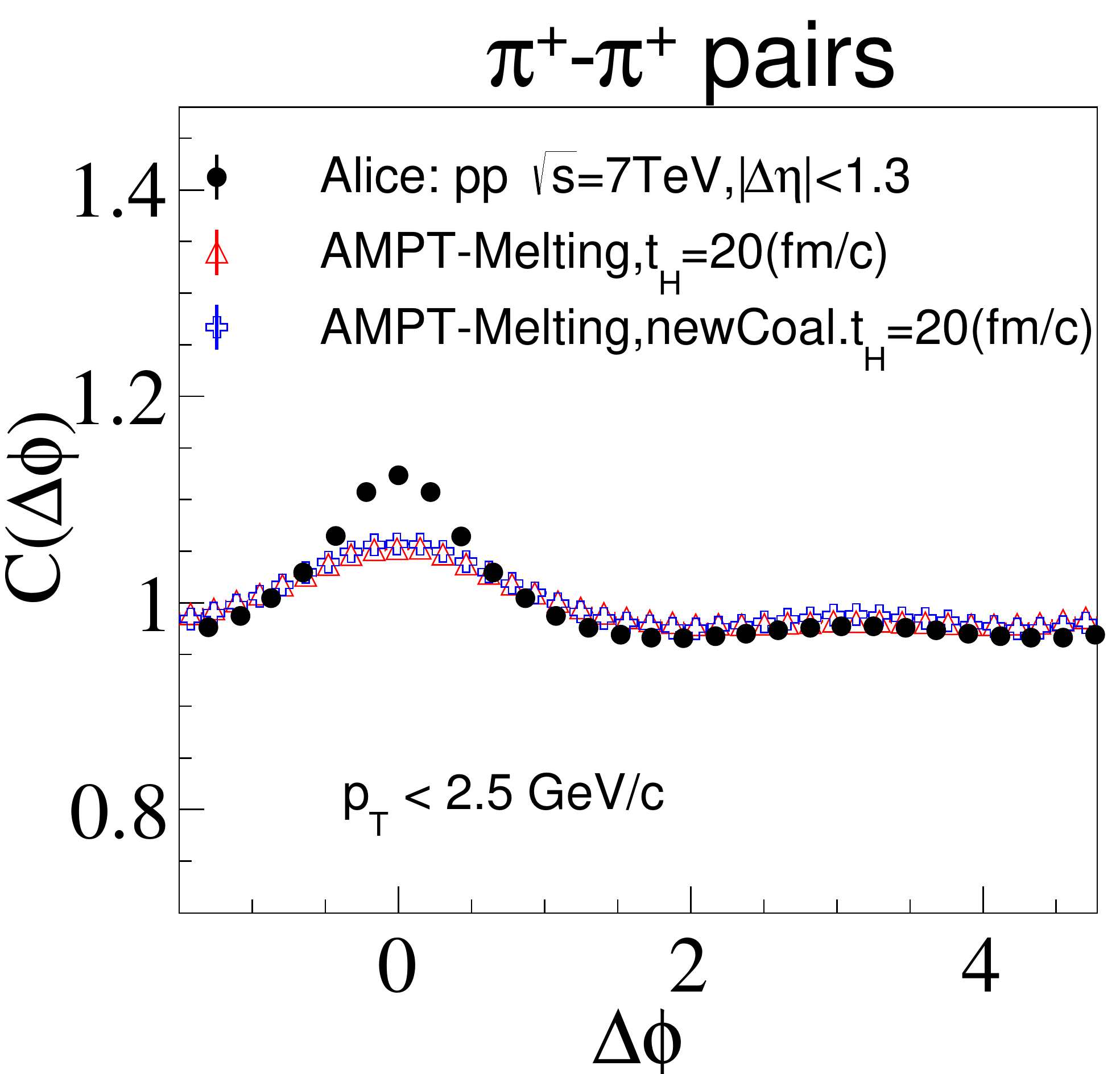}
	\includegraphics[scale=0.24]{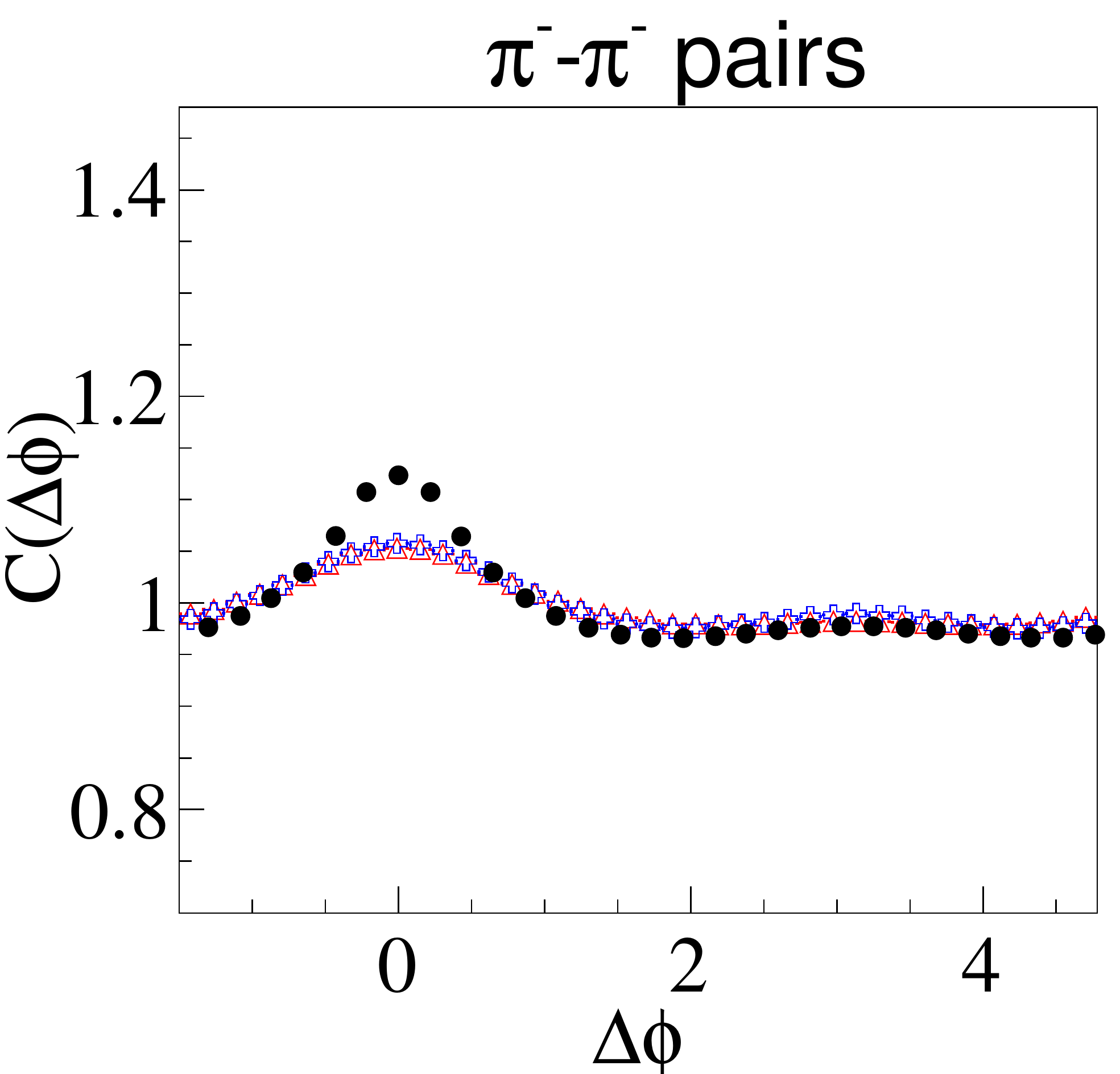}\\
	\includegraphics[scale=0.24]{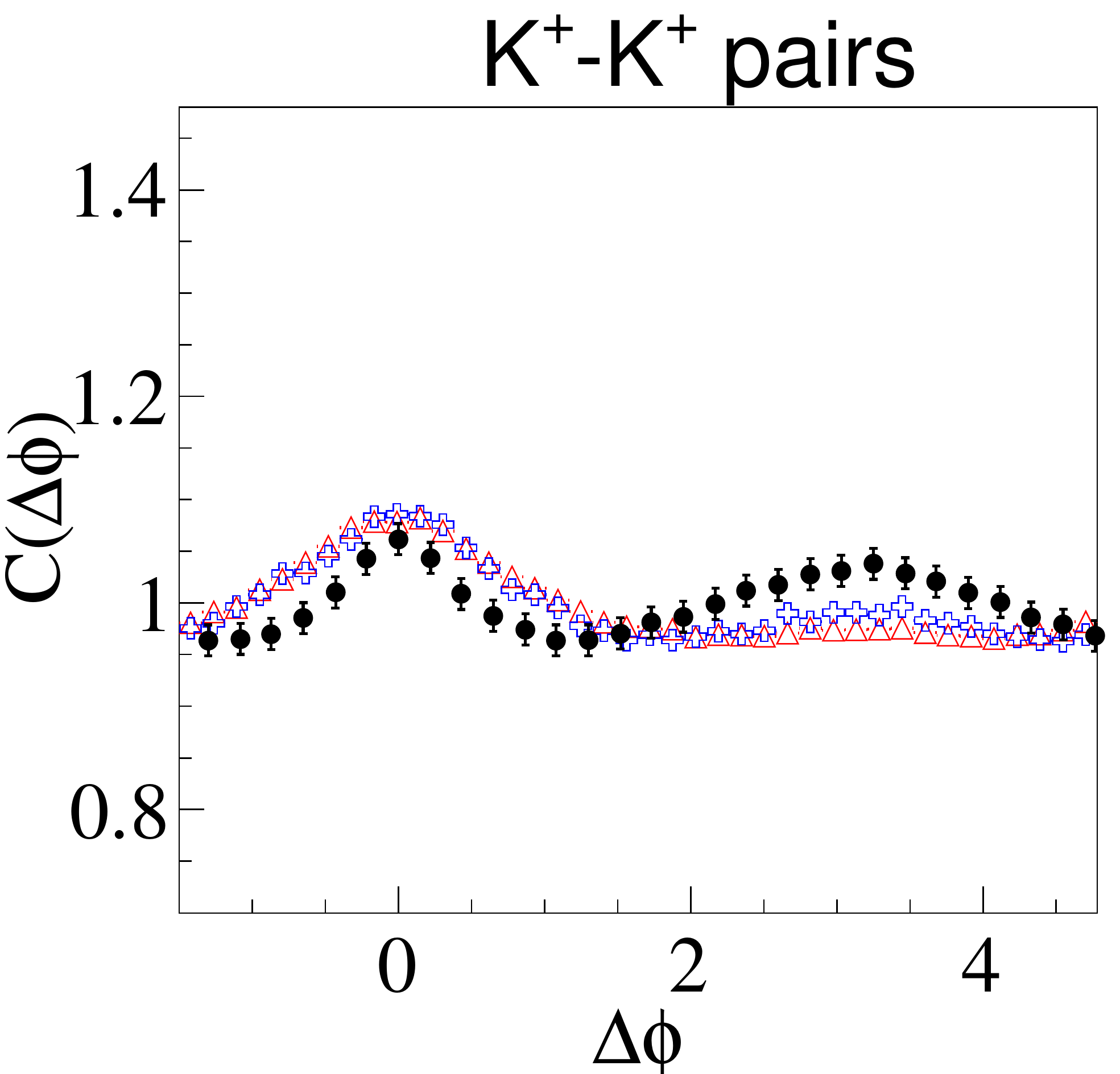}
	\includegraphics[scale=0.24]{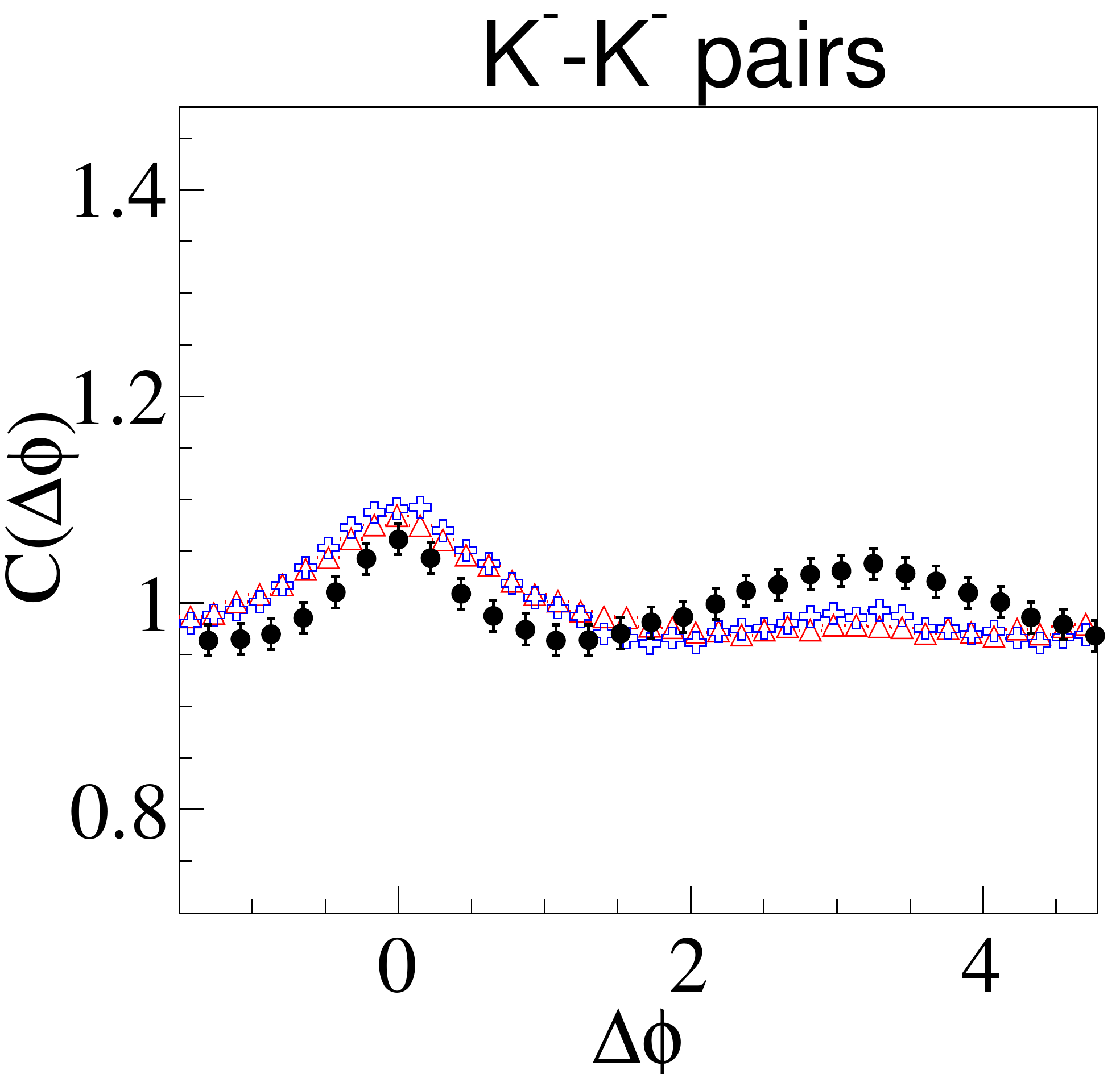}\\
	\includegraphics[scale=0.24]{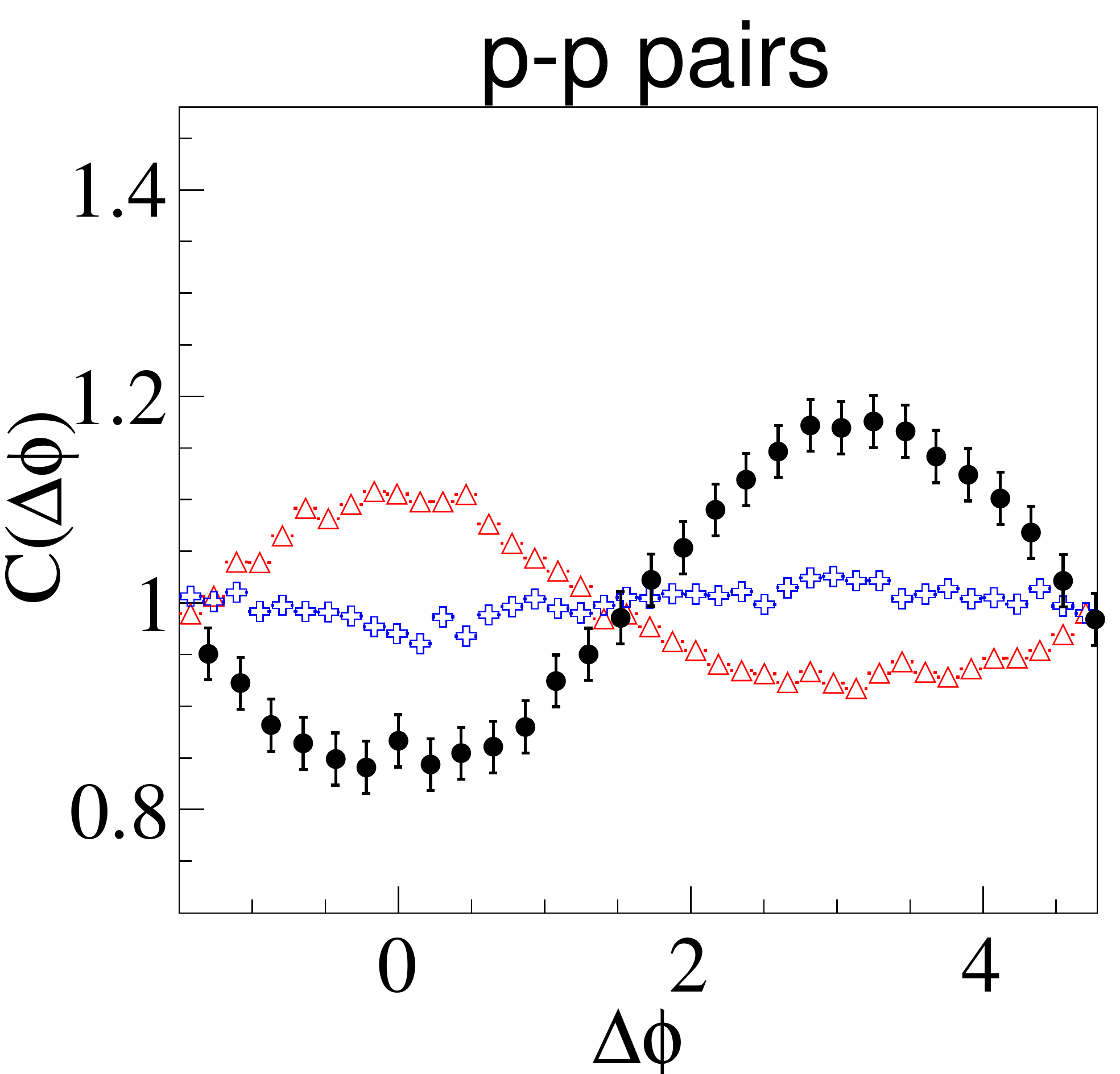}
	\includegraphics[scale=0.24]{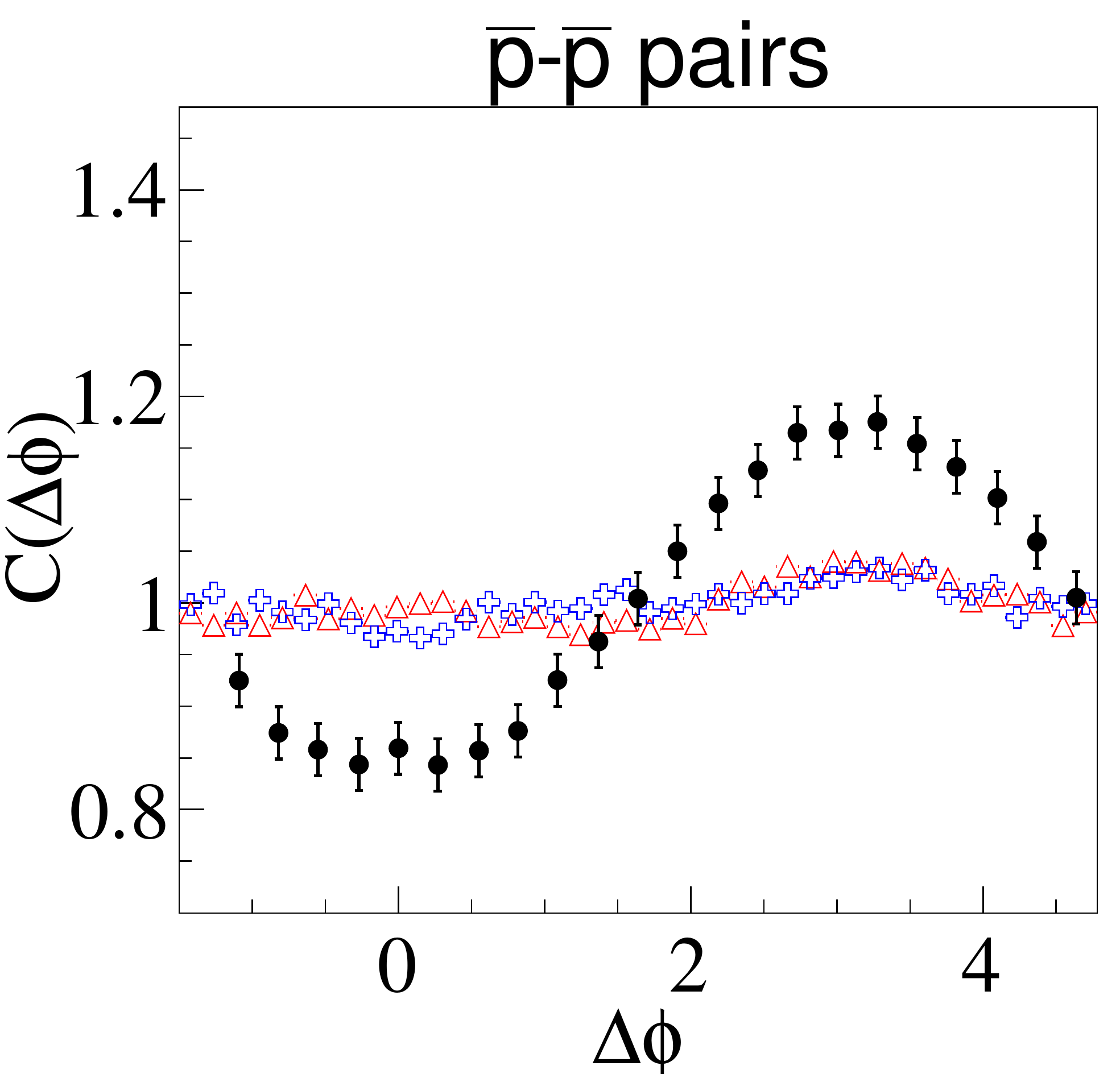}\\
	\includegraphics[scale=0.24]{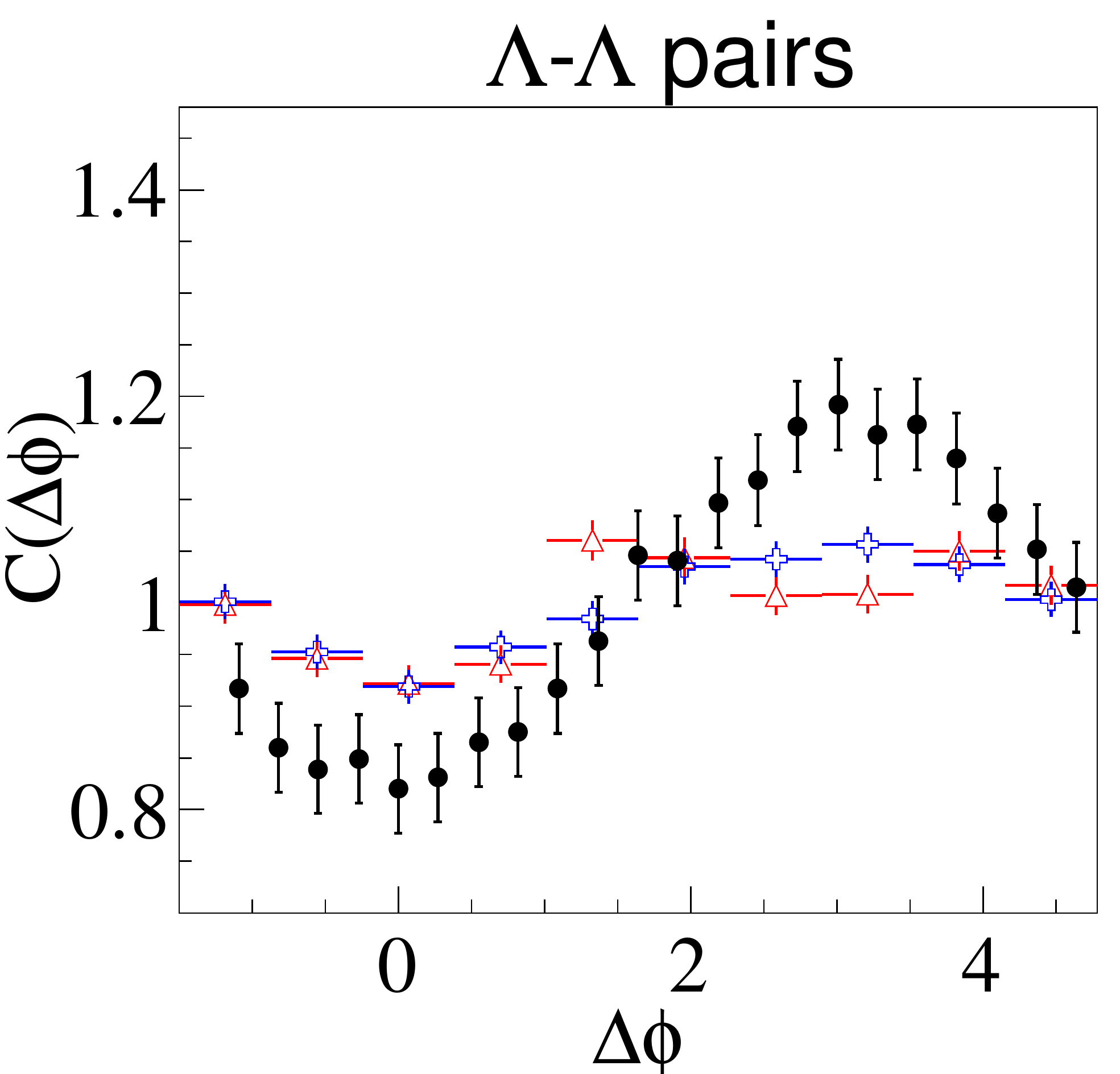}
	\includegraphics[scale=0.24]{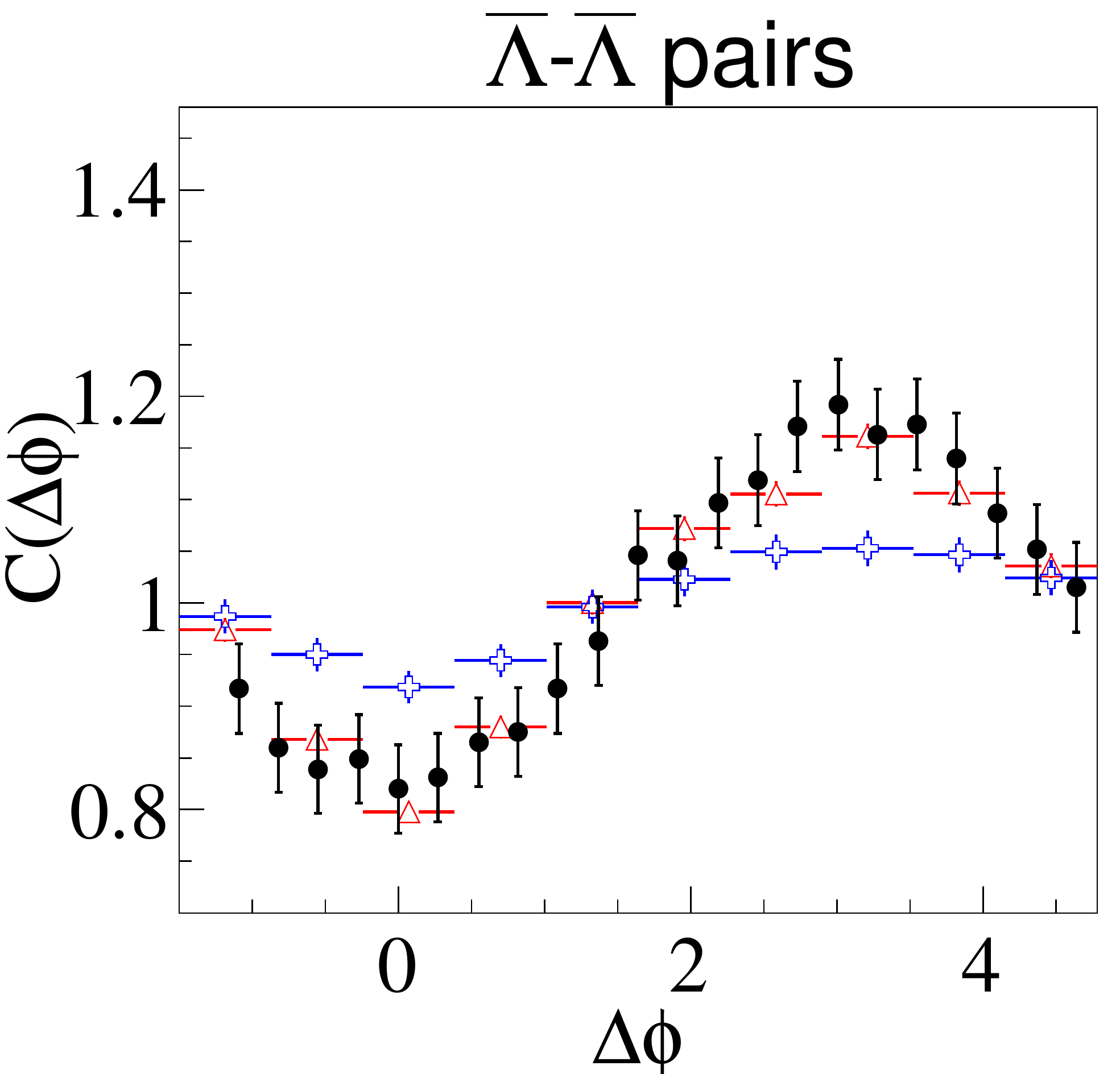}
	\caption{Comparison of two-particle correlations between the old and new quark coalescence in the string melting AMPT model.}
	\label{fig4:1D-CorF-like-sign-newCoal}
\end{figure}

The version of the AMPT-Melting model used until here has some limitations which may affect our results. For example, the current (i.e. ``old'') coalescence component in the AMPT version v2.26t5 forces the numbers of mesons, baryons, and antibaryons in an event to be separately conserved through the quark coalescence process, where only the net-baryon number needs to be conserved. The recent development on a new quark coalescence component~\cite{He:2017} in the AMPT model removes this forced separate conservations in the old quark coalescence model, and it has been shown to provide a better description of baryon productions at LHC energies~\cite{He:2017}. In Fig.~\ref{fig4:1D-CorF-like-sign-newCoal} we compare the correlations from the old and new quark coalescence. The effect between the difference quark coalescence model is tiny on meson-meson correlations. For baryon-baryon correlations, however, the new quark coalescence lead to different results, where the baryon-baryon correlation is now almost the same as the corresponding antibaryon-antibaryon correlation (as expected at this high energy).

\subsection{The transverse momentum dependence of two particle correlations in AMPT model} 
\begin{figure}[!htb]	
	\includegraphics[scale=0.32]{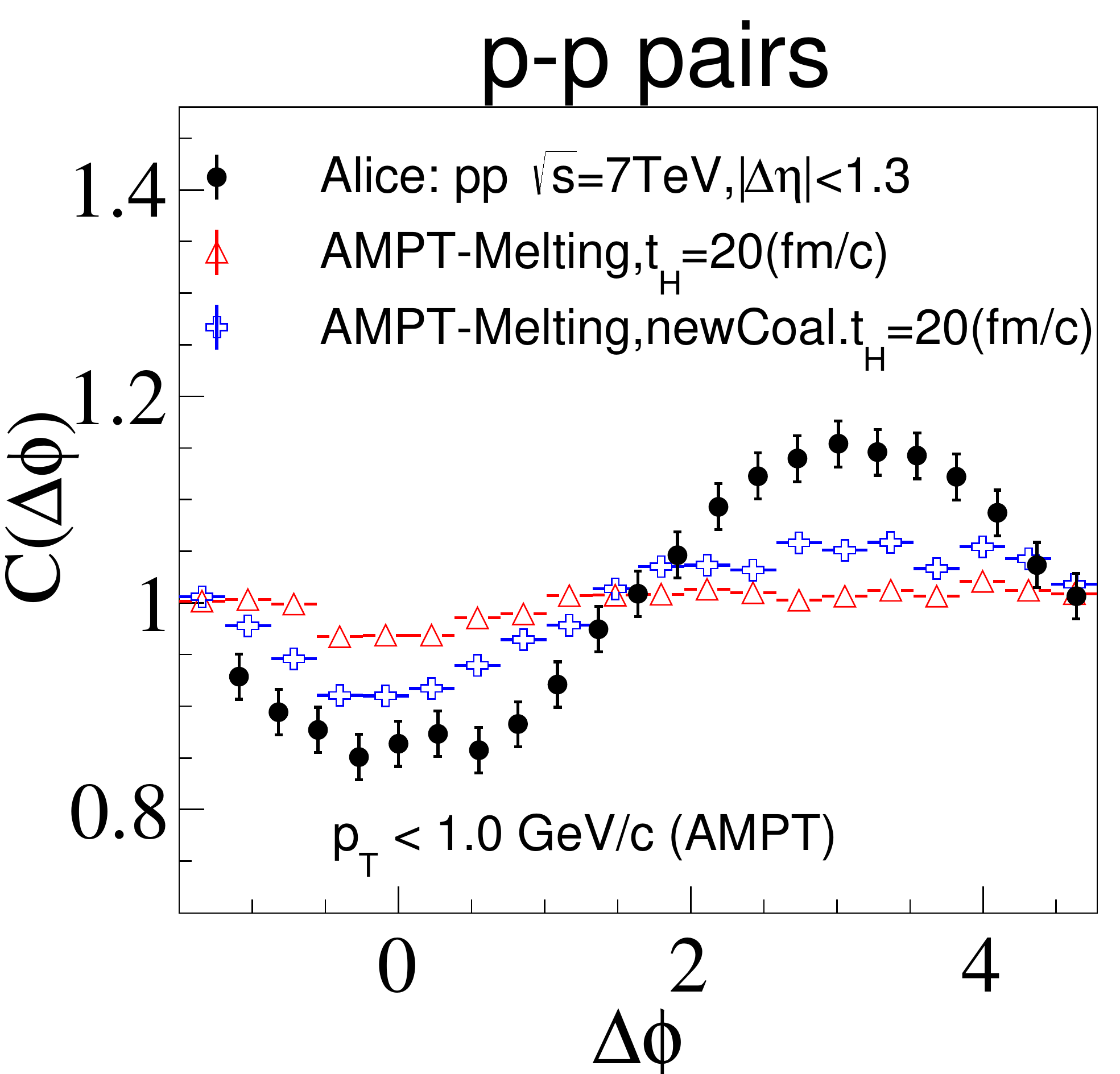}
	\includegraphics[scale=0.32]{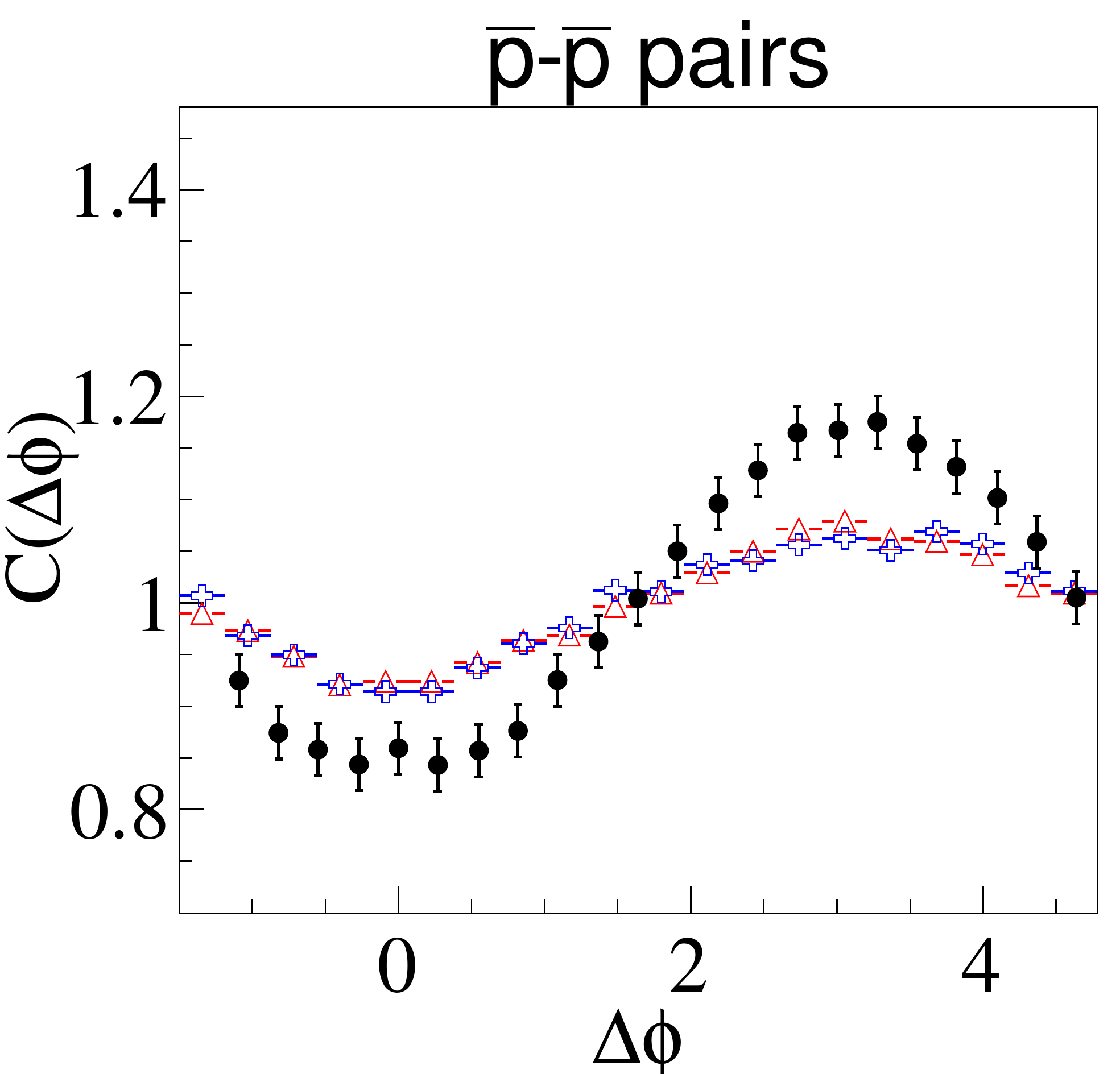}
	\includegraphics[scale=0.32]{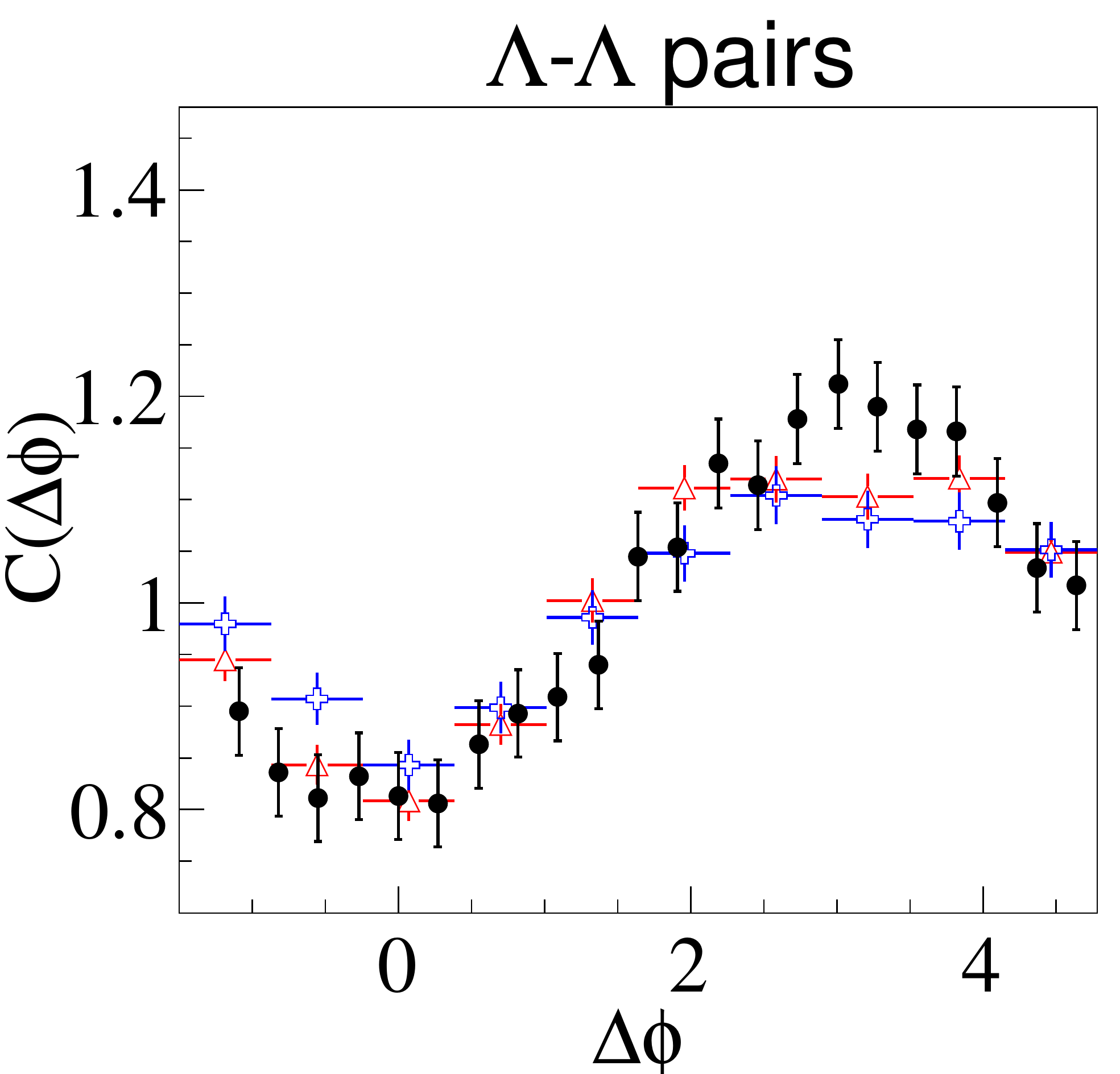}
	\includegraphics[scale=0.32]{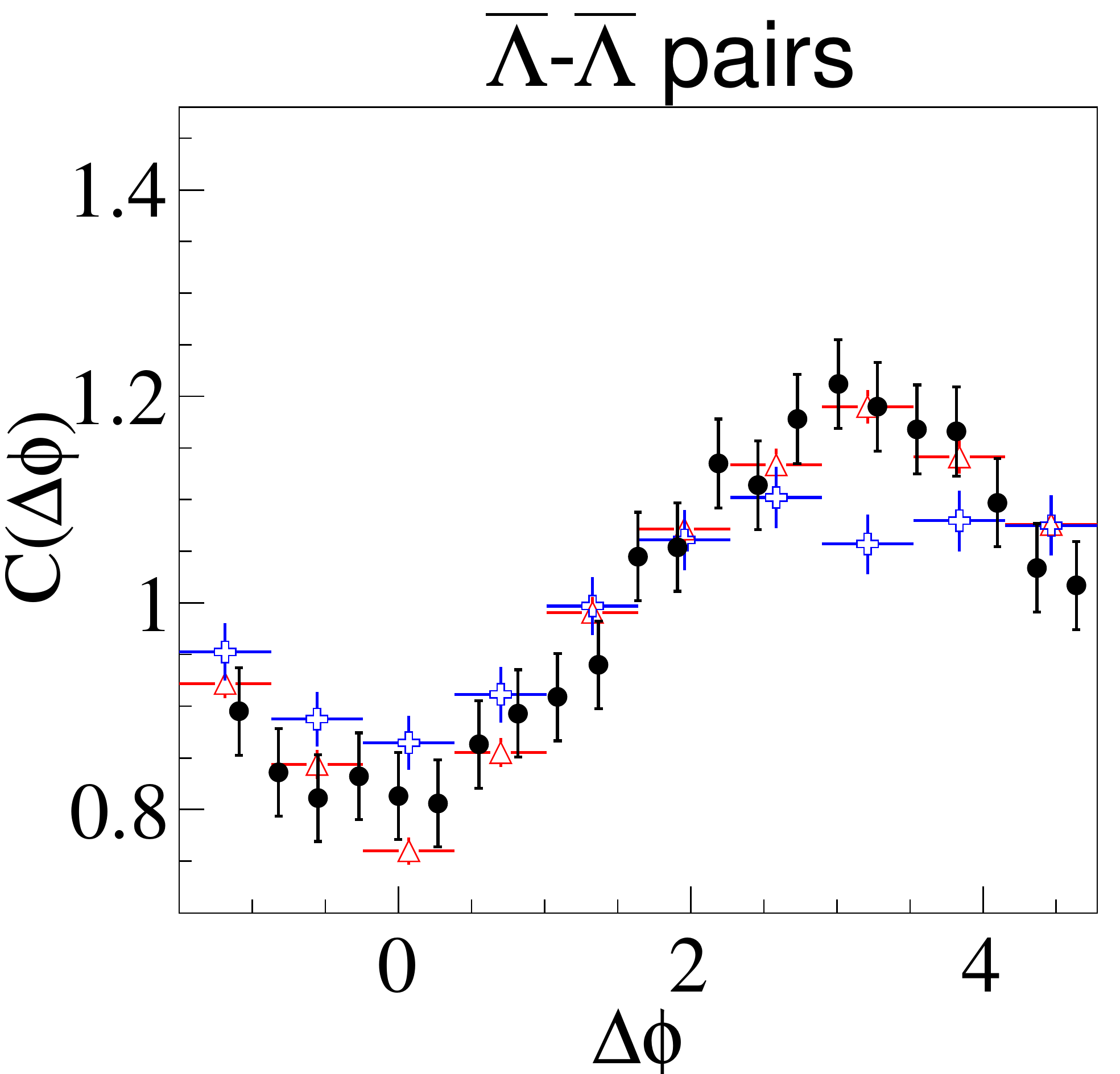}
	\caption{
One-dimensional $\Delta\phi$ correlation functions of p-p, $\bar{p}$-$\bar{p}$, $\Lambda$-$\Lambda$ and $\bar{\Lambda}$-$\bar{\Lambda}$ for $p_T <$ 1.0 GeV/c from the AMPT model. Open symbols represent AMPT calculations with different configurations as illustrated in the figure. Solid points are experiment data~\cite{J.Adam:2017}. }
	\label{fig5:CF-lowpt}
\end{figure}

The two-particle angular correlations may depend on the transverse momentum. We investigate this effect by studying the baryon-baryon correlation functions of baryons with $p_T<$ 1.0 GeV/c in the model calculations, shown in Fig.~\ref{fig5:CF-lowpt}. In comparison to the model results for the full $p_T$ window [c.f. Fig.~\ref{fig4:1D-CorF-like-sign}], clear differences are found. An anti-correlation structure is observed, and proton-proton correlations from the new quark coalescence model are closer to the experimental data.

\subsection{The electric charge dependence of two particle correlations in AMPT model}
\begin{figure}[!htb]	
	\includegraphics[scale=0.32]{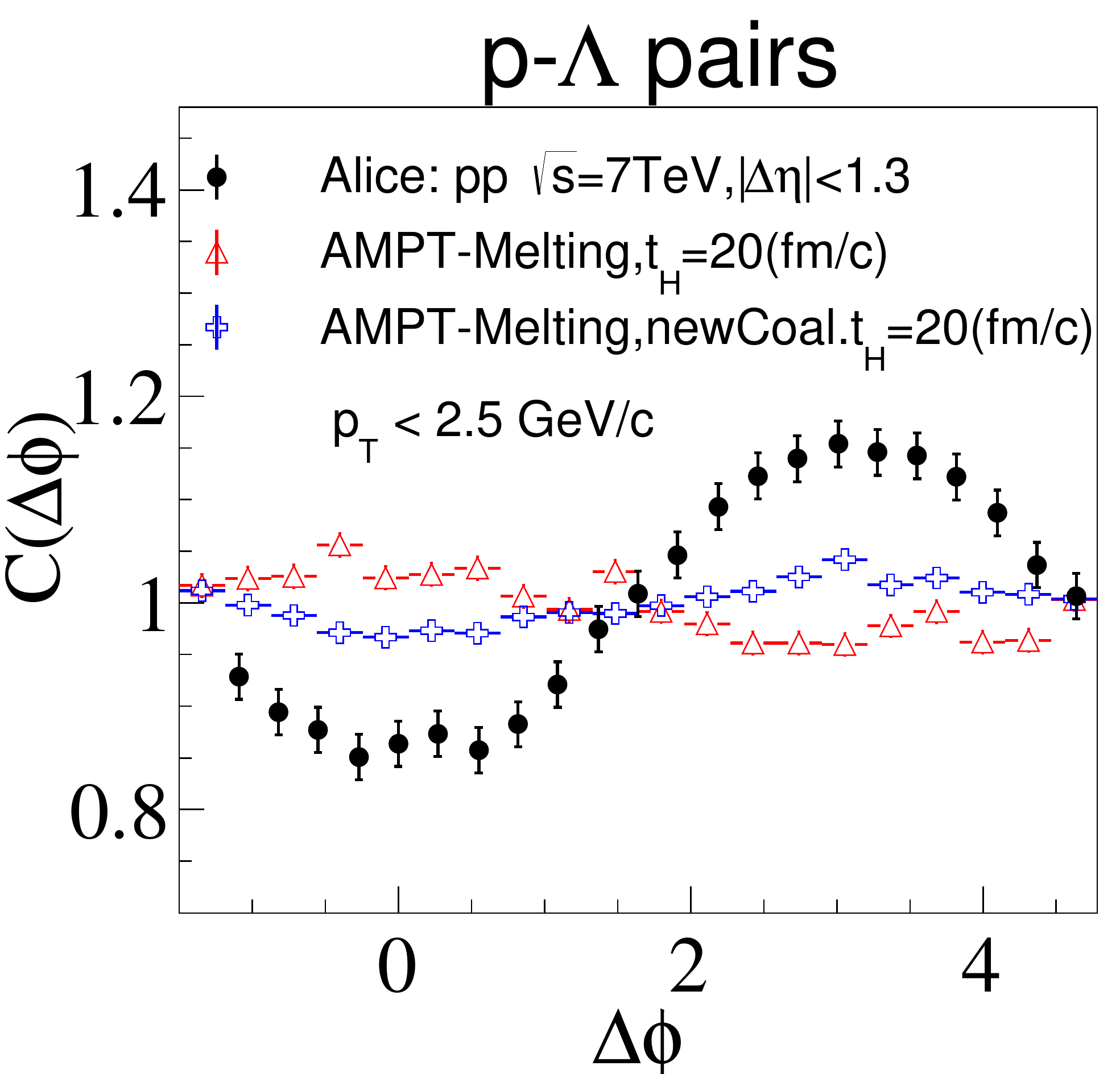}
	\includegraphics[scale=0.32]{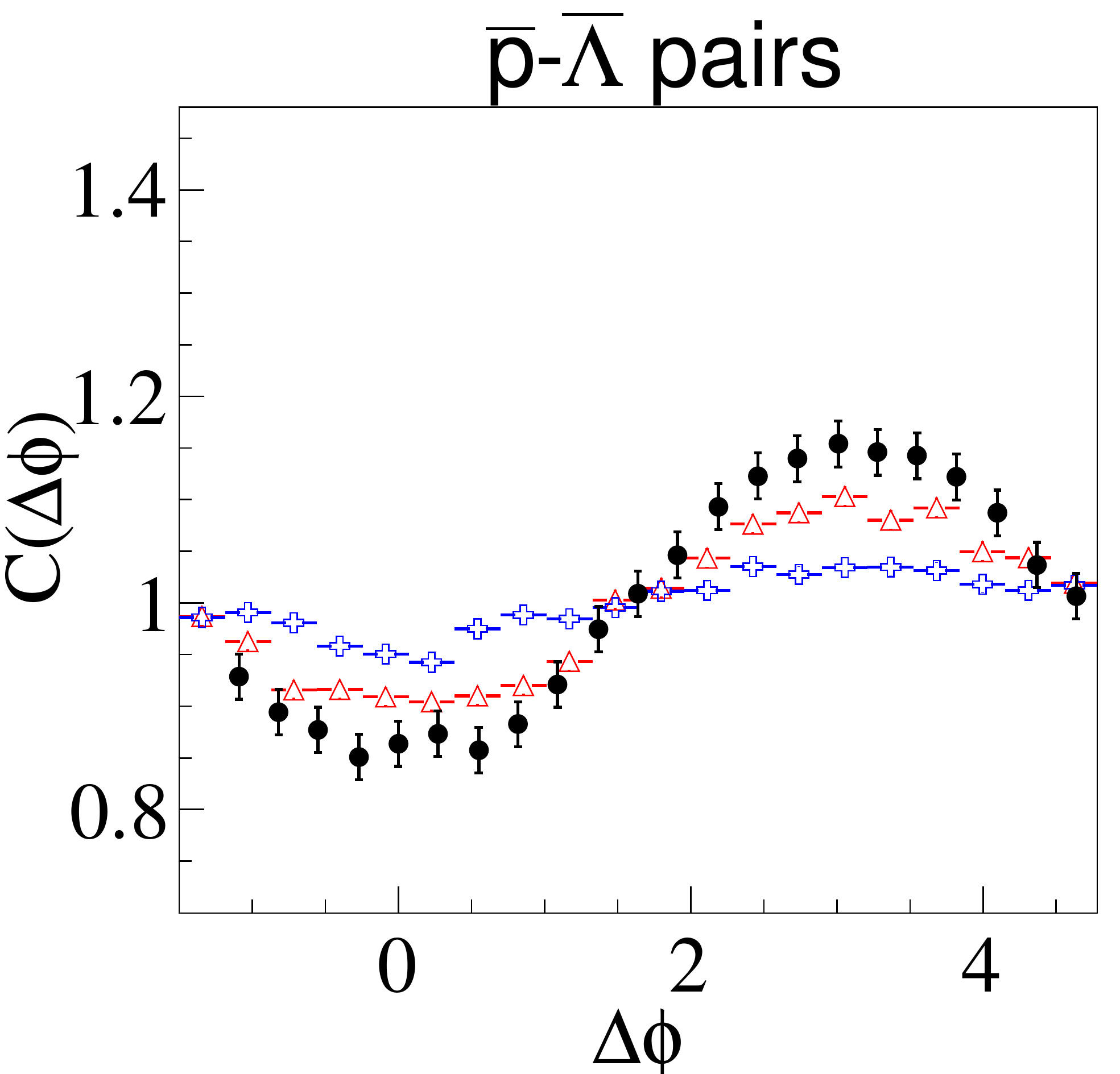}
	\includegraphics[scale=0.32]{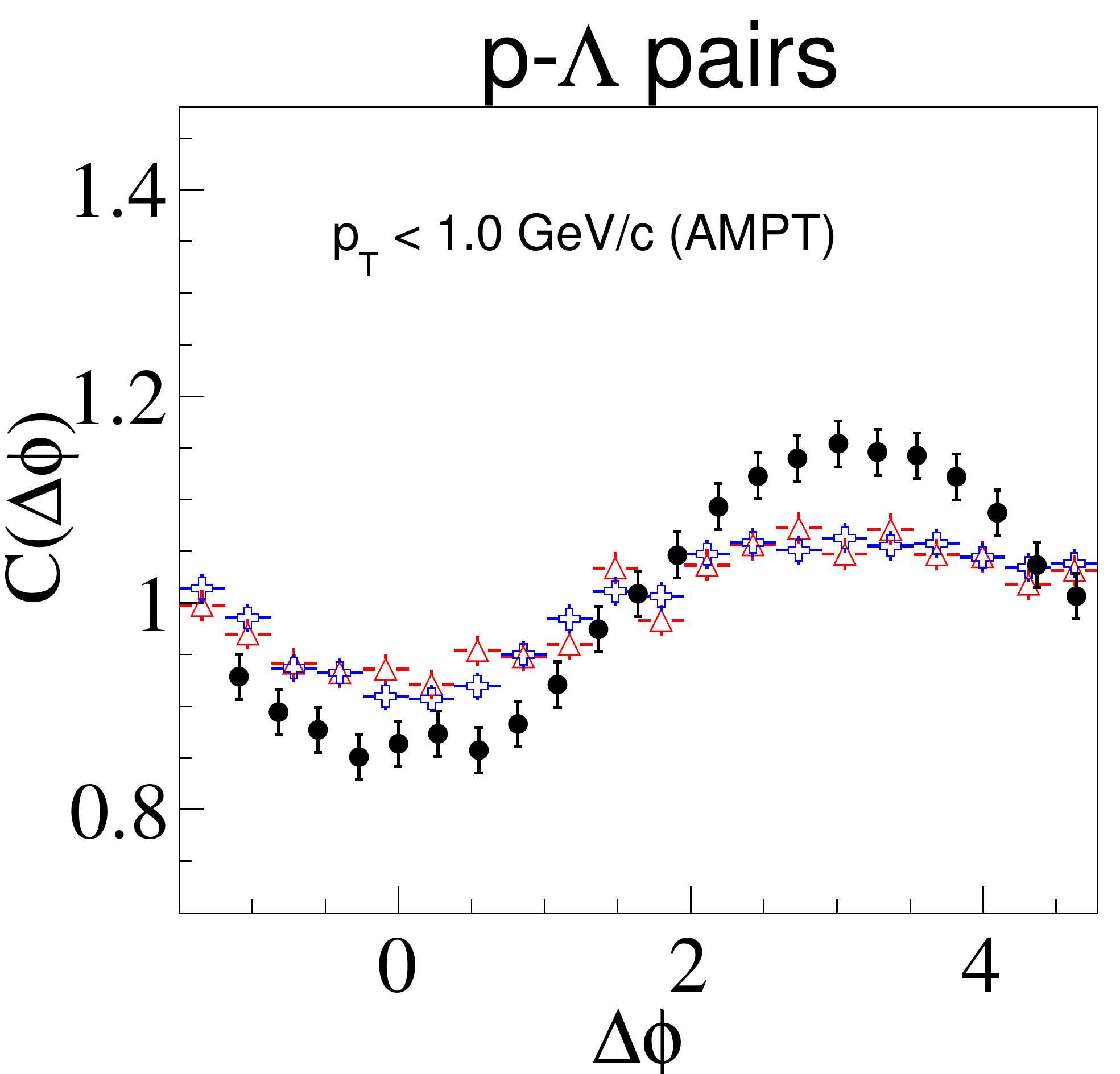}
	\includegraphics[scale=0.32]{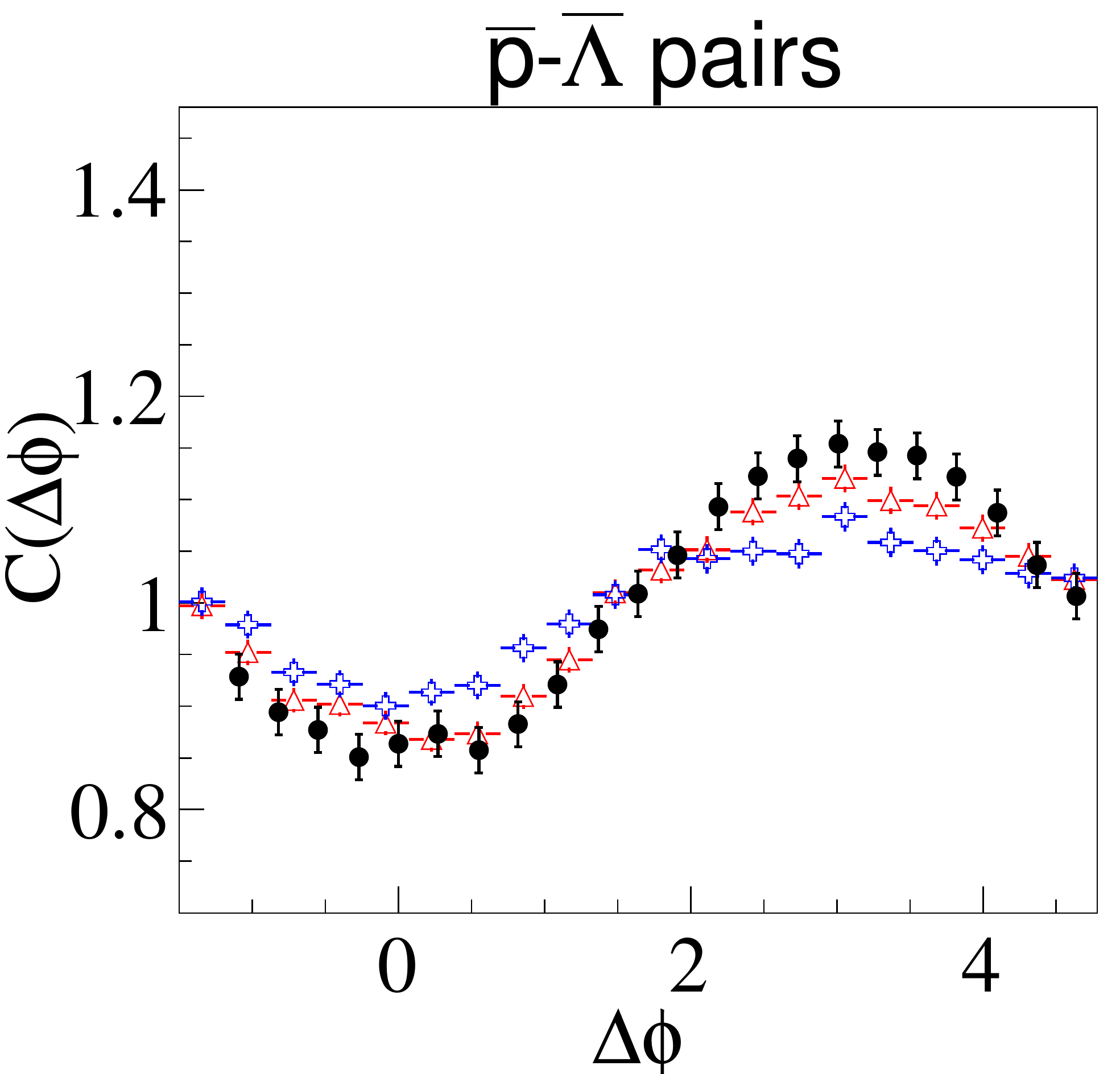}
	\caption{
One-dimensional $\Delta\phi$ correlation functions of p-$\Lambda$, $\bar{p}$-$\bar{\Lambda}$ in comparison with the experimental data for $p_T<$ 2.5 GeV/c (upper panels) and $p_T<$ 1.0 GeV/c (lower panels). Open symbols represent AMPT calculations with different configurations as illustrated in the figure. Solid points are experiment data~\cite{J.Adam:2017}. }
	\label{fig6:CF-p-la}
\end{figure}

The two-particle angular correlations may also depend on the electric charge of the pairs. Figure~\ref{fig6:CF-p-la} presents results from the AMPT-Melting model on p-$\Lambda$ and $\bar{p}$-$\bar{\Lambda}$ correlations in $pp$ collisions at LHC energies. It is seen that the shape of the correlation function is similar to the results present in Fig.~\ref{fig4:1D-CorF-like-sign-newCoal} and Fig.~\ref{fig5:CF-lowpt}. Our results are consistent with the experimental findings, where the depression is a characteristic attribute connected solely to the baryonic nature of a particle~\cite{J.Adam:2017}.
 
\subsection{The parton cross section dependence of two particle correlations in AMPT model}
\begin{figure}[!htb]
\includegraphics[scale=0.28]{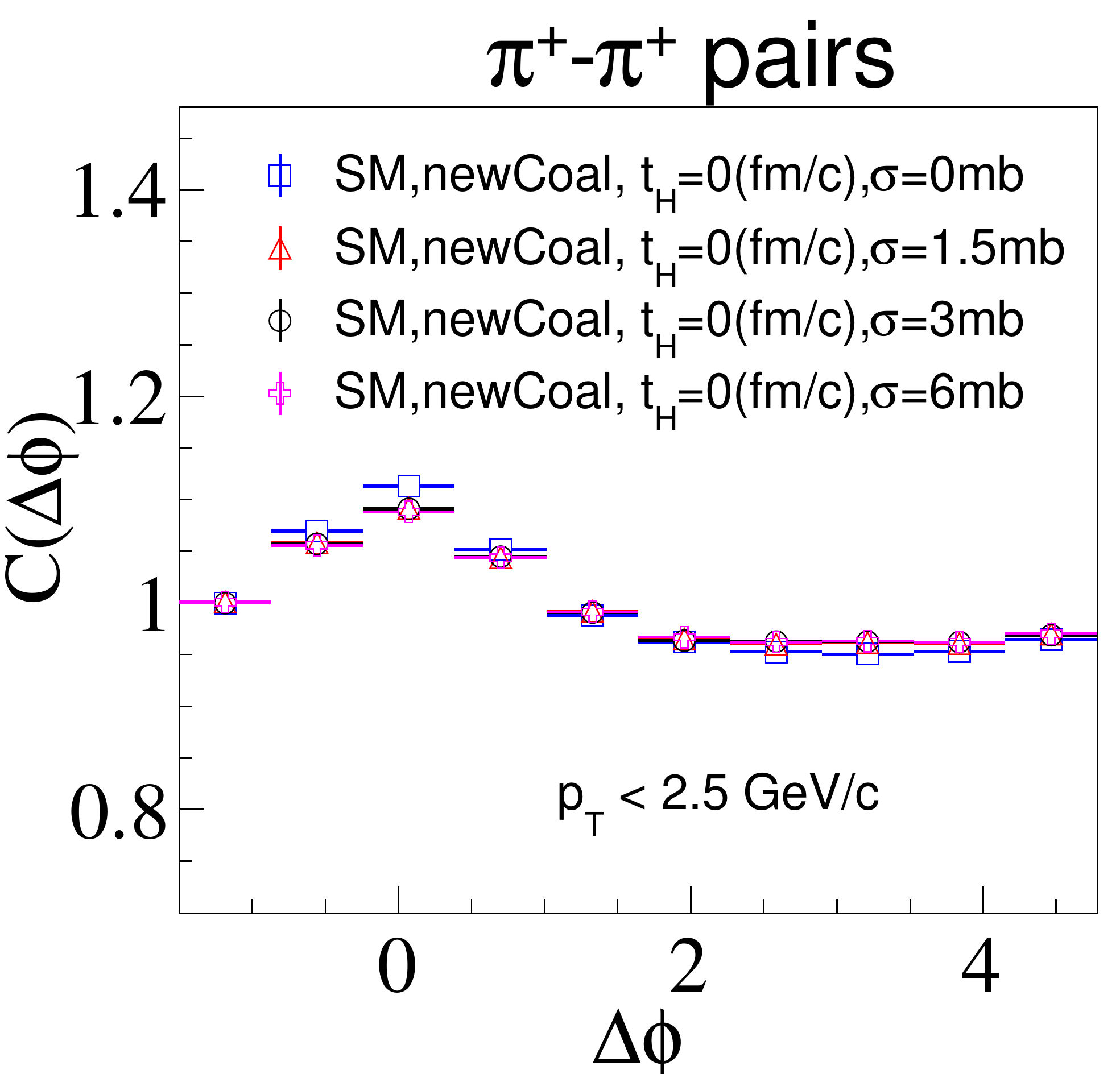}
\includegraphics[scale=0.28]{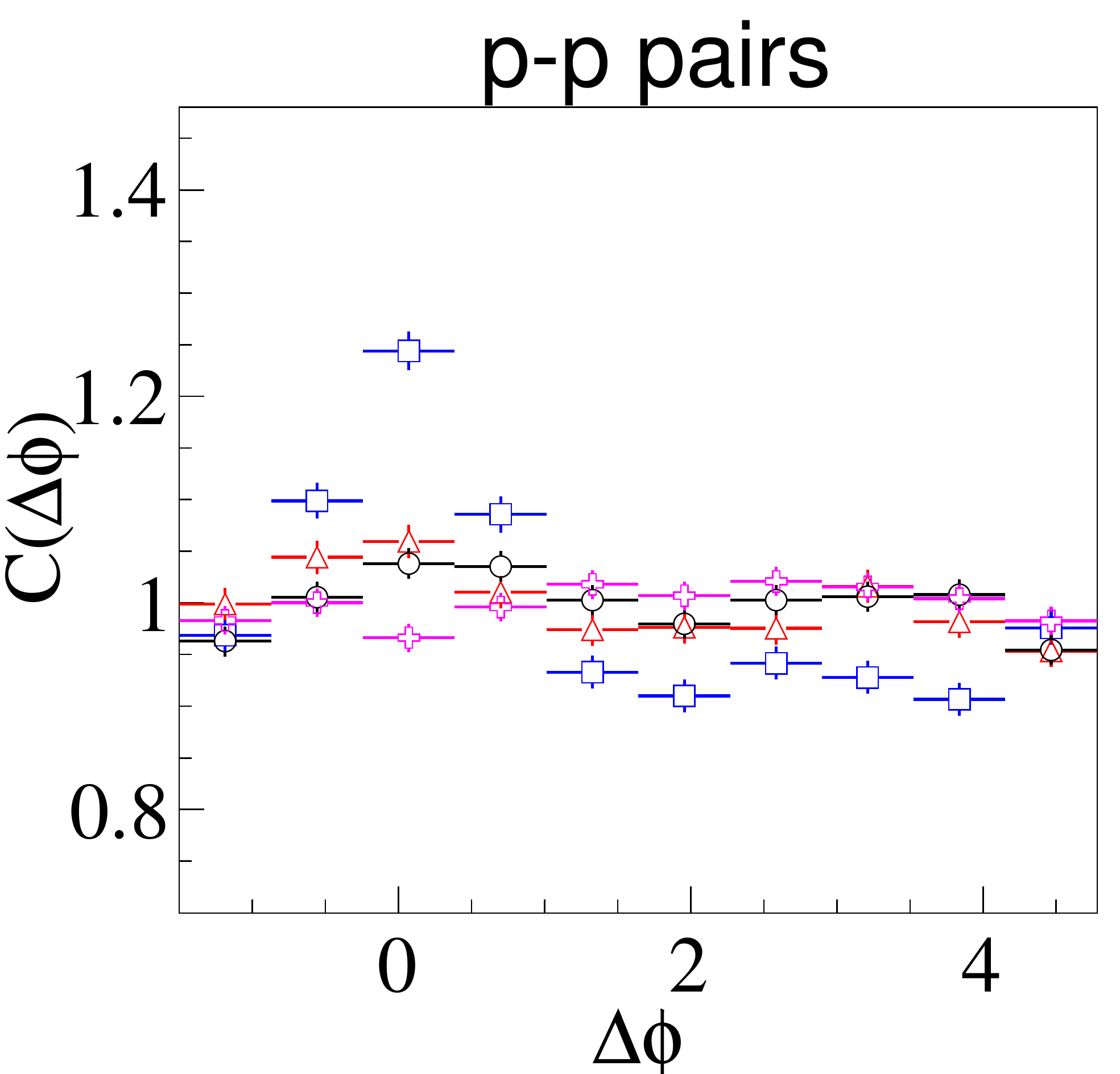}
\includegraphics[scale=0.28]{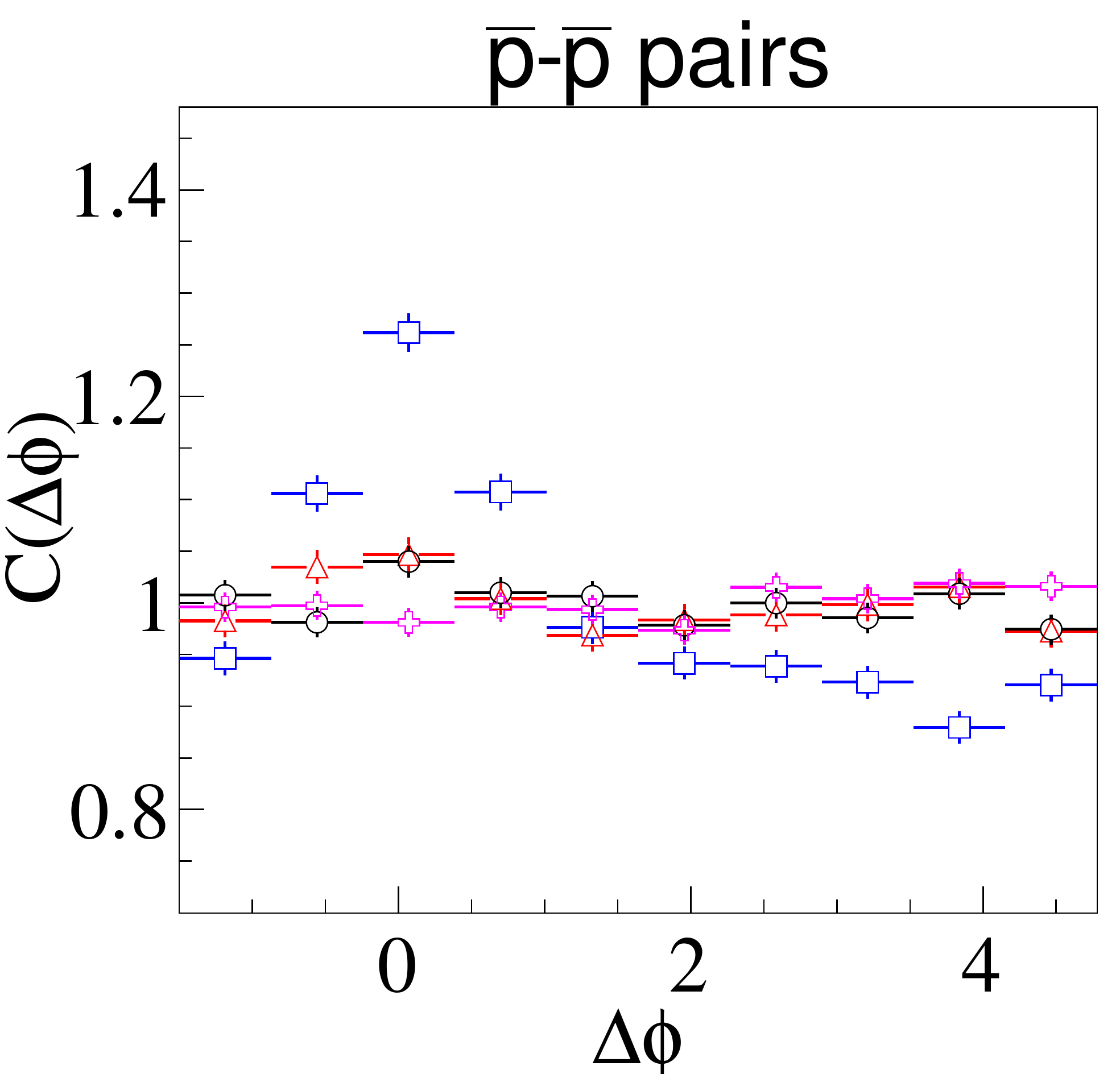}
\caption{Two-particle correlations from AMPT-Melting with the new coalescence at different parton cross sections.}
\label{fig7:CF-cross-section} 
\end{figure}

Next, we address the parton scattering effect on the two-particle correlations by using different parton scattering cross sections in the string melting AMPT model, 
in order to investigate the separate contributions from parton cascade and from quark coalescence hadronization. 
We compare the results for 0 mb, 1.5 mb, 3 mb and 6 mb in Fig.~\ref{fig7:CF-cross-section}, where the 0 mb results represent the hadronization contribution only. 
The parton cross section is seen to have a small effect for $\pi$-$\pi$ correlations but a large effect for both proton-proton and anti-proton-anti-proton correlations; this suggests that parton interactions play an important role in baryon pair correlations at LHC energies.

\subsection{The system size dependence of two particle correlations in AMPT model}
\begin{figure}[!htb]
\includegraphics[scale=0.32]{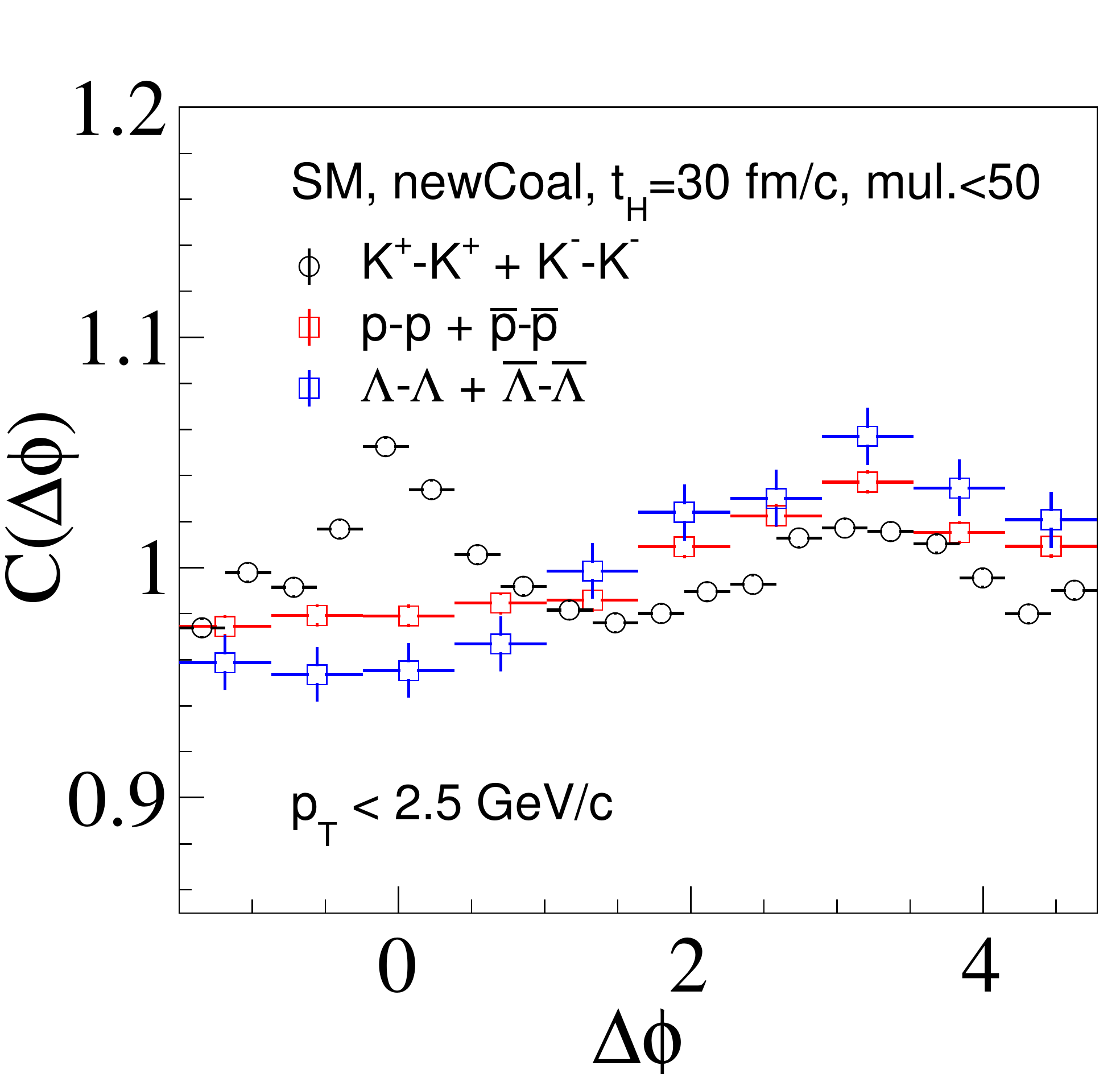}
\includegraphics[scale=0.32]{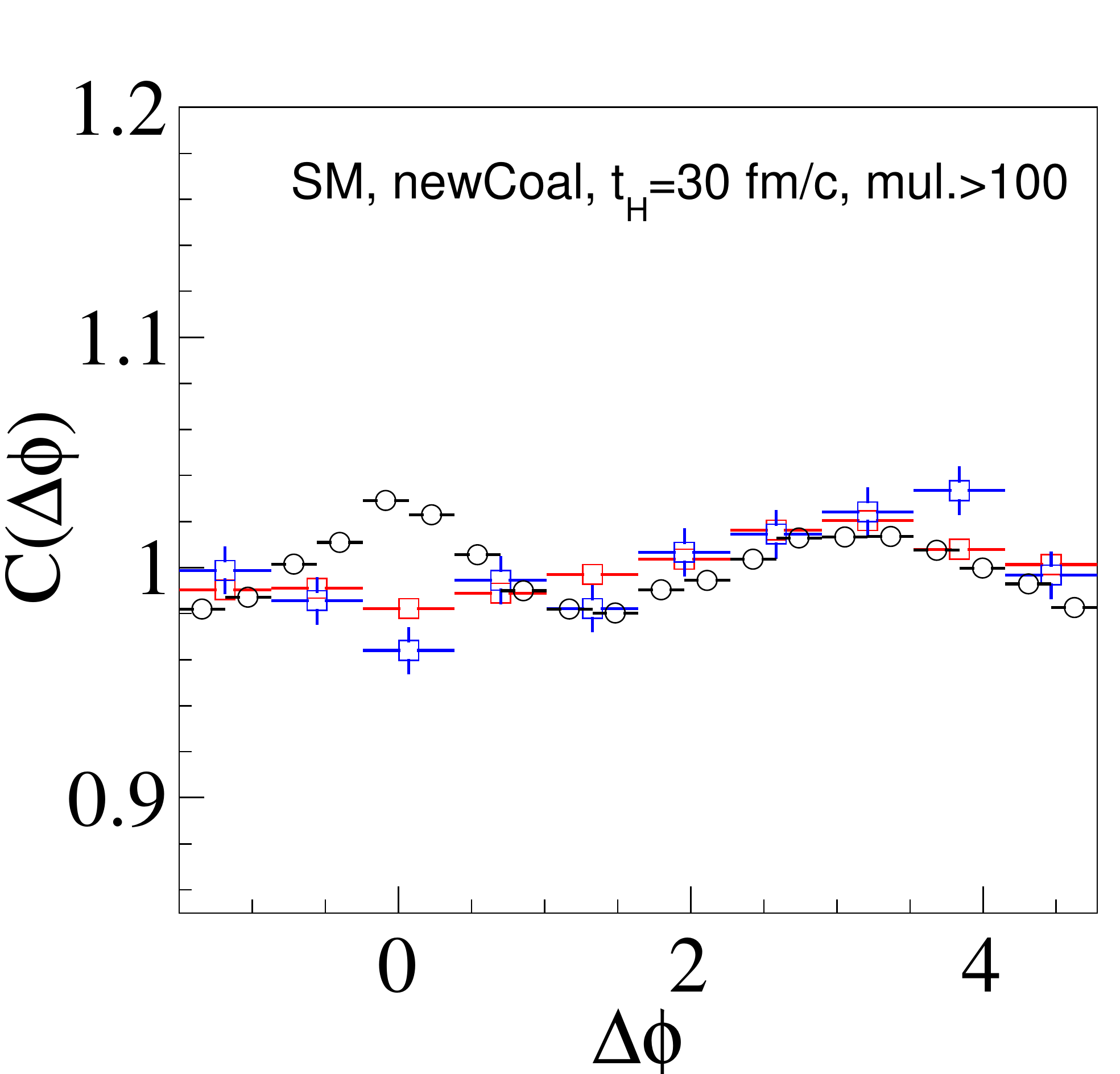}
\caption{Two-particle correlations for the multiplicity interval $<$50 (left panel) and $>$100 in p-Pb collisions at $\sqrt{s}= 5.02$ TeV from AMPT-Melting with the new coalescence.}
\label{fig8:CF-p-Pb} 
\end{figure}

Finally, we extend the study to p-Pb collisions to see whether such a depletion structure of correlations will be present in small systems from $pp$ to p-Pb collisions. Figure~\ref{fig8:CF-p-Pb} shows the same-charge particle pair correlations for mesons and for  baryons in the string melting AMPT with the new quark coalescence. The correlations are shown for a low multiplicity interval and a high multiplicity interval separately, where the parton stage lifetime may be different. We see the usual correlations for the meson pairs but a clear depression on the near side for the baryon pairs. These results indicate that such a depression structure of low $p_T$ baryon pair correlations are present in both $pp$ and p-Pb collisions at LHC energies.

 \section{conclusion}
We have carried out a detailed study of two-particle angular correlations in $pp$ and p-Pb collisions at LHC energies in the framework of a multi-phase transport model, with the focus on understanding the origin of anti-correlation between baryon pairs observed in the experiment. We find that mini-jet and hadronic scatterings are both important components in order to describe the experimental data, especially for the meson pairs. In addition, only the string melting AMPT model can qualitatively describe the near side depression in the angular correlations of baryon pairs, which suggests that quark coalescence and parton scatterings are essential to describe the particle productions in $pp$ collisions at LHC energies. The new quark coalescence model for string melting AMPT improves the description on experimental data. By comparing the correlation results with difference parton scattering cross sections, it is also clear that parton scatterings are important for baryon pair correlations at LHC energies. Extension to p-Pb collisions with the AMPT model predicts similar baryon-baryon correlations as observed in $pp$ collisions.
 
 \section{acknowledgements}
We gratefully acknowledge discussions with Dr. G. L. Ma. This work was supported in part by the Major State Basic Research Development Program in China under Contract No. 2014CB845400 and No. 2015CB856904, the National Natural Science Foundation of China under contract Nos. 11775288, 11421505, 11628508 and 11520101004.

%\end{CJK*}
\end{document}